\documentclass[11pt]{article}
\usepackage[utf8]{inputenc}
\usepackage[T1]{fontenc}
\usepackage[normalem]{ulem}
\pdfoutput = 1

\makeatletter
\DeclareFontFamily{OMX}{MnSymbolE}{}
\DeclareSymbolFont{MnLargeSymbols}{OMX}{MnSymbolE}{m}{n}
\SetSymbolFont{MnLargeSymbols}{bold}{OMX}{MnSymbolE}{b}{n}
\DeclareFontShape{OMX}{MnSymbolE}{m}{n}{
    <-6>  MnSymbolE5
   <6-7>  MnSymbolE6
   <7-8>  MnSymbolE7
   <8-9>  MnSymbolE8
   <9-10> MnSymbolE9
  <10-12> MnSymbolE10
  <12->   MnSymbolE12
}{}
\DeclareFontShape{OMX}{MnSymbolE}{b}{n}{
    <-6>  MnSymbolE-Bold5
   <6-7>  MnSymbolE-Bold6
   <7-8>  MnSymbolE-Bold7
   <8-9>  MnSymbolE-Bold8
   <9-10> MnSymbolE-Bold9
  <10-12> MnSymbolE-Bold10
  <12->   MnSymbolE-Bold12
}{}

\let\llangle\@undefined
\let\rrangle\@undefined
\DeclareMathDelimiter{\llangle}{\mathopen}    {MnLargeSymbols}{'164}{MnLargeSymbols}{'164}
\DeclareMathDelimiter{\rrangle}{\mathclose}               {MnLargeSymbols}{'171}{MnLargeSymbols}{'171}
\makeatother

\usepackage{amsmath}
\usepackage{amssymb}
\usepackage[english]{babel}
\usepackage{bbm}
\usepackage{bbold}
\usepackage{braket}
\usepackage{caption}
\usepackage{color}
    \definecolor{darkgreen}{rgb}{0,0.5,0}
    \definecolor{darkred}{rgb}{0.5,0,0}
    \definecolor{darkblue}{rgb}{0,0,0.6}
    \definecolor{purple}{rgb}{0.4,.2,0.7}
\usepackage{comment}
\usepackage{csquotes}
\usepackage[mathcal]{eucal}
\usepackage{graphicx}
\usepackage[hyperfootnotes = true, colorlinks = true, linkcolor = darkblue, citecolor = purple]{hyperref}
\usepackage{mathabx}
\usepackage{mathtools}
\usepackage{pdfsync}
\usepackage{slashed}
\usepackage{url}
\usepackage{upgreek}
\usepackage[normalem]{ulem}

\numberwithin{equation}{section}

\usepackage[margin = 2.2cm]{geometry}
\usepackage[ragged]{footmisc}
\renewcommand{\d}{\mathrm{d}}
\renewcommand{\i}{\mathrm{i}}

\newcommand{\eff}{\mathrm{eff}}

\newcommand{\cN}{\mathcal{N}}

\DeclareMathOperator{\Tr}{Tr}
\DeclareMathOperator{\tr}{tr}

\DeclareMathOperator{\asinh}{arsinh}
\DeclareMathOperator{\acosh}{arcosh}

\newcommand{\ev}[1]{\left\langle #1 \right\rangle}

\newcommand{\ns}{\mathrm{NS}}
\newcommand{\nstilde}{{\widetilde{\mathrm{NS}}}}
\renewcommand{\r}{\mathrm{R}}
\newcommand{\gh}{\text{gh}}

\newcommand{\dsfixedquantity}{{\kappa}}

\usepackage{subfiles}

\begin{document}

\thispagestyle{empty}
\begin{center}
    ~\vspace{5mm}
    
    {\Large \bf 

        Normalization of ZZ instanton amplitudes in type 0B minimal superstring theory
    }
    
    \vspace{0.4in}
    
    {\bf Vivek Chakrabhavi,$^1$ Dan Stefan Eniceicu,$^2$ Raghu Mahajan,$^2$ and Chitraang Murdia.$^{1}$}

    \vspace{0.4in}

    $^1$ Department of Physics and Astronomy, University of Pennsylvania, Philadelphia, PA 19104, USA \vskip1ex
    $^2$ Department of Physics, Stanford University, Stanford, CA 94305-4060, USA 
    \vspace{0.1in}
    
    {\tt vivekcm@sas.upenn.edu, eniceicu@stanford.edu, raghumahajan@stanford.edu, murdia@sas.upenn.edu}
\end{center}

\vspace{0.4in}

\begin{abstract}
We study ZZ instanton corrections in the $(2,4k)$ $\mathcal{N}=1$ minimal superstring theory with the type 0B GSO projection, which becomes
the type 0B $\mathcal{N}=1$ super-JT gravity in the $k \to \infty$ limit.
Each member of the $(2,4k)$ family of theories has two phases distinguished by the sign of the Liouville bulk cosmological constant.
The worldsheet method for computing the one-loop normalization constant multiplying the instanton corrections gives an ill-defined answer in both phases.
We fix these divergences using insights from string field theory and find finite, unambiguous results.
Each member of the $(2,4k)$ family of theories is dual to a double-scaled one-matrix integral, where the double-scaling limit can be obtained starting either from a unitary matrix integral with a leading one-cut saddle point, or from a hermitian matrix integral with a leading two-cut saddle point.
The matrix integral exhibits a gap-closing transition, which is the same as the double-scaled Gross-Witten-Wadia transition when $k=1$.
We also compute instanton corrections in the double-scaled matrix integral for all $k$ and in both phases, and find perfect agreement with the string theory results.

\end{abstract}

\pagebreak

\tableofcontents

\pagebreak

\section{Introduction}
Many interesting string theory observables like scattering amplitudes contain contributions of order $e^{-1/g_s}$ arising from D-instantons \cite{Polchinski:1994fq, Shenker:1990uf, David:1990sk}.
However, worldsheet perturbation theory around D-instantons is often ill-defined --- the resulting answers contain undetermined constants that need to be fixed using certain external assumptions, like the existence of a gauge theory dual or superstring duality. 

Recently, there has been much progress in precise computations of non-perturbative effects in two-dimensional string theory arising from ZZ branes \cite{Zamolodchikov:2001ah}, pioneered by the work of Balthazar, Rodriguez and Yin \cite{Balthazar:2019rnh},
and of Sen \cite{Sen:2019qqg, Sen:2020eck, Sen:2021qdk}.
A remarkable insight of these computations is that one needs string field theory\footnote{See Ref.~\cite{Sen:2024nfd} for a comprehensive review highlighting modern developments.} to fix the ambiguities and divergences that arise in the worldsheet computations \cite{Sen:2019qqg, Sen:2020eck, Sen:2021qdk}.
After fixing the calculations using insights from string field theory, one finds perfect agreement with the corresponding results from the dual matrix quantum mechanics \cite{Balthazar:2019rnh, Balthazar:2019ypi, Sen:2019qqg, Sen:2020eck, Sen:2021qdk}, which serves as an important check of the string field theory methods.
These results have been extended to the type 0B supersymmetric version of two-dimensional string theory \cite{Balthazar:2022apu, Chakravarty:2022cgj, Sen:2022clw}, and to include sine-Liouville deformations in the bosonic version \cite{Alexandrov:2023fvb}.

Toy models are valuable because the lessons that they teach us can turn out to be very general.
It is testament to this philosophy that the original ideas in Refs.~\cite{Sen:2019qqg, Sen:2020eck, Sen:2021qdk}, which were developed in the context of two-dimensional string theory, have been generalized to tackle various nonperturbative effects in critical 10-d superstring theory and its compactifications \cite{Sen:2021tpp, Sen:2021jbr, Alexandrov:2021shf, Alexandrov:2021dyl, Alexandrov:2022mmy, Agmon:2022vdj}, leading to highly nontrivial checks of superstring dualities.
For instance, Ref.~\cite{Sen:2021tpp} computed the one-loop constant multiplying the D-instanton correction to the 10-d flat space four-graviton amplitude in type IIB string theory from first principles, and found precise agreement with the S-duality prediction encoded in the formula of Green and Gutperle \cite{Green:1997tv}.

As far as the fundamental peculiarity of string perturbation theory around D-instantons is concerned, one can simplify the problem even further and consider minimal string theory \cite{Seiberg:2004at} instead of two-dimensional string theory. 
Like the $c=1$ theory, the worldsheet description of the minimal string theory consists of a Liouville CFT and the $bc$ ghost system, but the matter sector is a minimal model with $c<1$ instead of a $c=1$ boson --- hence the added simplicity.
The normalization of ZZ instanton contributions to the partition function also suffers from divergences in these simpler theories \cite{Martinec:2003ka, Kutasov:2004fg}, and can be fixed using the same string field theory ideas, as shown in Refs.~\cite{Eniceicu:2022nay, Eniceicu:2022dru}.
The resulting answers agree perfectly with those of the dual matrix integrals.

Emphasizing the value of simplifying toy models, we note in the present context that an apparent mismatch found between two-dimensional string theory and matrix quantum mechanics in Ref.~\cite{Sen:2020eck} was resolved by first analyzing the analogous problem in the simpler minimal string/matrix integral setting \cite{Eniceicu:2022xvk}.
Thus, we believe that it is valuable to explore low-dimensional string theory as much as possible. 

Continuing the above thread of ideas, in this work, we analyze the normalization of instanton corrections to the partition function of the type 0B minimal superstring \cite{Klebanov:2003wg, Seiberg:2003nm, Seiberg:2004ei}, exclusively focusing on the $(2,4k)$ series.
Here, $k$ is any positive integer, and the Liouville sector becomes semi-classical in the $k \to \infty$ limit.
Apart from the presence of worldsheet supersymmetry, a qualitatively new feature of these theories compared to the bosonic minimal string is that they have two different phases depending on the sign of the Liouville cosmological constant $\mu$.
The annulus diagrams between super-ZZ branes are divergent and need to be treated using insights from string field theory.
We analyze and fix all putative divergences in both phases, getting finite and unambiguous results for the normalization of non-perturbative corrections. (See also \cite{Kawai:2004pj, Fukuma:2007qz} for some previous work.)

The string theories considered in this paper are dual to a double-scaled one-matrix integral in the $\upbeta = 2$ Dyson class, where the double-scaling limit can be taken either using a one-cut unitary matrix integral, or using a two-cut Hermitian matrix integral \cite{Klebanov:2003wg, Seiberg:2003nm}.
These matrix integrals also exhibit two phases, separated by a gap-closing phase transition in the eigenvalue density, the simplest example of which, at $k=1$, is the double-scaled version of the Gross-Witten-Wadia transition \cite{Gross:1980he, Wadia:1980cp, Wadia:2012fr}.
We generalize the matrix computation of non-perturbative corrections to the partition function from the $k=1$ case \cite{Marino:2008ya, Ahmed:2017lhl, Eniceicu:2023cxn} to general $k$ in both phases, and find perfect agreement with the dual string theory results.

The structure of the paper is as follows: in section \ref{sec:matrixintegral}, we introduce the relevant matrix integrals, specify the spectral curves that are needed to match with the conformal background of the dual string theory and compute the leading instanton corrections to the partition function in both phases.
Many of the details are deferred to appendix \ref{app:weak} for the gapped phase and to appendix \ref{app:strong} for the ungapped phase.
Section \ref{sec:reviewbdrystates} is a review of the necessary string theory worldsheet technology which includes the description of boundary states in $\mathcal{N}=1$ Liouville theory.
Section \ref{sftinput} explains the input from string field theory that will be used in later sections to fix the artificial divergences in the annulus amplitudes.
In section \ref{sec:stringgapped}, we compute the instanton corrections on the string theory side of the duality in the gapped phase, and in
section \ref{sec:stringungapped}, we do the same for the ungapped phase.
We conclude in section \ref{sec:discussion} with a summary and brief discussion.
We comment on the relationship with super-JT gravity which arises in the $k \to \infty$ limit and on the difficulty of approaching this limit in the gapped phase.

\section{Matrix Integral Computations}
\label{sec:matrixintegral}

We will be interested in double-scaled matrix integrals that are dual to $\mathcal{N}=1$ minimal superstring theory with the type 0B GSO projection.
The double-scaled matrix integrals were studied in
Refs.~\cite{Periwal:1990gf, Periwal:1990qb, Nappi:1990bi, Crnkovic:1990mr, Crnkovic:1992wd, Hollowood:1991xq}\footnote{The double-scaled matrix integrals that are dual to type 0A minimal superstring theory arise from a complex matrix $M$ and a potential of the form $V(M^\dagger M)$. They were studied in the past in Refs.~\cite{Morris:1990cq, Dalley:1991qg, Dalley:1991vr, Dalley:1991yi, Dalley:1992br}. More recent non-perturbative studies are Refs.~\cite{Johnson:2020heh,Johnson:2020exp,Johnson:2020mwi}.}, and the duality to minimal superstring theory was established in Refs.~\cite{Klebanov:2003wg, Seiberg:2003nm}.
These developments happened almost in parallel to the discovery of the duality between two-dimensional type 0 string theory and matrix quantum mechanics \cite{Douglas:2003up, Takayanagi:2003sm}.

We will focus on the $(2,4k)$ family of type 0B minimal superstring theories, where $k$ is a positive integer.
The matrix duals of these theories consist of integrals over the entries of a single matrix.
The double-scaling limit can be obtained starting either with a two-cut Hermitian matrix integral or with a one-cut unitary matrix integral.

Consider the following matrix integral over $N \times N$ unitary matrices $U$,
\begin{equation}
    Z_k(N,g) := \int \frac{\d U}{\text{vol}~ U(N)} \,\exp \left( \frac{N}{g} W_k(U) \right) \, .
    \label{defzknt}
\end{equation}
Here, $g$ denotes the 't Hooft coupling and the potential is of the form,
\begin{equation}
    W_k(U) = \sum_{l=1}^{k} \frac{1}{l} \left( t^{+}_{l} \tr U^l + t^{-}_{l} \tr U^{-l} \right) \, .
    \label{defvku}
\end{equation}
Such unitary matrix integrals have a phase transition as a function of the 't Hooft coupling $g$. 
In the weak-coupling (small-$g$) phase, the support of the eigenvalue density --- often referred to as the ``cut'' --- is a proper subset of the unit circle, and in the strong-coupling (large-$g$) phase, the support of the eigenvalue density is the full circle.
When $k=1$, i.e.~when the potential is proportional to $\Tr U + \Tr (U^{-1})$, this integral is related to the path integral of 2D lattice QCD, and the phase transition is the Gross-Witten-Wadia transition \cite{Gross:1980he, Wadia:1980cp, Wadia:2012fr}.

In the double-scaling limit, we send $N\rightarrow\infty$ while tuning the parameters $t_l^\pm$ and $g$ in the potential to approach the phase transition point in a coordinated way in order to land on the desired spectral curve.
Let us now specify the spectral curves that we need in order to match to the conformal background of $(2,4k)$ type 0B minimal superstring theory.

If we approach the double-scaling limit from the weak-coupling phase, the spectral curve is the two-sheeted Riemann surface \cite{Seiberg:2003nm} defined by
\begin{align}
    y^2 = (1-x^2) \, \left( U_{2k-1}(x) \right)^2\, ,
    \label{spectralcurveweak}
\end{align}
where $U$ denotes the Chebyshev polynomial of the second kind.\footnote{It is understood throughout the paper that $y$ and $V_\eff$ carry an overall large multiplicative factor appropriate for the disk topology (labeled $\kappa$ in appendices \ref{app:weak} and \ref{app:strong}, with $\kappa \sim e^{S_0} \sim g_s^{-1}$). 
We will not write this overall factor explicitly in the main text.}
One should think of $x$ as parameterizing the locations of the eigenvalues of the matrix after the double-scaling limit has been taken.
The density of states is proportional to the branch cut discontinuity in $y$, 
\begin{equation}
\label{dosweak}
    \rho(x) \, \propto \, \sinh \left( 2k \acosh |x| \right) \, , 
    \qquad
    x \in (-\infty,-1] \cup [1,\infty) \, .
\end{equation}

Instead, if we approach the double-scaling limit from the strong-coupling phase, we get a spectral curve that is a reducible algebraic curve, i.e.~one whose defining equation is a product of two factors.
This is related to the fact that in the strong-coupling phase of unitary matrix integrals, where the support of the eigenvalue density is the full circle, there are two resolvent functions, defined starting from outside and inside the unit circle, respectively, and which are not analytic continuations of each other.
Explicitly, the spectral curve is \cite{Seiberg:2003nm}
\begin{align}
\label{spectralcurvestrong}
    y^2 &= - T_{2k}(\i x)^2\, ,
\end{align}
where $T$ denotes the Chebyshev polynomial of the first kind, and thus, the two components are
\begin{equation}
    y^\pm=\pm\i(-1)^k\,T_{2k}(\i x)\,.
\end{equation}
The factor of $(-1)^k$ is convention dependent.
The density of states is supported on the entire real axis in the complex $x$-plane and is proportional to $\i (y^- - y^+)$:
\begin{align}
    \rho(x) \, \propto \, \cosh (2k \, \asinh (x))\, ,\quad x \in \mathbb{R}\, .
\end{align}
In the $k \to \infty$ limit, the duality we are discussing becomes the duality between type 0B $\mathcal{N}=1$ super-JT gravity and a corresponding matrix integral \cite{Stanford:2019vob}.
See also Refs.~\cite{Johnson:2021owr, Rosso:2021orf, Mertens:2020pfe}.
In this limit, the density of states is $\rho(x) \,\propto \, \cosh(x)$, after a $k$-dependent rescaling of $x$.\footnote{
In the context of super-JT gravity, it is usual to discuss the density of states of the matrix $H$, where $H = Q^2$. 
In the dual random matrix ensemble, $Q$ is thought of as a hermitian supercharge without any particular sub-block structure, and $H$ is the Hamiltonian.
The variable $x$ is the eigenvalue of $Q$, and so the eigenvalue of $H$ is $x^2$.
For the type 0B version, the two-dimensional spin manifolds that are summed over in the path integral are not weighted by any spin-structure dependent topological weighting factor \cite{Stanford:2019vob}.
The corresponding $\mathcal{N}=1$ supersymmetric SYK model \cite{Fu:2016vas} has an \emph{odd} total number of fermions.
}

We want to study non-perturbative contributions to the partition function (\ref{defzknt}),
\begin{align}
    Z &= Z^{(0)} + Z^{\text{np}} + \ldots \\
    \frac{Z^{\text{np}}}{Z^{0}} &= e^{-T} \, \mathcal{N} \, \left( 
    1 + O (g_s)
    \right)\, .
\end{align}
Here, $Z^{(0)}$ is the perturbative contribution to the partition function and $Z^{\text{np}}$ is a particular non-perturbative contribution from a different saddle point, with an order-one number of eigenvalues moved out of the cut.
The quantity $T$ is the difference between on-shell actions of the perturbative saddle point and the second saddle point under consideration.
The quantity $\mathcal{N}$ is the one-loop Gaussian integral around the second saddle point.
On the string theory side, $T = O(g_{s}^{-1})$ is the action of the (collection of) ZZ instantons that correspond to the particular matrix saddle point. 
It is computed using an empty disk diagram, while $\mathcal{N}$ is the exponential of the (appropriate sum of) annulus diagrams.

Below, we will be interested in considering ratios of quantities in which the $g_s$-dependence drops out, since the relative normalization between the genus counting parameters on the matrix side and the string theory side is arbitrary.
Considering $g_s$-independent ratios allows us to compare pure order-one numbers on both sides of the duality.

\subsection{Instanton effects in the gapped phase of the matrix integral}

Before double-scaling, the gapped (weak-coupling) phase corresponds to the phase of the large-$N$ unitary matrix integral where the eigenvalues cover only part of the unit circle.

\begin{figure}[t]
    \centering
    \includegraphics[width=0.36\textwidth]{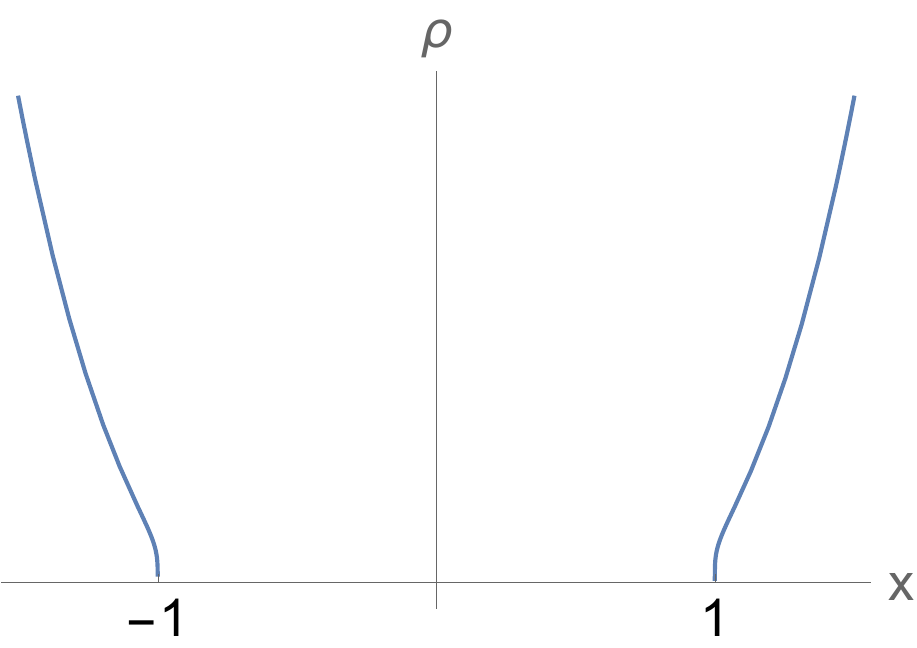}\hspace{0.15in}
    \includegraphics[width=0.44\textwidth]{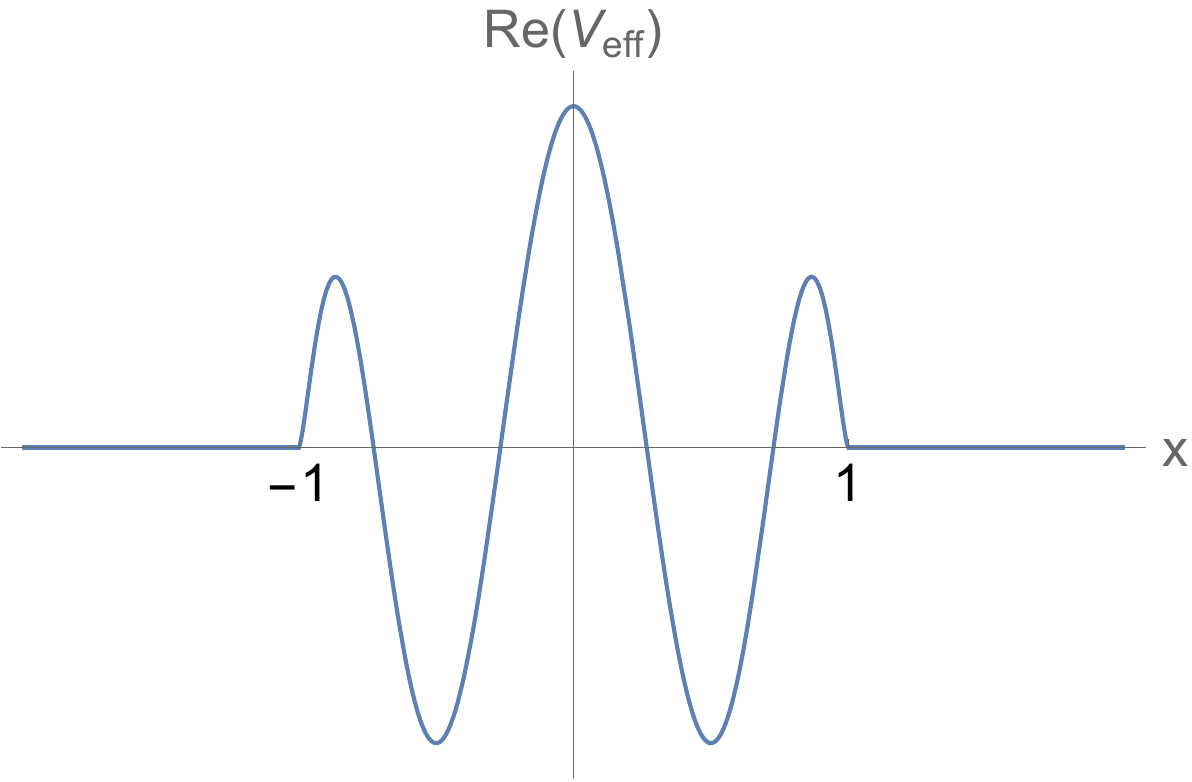}
    \caption{The density of states (schematic) and the effective potential for the $k=3$ model in the gapped phase. The two cuts lie along the intervals $(-\infty, -1]$ and $[1,\infty)$ where the density of eigenvalues is supported. The extrema of $V_\eff$ represent the locations of the $2k-1$ one-eigenvalue instantons, with one of them located at the $\mathbb{Z}_2$ symmetric point $x=0$, and the others occurring in $(k-1)$ pairs related by the $\mathbb{Z}_2$ reflection symmetry $x \to -x$.
    The D-branes corresponding to these extrema are labeled by the degenerate Ramond Kac labels $(1,n)$ of the $(2,4k)$ minimal superstring.
    Here, $n$ is an even integer with $2 \leq n \leq 4k-2$. 
    The instanton located at $x=0$ corresponds to the ``Ramond sector ground state'' with Kac label $(1,2k)$.
    }
    \label{fig:weakcouplingspectralcurve}
\end{figure}

After double-scaling, the spectral curve of the matrix integral that is dual to the conformal background of the $(2,4k)$ type 0B minimal superstring in the two-cut phase is given in (\ref{spectralcurveweak}).
The support of the eigenvalue distribution is $(-\infty,-1] \cup [1,\infty)$, where $y^2 < 0$.
The zeroes of the polynomial $U_{2k-1}(x)$ in the interval $(-1,1)$ represent the locations of the eigenvalue instantons.
Explicitly, there are $2k-1$ such locations,
\begin{align}
    x_n = \cos \frac{\pi n}{4k} \,, \quad n \in \{2,4,6,\ldots, 4k-2\}\, .
    \label{xnweak}
\end{align}
Note that $x_{2k} = 0$ is the instanton located at the $\mathbb{Z}_2$-symmetric point, and the other $2k-2$ instantons occur in pairs with $x_n = -x_{4k-n}$ (see figure \ref{fig:weakcouplingspectralcurve}).

The effective potential --- the potential an eigenvalue feels due to the combination of the external potential $W_k$ and the eigenvalue repulsion associated to the Vandermonde determinant --- is proportional to $\int y \, \d x$, and we pick the additive constant by imposing that the effective potential vanishes on the eigenvalue cut.
We find\footnote{Note that here, the square root is defined such that $\sqrt{1-x^2}=-|1-x^2|^{1/2}$ for $x\in[-1,1]$. See appendix \ref{app:weak} for details.}
\begin{align}
    V_\eff(x) &= 
    \frac{2}{4k^2-1} \Big( 2k \, T_{2k}(x) - x \, U_{2k-1}(x) \Big) \sqrt{1-x^2} \, .
\label{veffweak}
\end{align}
We have dropped an overall proportionality constant which is not relevant for our purposes.
The values of the effective potential at the extrema are given by
\begin{align}
    V_\eff(x_n) = -(-1)^{\frac{n}{2}} \sin \frac{\pi n}{4k}\, \frac{4k}{4k^2-1}\, .
\end{align}
We record the following ratios that we will be able to match precisely with the ratios of tensions of the corresponding ZZ branes in minimal superstring theory,
\begin{align}
    \frac{V_\eff(x_n)}{V_\eff(x_{2k})} =
    (-1)^{\frac{n-2k}{2}} \sin \frac{\pi n}{4k}\, .
    \label{vratioweak}
\end{align}

To calculate the one-loop normalization associated to a general multi-instanton configuration, one needs the one-loop normalization associated to a single eigenvalue instanton of type $n$ which we label $\mathcal{N}_n$, and also the contribution arising from the interaction between an instanton of type $n$ and an instanton of type $n'\neq n$ which we label $\mathcal{C}_{n,n'}$ \cite{Eniceicu:2022dru}.
These contributions will respectively match the exponential of the cylinder diagram connecting a ZZ brane to itself and the exponential of the cylinder diagram connecting two different ZZ branes in minimal superstring theory.

On the matrix integral side, these contributions  can be computed by adapting techniques from the hermitian one-matrix integral case \cite{Marino:2008ya, Eniceicu:2022dru}.
For the normalization of the one-instanton contribution, we need to carefully evaluate the Gaussian integral about the saddle-point. 
Deferring the details to appendix \ref{app:weak}, we find the result
\begin{align}
    \mathcal{N}_n = \frac{1}{2\pi} \, \frac{\i}{2(1-x_n^2)} \, \sqrt{\frac{2\pi}{-V_\eff''(x_n)}}\, .
    \label{bn}
\end{align}
The second derivative of $V_\eff$ at its extrema can be calculated using (\ref{veffweak}),
\begin{align}
    V_\eff''(x_n) = (-1)^{\frac{n}{2}} \, 4k  \csc\frac{\pi n}{4k} \, .
\end{align}
Using this in (\ref{bn}), we find the relation 
\begin{align}
    \mathcal{N}_n \, V_\eff(x_n)^\frac{1}{2} = \frac{\i}{\sqrt{2\pi(16k^2-4)}} \, 
    \csc \frac{\pi n}{4k}
    \, .
    \label{matrixanswerweak}
\end{align}
A quick sanity check of this result is that, because of the $\mathbb{Z}_2$ symmetry $x \to -x$ in the spectral curve (\ref{spectralcurveweak}), we expect the answer to be invariant under $n \to 4k-n$, and that is indeed true.

Similarly, we can also compute the interaction term $\mathcal{C}_{n,n'}$.
The details are presented in appendix \ref{app:weak}, with the final result
\begin{align}
    \mathcal{C}_{n,n'} &= \left( 
    \frac{x_n - x_{n'}}{1-x_n x_{n'} + (1-x_n^2)^\frac{1}{2}(1-x_{n'}^2)^\frac{1}{2}}
    \right)^2\, .
\end{align}
Plugging in the locations of the instantons given in (\ref{xnweak}), we find
\begin{align}
    \mathcal{C}_{n,n'} &=  \frac{\sin^2 \frac{\pi(n-n')}{8k}}{\sin^2 \frac{\pi(n+n')}{8k}}\, .
    \label{crosscylinderweakmatrix}
\end{align}

\subsection{Instanton effects in the ungapped phase of the matrix integral}
Before double-scaling, the ungapped (strong-coupling) phase corresponds to the phase of the large-$N$ unitary matrix integral defined by (\ref{defzknt}) and (\ref{defvku}) where the eigenvalue density covers the entire unit circle.

After double-scaling, the spectral curve of the matrix integral that is dual to the conformal background of the $(2,4k)$ type 0B minimal superstring in the ungapped phase is given by (\ref{spectralcurvestrong}). The support of the eigenvalue distribution is the entire real axis. 
The effective potential has distinct expressions in the upper half-plane and in the lower half-plane, but each expression can be analytically continued to the other side. 
This fact makes it crucial to distinguish between different types of instantons. 

\begin{figure}[t]
    \centering
    \includegraphics[width=0.48\textwidth]{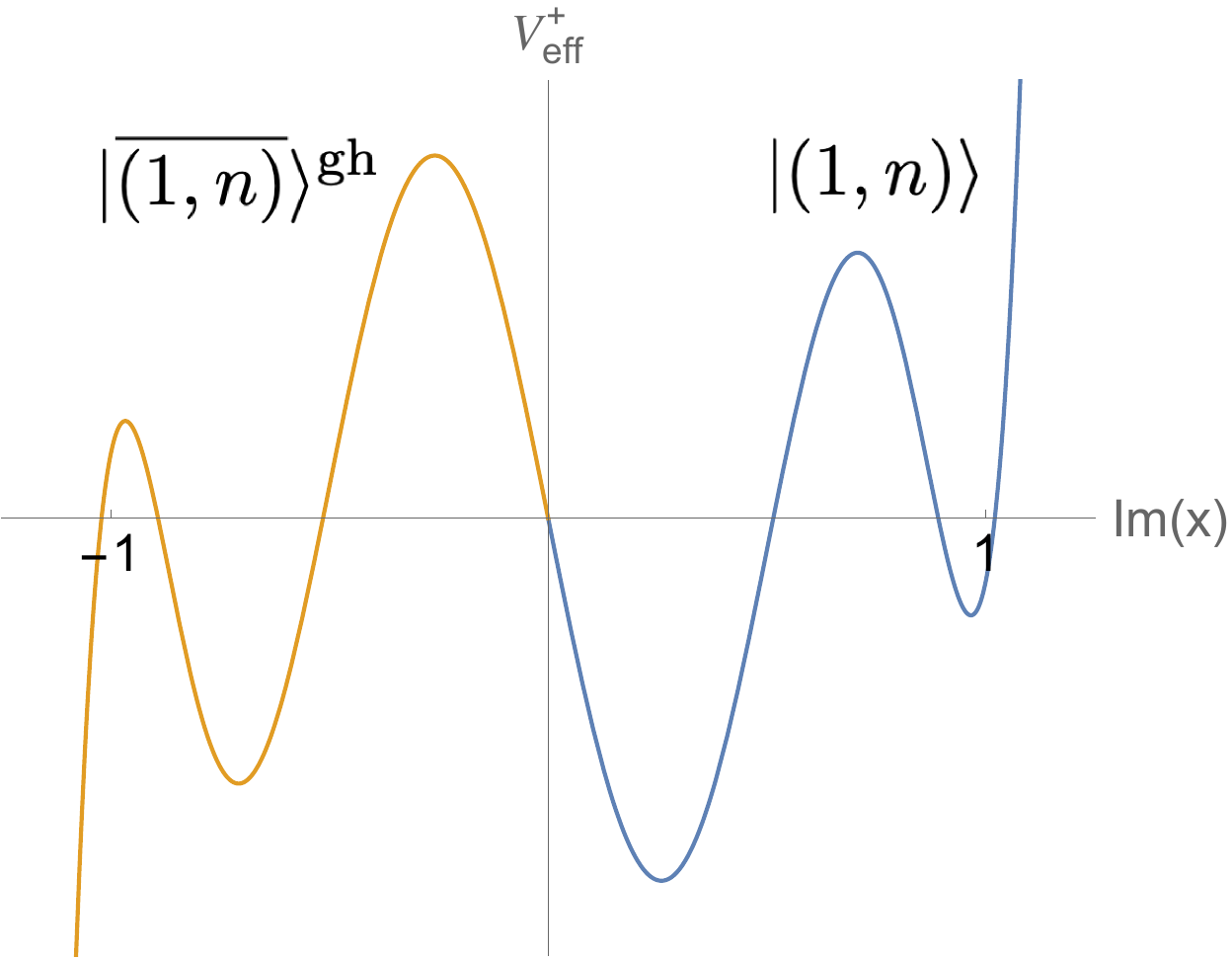}\hspace{0.2in}
    \includegraphics[width=0.48\textwidth]{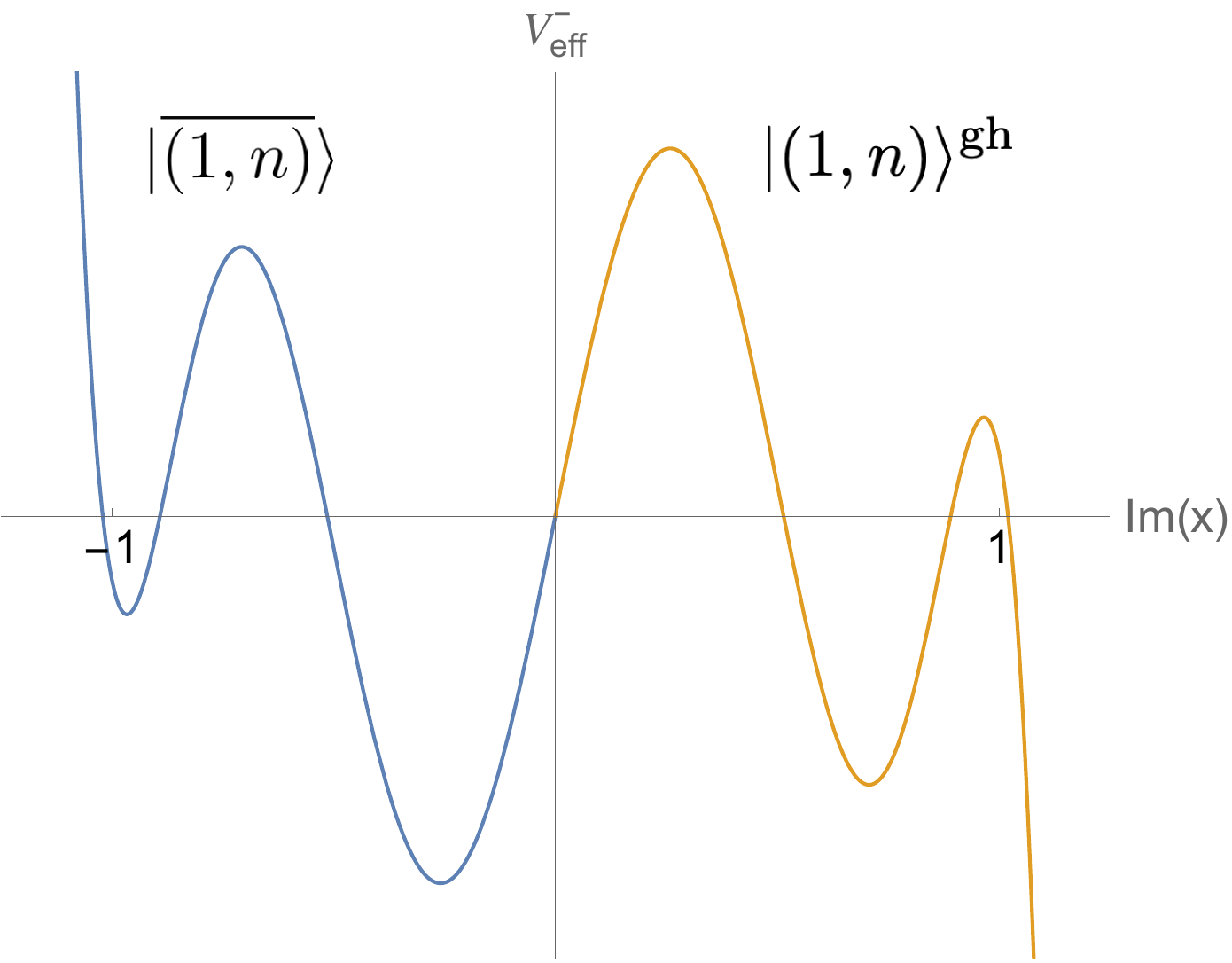}
    \caption{The double-scaled effective potentials $V_\eff^+$ and $V_\eff^-$ for the $k=3$ model in the ungapped, edgeless phase. 
    The eigenvalue density is supported along the entire real axis.
    Because the cut divides the plane into two disconnected parts, we now have two separate analytic functions.
    In the minimal superstring, the relevant ZZ branes are labeled by $(1,n)$ with $n$ an odd integer satisfying $1 \leq n \leq 2k-1$. The branes corresponding to $V^+_\eff$ and $V_\eff^-$ carry opposite Ramond charge.
    Further we have ``ghost'' branes, extrema on the orange part of the curves, which are the extrema that occur in regions after analytic continuation past the cut \cite{Marino:2022rpz}. 
    The boundary state of a ghost brane differs by an overall minus sign from the corresponding non-ghost brane \cite{Douglas:2003up, Okuda:2006fb, Erler:2012qn}.
    }
    \label{fig:strongcouplingspectralcurve}
\end{figure}

We will refer to an instanton associated with the effective potential defined above the real axis ``positively charged,'' and an instanton associated with the effective potential defined below the real axis ``negatively charged''. 
An instanton associated with the effective potential defined in its original domain will be called ``real'', while an instanton associated with the analytic continuation of its corresponding effective potential to the other side of the real axis will be called ``ghost''.\footnote{The negatively-charged instantons are indicated with a bar above their label, e.g.~$\overline{n}$, while the positively-charged instantons are indicated with an unbarred label $n$. The real instantons carry no superscript on their label $n$, while their ghost partners are indicated with a superscript on their label, i.e.~$n^\gh$. Note that the instanton labeled by $n^\gh$ will correspond to the same location as the instanton labeled by $n$, but will carry opposite charge.}
These are shown in figure \ref{fig:strongcouplingspectralcurve}.

The zeros of the polynomial $T_{2k}(\i x)$ represent the locations of the eigenvalue instantons. Explicitly, there are $k$ pairs of locations, where each pair comprises a point $x_n=x_{n^\gh}$ above and a point $x_{\overline{n}}=x_{\overline{n}^\gh}$ below the real axis,
\begin{equation}
    x_n=x_{n^\gh}=\i\sin\frac{\pi n}{4k},\qquad x_{\overline{n}}=x_{\overline{n}^\gh}=-\i\sin\frac{\pi n}{4k},\qquad n\in\{1,3,\dots,2k-1\}\,.
    \label{xnstrong}
\end{equation}
Up to an overall multiplicative constant which will be unimportant for our purposes, the effective potentials are
\begin{equation}
    V_{\eff}^\pm(x)=\pm(-1)^k\left(\frac{2k}{4k^2-1}T_{2k+1}(\i x)-\frac{\i x}{2k-1}T_{2k}(\i x)\right)\,.
\end{equation}
The values of the effective potentials at the extrema are related to the tensions of the corresponding ZZ branes in the dual string theory and are given by
\begin{align}
    V_{\eff}^+(x_n)=V_{\eff}^-(x_{\overline{n}})&=\frac{2k}{4k^2-1}(-1)^{(n+1)/2}\cos\frac{\pi n}{4k}\,,\\
    V_{\eff}^-(x_{n^\gh})=V_{\eff}^+(x_{\overline{n}^\gh}) &= - \frac{2k}{4k^2-1}(-1)^{(n+1)/2}\cos\frac{\pi n}{4k}\,.
\end{align}
We see that the values of the potentials corresponding to the ghost instantons are the opposites of the values of the potentials of their real counterparts. We note the following ratios which we will match precisely with the ratios of tensions of the corresponding ZZ branes in the minimal superstring,
\begin{equation}
    \frac{V_{\eff}^+(x_n)}{V_{\eff}^+(x_1)}=(-1)^{(n-1)/2}\;\frac{\cos\frac{\pi n}{4k}}{\cos\frac{\pi}{4k}}\,.\label{eqn:ungapped_tension_ratios}
\end{equation}

In contrast with the situation in the gapped phase, where the leading instanton contributions to the partition function consisted of single-eigenvalue tunneling, in the ungapped phase, the leading corrections are due to pairs of instantons \cite{Eniceicu:2023cxn} consisting of a positive-charge and a negative-charge instanton. We will defer the details of how to obtain these contributions to appendix \ref{app:strong}, and just note that the one-loop normalization factor associated to a pair of instantons located at $\xi_1$ and $\xi_2$ is
\begin{equation}
    \mathcal{N}_{\xi_1,\xi_2}=\frac{1}{2\pi}\frac{1}{\sqrt{V_{\eff}^+{''}(\xi_1)}}\frac{1}{\sqrt{V_{\eff}^-{''}(\xi_2)}}\frac{1}{(\xi_1-\xi_2)^2}\,.
\end{equation}
The second derivatives of $V_{\eff}^\pm$ are given by
\begin{equation}
    V_{\eff}^\pm{''}(x)=\mp2k\,(-1)^k\,U_{2k-1}(\i x)\,,
\end{equation}
which, at the instanton locations, evaluate to
\begin{align}
    V_{\eff}^+{''}(x_n) &= V_{\eff}^-{''}(x_{\overline{n}})=\frac{2k(-1)^{(n+1)/2}}{\cos\frac{\pi n}{4k}}\,,\\
    V_{\eff}^-{''}(x_{n^\gh}) &= V_{\eff}^+{''}(x_{\overline{n}^\gh})= - \frac{2k(-1)^{(n+1)/2}}{\cos\frac{\pi n}{4k}}\,.
\end{align}
Using these expressions, we note the following relations between the normalization factors associated with a pair of instantons and the corresponding effective potentials in the four cases of interest:
\begin{itemize}
    \item 
    Real positive-charge brane $(1,n)$ and real negative-charge brane $\overline{(1,n')}$:
\begin{equation}
    \mathcal{N}_{n,\,\overline{n}'}\,V_{\eff}^+(x_n)^{1/2}\,V_{\eff}^-(x_{\overline{n}'})^{1/2}=-\frac{1}{2\pi(16k^2-4)}\times\frac{ \cos \frac{\pi n}{4k} \cos \frac{\pi n'}{4k} }{\sin^2 \frac{\pi (n+n')}{8k} \cos^2 \frac{\pi(n-n')}{8k}} \,.
\label{eqn:ug_nm_rrd}
\end{equation}

For $n=n'$, this simplifies to
\begin{equation}
    \mathcal{N}_{n,\,\overline{n}} \,V_{\eff}^+(x_n)^{1/2}\,V_{\eff}^-(x_{\overline{n}})^{1/2}=-\frac{\cot^2 \frac{\pi n}{4k} }{2\pi(16k^2-4)}\,.
\label{eqn:ug_nm_rrs}
\end{equation}

\item Real positive-charge brane $(1,n)$ and ghost negative-charge brane $(1,n')^\gh$:
\begin{equation}
    \mathcal{N}_{n,\,n'{}^{\text{gh}}}\,V_{\eff}^+(x_n)^{1/2}\,V_{\eff}^-(x_{n'{}^\gh})^{1/2}=-\frac{1}{2\pi(16k^2-4)}\times\frac{\cos \frac{\pi n}{4k} \cos \frac{\pi n'}{4k}}{\cos^2 \frac{\pi (n+n')}{8k} \sin^2 \frac{\pi (n-n')}{4k}}\,.
\label{eqn:ug_nm_rg}
\end{equation}

\item Ghost positive-charge brane $\overline{(1,n)}^\gh$ and real negative-charge brane $\overline{(1,n')}$:
\begin{equation}
    \mathcal{N}_{\overline{n}^{\text{gh}},\,\overline{n}'}\,V_{\eff}^+(x_{\overline{n}^\gh})^{1/2}\,V_{\eff}^-(x_{\overline{n}'})^{1/2}=-\frac{1}{2\pi(16k^2-4)}\times\frac{\cos \frac{\pi n}{4k} \cos \frac{\pi n'}{4k}}{\cos^2 \frac{\pi (n+n')}{8k} \sin^2 \frac{\pi (n-n')}{4k}}\,. 
\label{eqn:ug_nm_gr}
\end{equation}

\item Ghost positive-charge brane $\overline{(1,n)}^\gh$ and ghost negative-charge brane $(1,n')^\gh$:
\begin{equation}
    \mathcal{N}_{\overline{n}^{\text{gh}},\,n'{}^{\text{gh}}}\,V_{\eff}^+(x_{\overline{n}^\gh})^{1/2}\,V_{\eff}^-(x_{n'{}^\gh})^{1/2}=-\frac{1}{2\pi(16k^2-4)}\times\frac{\cos \frac{\pi n}{4k} \cos \frac{\pi n'}{4k}}{\sin^2 \frac{\pi (n+n')}{8k} \cos^2 \frac{\pi(n-n')}{8k}}\,.
\label{eqn:ug_nm_gg}
\end{equation}

For $n=n'$, this simplifies to
\begin{equation}
    \mathcal{N}_{\overline{n}^{\text{gh}},\,n{}^{\text{gh}}}\,V_{\eff}^+(x_{\overline{n}^\gh})^{1/2}V_{\eff}^-(x_{n^\gh})^{1/2}=-\frac{\cot^2 \frac{\pi n}{4k}}{2\pi(16k^2-4)}\,.
\end{equation}
\end{itemize}
Note that (\ref{eqn:ug_nm_rg}) and (\ref{eqn:ug_nm_gr}) are only valid when $n \neq n'$.\footnote{Although we shall not pursue it here, we note that Ref.~\cite{Marino:2022rpz} handles the case where $n=n'$ by using a different contour prescription for the eigenvalue instantons to obtain a non-trivial answer.}

We have restricted to two-instanton configurations here. 
In appendix \ref{sec:appgeneralungapped}, we extend this analysis to general multi-instanton configurations.

\section{Review of boundary states in minimal superstring theory}
\label{sec:reviewbdrystates}

The worldsheet theory of minimal superstring theory is a $(1,1)$ SCFT and consists of an $\mathcal{N}=1$ super-Virasoro minimal CFT \cite{Friedan:1984rv}, the $\mathcal{N}=1$ super-Liouville theory \cite{Polyakov:1981re, Poghossian:1996agj, Rashkov:1996np, Nakayama:2004vk}, and the $bc$ and the $\beta\gamma$ ghosts \cite{Seiberg:2003nm}.
We will exclusively focus on the $(2,4k)$ family of minimal superstring theories with the type 0B GSO projection, where $k$ is a positive integer.

\paragraph{The $(2,4k)$ minimal SCFT.}
Let us give a very brief review of the properties of the $(2,4k)$ super-minimal model that we will need.
The central charge is
\begin{align}
    c_\text{Matter} &= \frac{3}{2} \left(1-\frac{2 (4 k-2)^2}{2 \cdot 4k}\right) = \frac{3}{2} - \frac{3(2k-1)^2}{2k} \, .
    \label{cmatter}
\end{align}
It contains $k$ NS primaries and $k$ R primaries.
The Kac labels of the NS primaries are $(1,1)$, $(1,3)$, $\ldots$, $(1,2k-1)$, while the Kac labels of the R primaries are $(1,2), (1,4), \ldots, (1,2k)$.
The NS operator with Kac label $(1,1)$ is the identity operator.
The R operator with Kac label $(1,2k)$ is the ``Ramond ground state'' with $h = \widetilde{h} = \frac{c}{24}$.
Below, we will need the identity characters for these minimal SCFTs \cite{Goddard:1986ee, Kiritsis:1986rv},
\begin{align}
    \chi^{\ns}_{\text{Matter}\,(1,1)}(q) &=  
    \sqrt{\frac{\theta_3(q)}{\eta(q)^{3}}} \sum_{j \in \mathbb{Z}} 
    \left( q^{\frac{(8k j + 2k - 1)^2}{16 k}} - q^{\frac{(8k j + 2k + 1)^2}{16 k}} 
    \right)\, , \label{chimatterns11} \\
    \chi^{\nstilde}_{\text{Matter}\,(1,1)}(q) &=  \sqrt{\frac{\theta_4(q)}{\eta(q)^{3}}} \sum_{j \in \mathbb{Z}} \left( q^{\frac{(8k j + 2k - 1)^2}{16 k}} + q^{\frac{(8k j + 2k + 1)^2}{16 k}} \right)\, .
    \label{chimatternstilde11}
\end{align}

\paragraph{The $\mathcal{N}=1$ Liouville theory.}
Now we briefly discuss the $\mathcal{N}=1$ Liouville sector of the worldsheet theory. 
See Ref.~\cite{Nakayama:2004vk} for a review.
The parameters of the $\mathcal{N}=1$ Liouville theory are chosen to make the total central charge of the worldsheet CFT zero:
\begin{align}
    b &= \sqrt{\frac{2}{4k}}\,,\\
    Q &= b^{-1} + b \,,\\
    c_\text{Liouville} &= \frac{3}{2} + 3 Q^2 = 
    \frac{3}{2} + \frac{3(2k+1)^2}{2k}\, .
    \label{cliouville}
\end{align}
Adding the central charge $-26$ for the $bc$ ghosts, $11$ for the $\beta\gamma$ ghosts to that of matter and Liouville CFTs given in (\ref{cmatter}) and (\ref{cliouville}), we indeed find zero.
The NS primaries of Liouville are exponential operators $e^{\alpha\phi}$ with scaling weights $h = \widetilde{h} =\frac{1}{2} \alpha(Q - \alpha)$.
The delta function normalizable states have $\alpha = \frac{Q}{2} + \i P$ with $P \in \mathbb{R}_{\geq 0}$ and with scaling weights $h = \widetilde{h} = \frac{Q^2}{8} + \frac{P^2}{2}$.
The R primaries are given by a Liouville exponential times the spin or disorder operators and have scaling weights given by 
$h = \widetilde{h} =\frac{1}{16} + \frac{1}{2} \alpha(Q - \alpha) = \frac{1}{16} + \frac{Q^2}{8} + \frac{P^2}{2}$, with the final expression holding for $\alpha = \frac{Q}{2} + \i P$.

The $\mathcal{N}=1$ Liouville cosmological constant $\mu$ can take either sign.
The phase transition that we discussed in the matrix integral happens at $\mu = 0$.
We will work with conventions such that $\mu>0$ corresponds to the two-cut or the gapped phase, and $\mu<0$ corresponds to the ungapped phase.

\subsection{Two-dimensional SCFTs on the annulus}
In this subsection, we will review the construction of boundary states and annulus characters for theories with $\mathcal{N}=1$ super-Virasoro symmetry.
Our treatment will necessarily have to be brief; we refer the reader to Refs.~\cite{Recknagel:2013uja, Gaberdiel:2000jr, Nepomechie:2001bu} for more detailed expositions.

\paragraph{Characters of non-degenerate modules.}
We begin by reviewing the characters of non-degenerate modules of the $\mathcal{N}=1$ super-Virasoro algebra with central charge $c$.
For the NS algebra, let us parametrize the weight using a ``momentum'' label $P$ via the relation $h = \frac{P^2}{2} + \frac{c}{24} - \frac{1}{16}$.
Then, 
\begin{align}
\chi^\ns_h(q) &= q^{\frac{P^2}{2}} \, \sqrt{\frac{\theta_3(q)}{\eta(q)}} \frac{1}{\eta(q)}
= q^{h - \frac{c}{24}} \left( 
1+q^\frac{1}{2}+q+2 q^{3/2}+3 q^2+O\left(q^{5/2}\right)
\right)
\,,\\
\chi^\nstilde_h(q) &= q^\frac{P^2}{2} \, \sqrt{\frac{\theta_4(q)}{\eta(q)}} \frac{1}{\eta(q)}
= q^{h - \frac{c}{24}}
\left(
1-q^\frac{1}{2}+q-2 q^{3/2}+3 q^2+O\left(q^{5/2}\right)
\right)
\, .
\end{align}
The character $\chi^\ns_h$ is the partition function while $\chi^\nstilde_h$ is the index, summing over all the states in the module with highest weight $h$.
Explicitly, the first few states that are being counted are $\ket{h}$, $G_{-\frac{1}{2}} \ket{h}$, $L_{-1}\ket{h}$, $G_{-\frac{3}{2}}\ket{h}$, $L_{-1} G_{-\frac{1}{2}}\ket{h}$, and so on.
The states with an odd number of $G$'s acting on them are counted with a minus sign in $\chi^\nstilde_h$.

For the R algebra, let us parametrize the weight using a ``momentum'' label $P$ via the relation $h = \frac{P^2}{2} + \frac{c}{24}$.
Then,
\begin{align}
\chi^\r_h(q) &= q^\frac{P^2}{2} \, \sqrt{\frac{\theta_2(q)}{2\eta(q)}} \frac{1}{\eta(q)}
= q^{h - \frac{c}{24}} \left( 1+2 q+4 q^2+8 q^3+14 q^4+O\left(q^{5} \right)\right)\, .
\end{align}
Explicitly, the first few states that are being counted are $\ket{h}$, $G_{-1}\ket{h}$, $L_{-1}\ket{h}$, $G_{-2}\ket{h}$, $L_{-2}\ket{h}$, $L_{-1}L_{-1}\ket{h}$, $L_{-1}G_{-1}\ket{h}$, and so on.
Note that, in our conventions, the R character has integer coefficients (as opposed to certain other conventions that have an overall $\sqrt{2}$ factor).
For a single copy of the R algebra, the relation $G_0^2 = L_0 - \frac{c}{24}$ defines a one-dimensional Clifford algebra and admits a one-dimensional representation. Instead, if we consider representations of the tensor product of the holomorphic and the anti-holomorphic copies of the $\mathcal{N}=1$ Ramond super-Virasoro algebra, then $G_0$ and $\widetilde{G}_0$ generate a two-dimensional Clifford algebra, whose smallest nontrivial representation is two-dimensional.
Thus, the RR characters will be $2\, \chi^\r_h(q)\chi^\r_{\tilde{h}}(\bar{q}) $, with an additional overall factor of two.

\paragraph{Characters of degenerate modules in $\mathcal{N}=1$ Liouville theory.}
Degenerate characters in the \mbox{$\mathcal{N}=1$} Liouville theory are labeled by a pair of positive integers $(m,n)$, corresponding to
\begin{equation}
    P^2 = - \frac{1}{4} (m b^{-1} + n b)^2.
\end{equation}
NS degenerate characters have $m+n$ odd, while R degenerate  characters have $m+n$ even.
The null state appears at level $mn/2$, which is a half-integer for NS modules with $m$, $n$ both odd.
The degenerate characters are
\begin{align}
    \chi^\ns_{(m,n)}(q) &= q^{-\frac{1}{8}(m b^{-1} + n b)^2}  ( 1 - q^\frac{mn}{2}) \,  \sqrt{\frac{\theta_3(q)}{\eta(q)}} \frac{1}{\eta(q)}\,,\quad \text{$m+n$ even}\, , \label{chinsmn}\\
    \chi^\r_{(m,n)}(q) &= q^{-\frac{1}{8}(m b^{-1} + n b)^2}  ( 1 - q^\frac{mn}{2}) \,  \sqrt{\frac{\theta_2(q)}{2\eta(q)}} \frac{1}{\eta(q)}\,,\quad \text{$m+n$ odd}\, \label{chirmn}.
\end{align}
When computing $\chi^\nstilde_{(m,n)}(q)$,  we should note the fact that, when $m$ and $n$ are both odd, the null descendant appears at a half-integer level and so we should change the sign of the subtraction:
\begin{align}
    \chi^\nstilde_{(m,n)}(q) = q^{-\frac{1}{8}(m b^{-1} + n b)^2}  ( 1 - (-1)^{mn}  q^\frac{mn}{2}) \,  \sqrt{\frac{\theta_4(q)}{\eta(q)}} \frac{1}{\eta(q)}\,,\quad \text{$m+n$ even}\, \label{chinstildemn} .
\end{align}

\paragraph{Boundary state conditions in 2D SCFTs.}
In addition to the condition $(L_n - \widetilde{L}_{-n})\vert \cdot \rrangle = 0$, boundary states in 2D SCFTs have to satisfy $(G_r +\i \,\eta\, \widetilde{G}_{-r})\vert \cdot \rrangle = 0$, where $\eta \in \{+1,-1\}$.
The states with different values of $\eta$ are distinct, so we can build two Ishibashi states $\vert h, + \rrangle$ and $\vert h, - \rrangle$  from each super-Virasoro module.
The cylinder between these two Ishibashi states built on the same module depends on whether the $\eta$ label on the two boundaries is the same or not.
If the $\eta$ label is the same on the two boundaries, the cylinder evaluates to the partition function over the module, and if the $\eta$ label is different on the two boundaries, the cylinder evaluates to the index of the module.
\begin{align}
    \llangle \ns, h, \eta \, \vert \, e^{-\frac{\pi}{t} H_\text{closed}} \, \vert \, 
    \ns, h, \eta' \rrangle &= \begin{cases}
        \frac{1}{2}\chi^\ns_h (e^{-2\pi/t}) \quad \text{if $\eta = \eta'$} \,,\\
        \frac{1}{2}\chi^\nstilde_h (e^{-2\pi/t}) \quad \text{if $\eta = -\eta'$} \,.
    \end{cases} \label{nsishibashigeneral}\\
    \llangle \r, h, \eta \, \vert \, e^{-\frac{\pi}{t} H_\text{closed}} \, \vert \, 
    \r, h, \eta' \rrangle &= \begin{cases}
        \chi^\r_h (e^{-2\pi/t}) \quad \text{if $\eta = \eta'$} \,,\\
        0 \quad\hspace{0.6in} \text{if $\eta = -\eta'$} \,.
    \end{cases}
\end{align}
The factors of $\frac{1}{2}$ on the right side of (\ref{nsishibashigeneral}) are conventional and we have put them for bookkeeping the factors of $\frac{1}{2}$ that arise in the GSO projection.

\subsection{Branes in \texorpdfstring{$\mathcal{N}=1$}{\mathcal{N}=1}  Liouville CFT}

It is convenient to start by discussing two ZZ branes $\vert (1,1) \rangle$ and $\vert \overline{(1,1)} \rangle$ in super Liouville theory. 
They have the following expansion as a sum over Ishibashi states \cite{Ahn:2002ev, Fukuda:2002bv, Irie:2007mp}:
\begin{align}
    \ket{(1,1)} &= \int_{0}^\infty \d P' \, \Big( 
    \Psi^\ns_{1,1}(P') \,\vert \text{NS}, P', + \rrangle
    + 
    \Psi^\text{R}_{1,1}(P') \,\vert \text{R}, P', + \rrangle \Big) \,,\\
    \ket{\overline{(1,1)}} &= \int_{0}^\infty \d P' \, \Big( 
    \Psi^\ns_{1,1}(P') \,\vert \text{NS}, P', + \rrangle
    -
    \Psi^\text{R}_{1,1}(P') \,\vert \text{R}, P', + \rrangle \Big) \,,\\
    \Psi^\ns_{1,1}(P') &= \mu^{-\i P' b^{-1}} \, \frac{\sqrt{2}\,\pi}{P'\,\Gamma \left( -\i P' b^{-1} \right)\,\Gamma \left( -\i P' b \right)} \label{psi-ns-11} \, ,\\
    \Psi^\text{R}_{1,1}(P') &= \mu^{-\i P' b^{-1}} \,  \frac{\sqrt{2}\,\pi}{\Gamma \left( \frac{1}{2}-\i P' b^{-1} \right)\,\Gamma \left( \frac{1}{2}-\i P' b \right)}\, 2^\frac{1}{4} \, .
\end{align}
The $\ket{(1,1)}$ state has the property that if we compute a cylinder between it and an FZZT state $\ket{ \text{NS}, P}$ or $\ket{ \text{R}, P}$, the open string channel only contains the states from the single module with highest weight labeled by $P$ and with $(-1)^F = +1$.
On the other hand, the $\ket{\overline{(1,1)}}$ state has the property that if we compute a cylinder between it and the state $\ket{ \text{NS}, P}$ or $\ket{ \text{R}, P}$, the open string channel only contains the states from the single module with highest weight labeled by $P$ and with $(-1)^F = -1$.
Explicitly, the FZZT boundary states are given by \cite{Ahn:2002ev, Fukuda:2002bv, Seiberg:2003nm, Irie:2007mp}
\begin{align}
    \ket{\text{NS}, P} = 
    \int_{0}^\infty \d P' &\left( 
    \Psi^\ns_{P}(P') \,\vert \text{NS}, P', + \rrangle
    + 
    \Psi^{\r +}_{P}(P') \,\vert \text{R}, P', + \rrangle 
    \right)\,, \label{opennsp}
    \\
    \ket{\overline{\text{NS}, P}} = 
    \int_{0}^\infty \d P' &\left( 
    \Psi^\ns_{P}(P') \,\vert \text{NS}, P', + \rrangle
    -
    \Psi^{\r +}_{P}(P') \,\vert \text{R}, P', + \rrangle 
    \right)\,, \label{opennsbarp} \\
    \ket{\r, P} = \sqrt{2}\int_{0}^\infty \d P' \, &\left(
    \Psi_P^\ns(P') \,\vert \ns, P', - \rrangle + 
    \Psi_P^{\r -}(P') \,\vert \r, P', - \rrangle \right)\, , \label{openrp} \\
    \ket{\overline{\r, P}} = \sqrt{2}\int_{0}^\infty \d P' \, &\left(
    \Psi_P^\ns(P') \,\vert \ns, P', - \rrangle - 
    \Psi_P^{\r -}(P') \,\vert \r, P', - \rrangle \right)\, , \label{openrbarp}
\end{align}
where the three wavefunctions $\Psi^\ns_P(P')$, $\Psi^{\r+}_{P}(P')$ and $\Psi^{\r-}_{P}(P')$ are determined via
\begin{align}
    \Psi^\ns_P(P') \, \Psi^\ns_{1,1}(P')^*
    &= \cos(2\pi P P') \,,\label{psi-ns}\\
    \Psi^{\r +}_{P}(P')\, \Psi^{\r}_{1,1}(P')^* &= \sqrt{2}\,\cos(2\pi P P')  \, ,
    \label{psirplus}\\
    \Psi^{\r -}_{P}(P')\, \Psi^{\r}_{1,1}(P')^* &= \sqrt{2}\,\sin(2\pi P P')  \, .\label{psirminus}
\end{align}
These wavefunctions ensure that
\begin{align}
    \braket{(1,1) | e^{- \frac{\pi}{t} H_\text{closed}} | \ns, P} &= 
    \frac{1}{2} \left(
   \chi^\ns_P(e^{-2\pi t}) 
    +
    \chi^\nstilde_P(e^{-2\pi t}) 
   \right)\,, \label{11nsoverlap} \\
   \braket{ \overline{(1,1)} | e^{- \frac{\pi}{t} H_\text{closed}} | \ns, P} &= 
    \frac{1}{2} \left(
   \chi^\ns_P(e^{-2\pi t}) 
    -
    \chi^\nstilde_P(e^{-2\pi t}) 
   \right)\, ,\label{11nsbaroverlap}\\
   \braket{(1,1) | e^{- \frac{\pi}{t} H_\text{closed}} | \r, P} &= \chi^\r_P(e^{-2\pi t})\, ,
   \label{11roverlap}
\end{align}
and other such relations.
We have chosen the normalization of the wavefunctions so that the NS-Ishibashi part of the state in (\ref{opennsp}) yields $\frac{1}{2} \chi^\ns_P(e^{-2\pi t})$ on the right side of (\ref{11nsoverlap}).
Similarly, the R-Ishibashi part of the state in (\ref{opennsp}) yields $\frac{1}{2} \chi^\nstilde_P(e^{-2\pi t})$ on the right side of (\ref{11nsoverlap}).
The Ramond character in (\ref{11roverlap}) arises purely from the NS-Ishibashi part of the state in (\ref{openrp}).

The discrete $(m,n)$ ZZ states can be obtained by combining the FZZT states of the pair of momenta appearing in each of the character formulas (\ref{chinsmn}), (\ref{chinstildemn}), (\ref{chirmn}), namely $P_{m,n}$ and $P_{m,-n}$, where
\begin{align}
   P_{m,n} := \frac{\i}{2} (m b^{-1} + n b) \, .
   \label{defpmn}
\end{align}
If $m+n$ is even, the $(m,n)$ degenerate state is in the NS sector, so we have to appropriately combine the states (\ref{opennsp}) and (\ref{opennsbarp}) as follows:
\begin{align}
    \ket{(m,n)} = \begin{cases}
        \ket{\ns, P_{m,n}} - \ket{\ns, P_{m,-n}} \quad \text{if $m$ and $n$ are both even}\,, \\
        \ket{\ns, P_{m,n}} - \ket{\overline{\ns, P_{m,-n}}} \quad \text{if $m$ and $n$ are both odd}\,.
    \end{cases}
\end{align}
This gives rise to the following states:
\begin{align}
    \ket{(m,n)} &= \int_{0}^\infty \d P' \left( 
    \Psi^\ns_{m,n}(P') \,\vert \text{NS}, P', + \rrangle + 
    \Psi^\r_{m,n}(P') \,\vert \text{R}, P', + \rrangle \right) \,,
    \label{cardy-mn}
    \\
    \ket{\overline{(m,n)}} &= \int_{0}^\infty \d P' \left( 
    \Psi^\ns_{m,n}(P') \,\vert \text{NS}, P', + \rrangle -
    \Psi^\r_{m,n}(P') \,\vert \text{R}, P', + \rrangle \right) \,,
    \label{cardy-mn-bar}
    \\
    \Psi^\ns_{m,n}(P') \Psi^\ns_{1,1}(P')^* &= 
    2 \sinh(\pi P' m b^{-1})\sinh(\pi P' n b)
     \label{psi-ns-mn} \, ,\\
    \Psi^\text{R}_{m,n}(P') \Psi^\r_{1,1}(P')^* &= 
    \begin{cases}
        2 \sqrt{2} \sinh(\pi P' m b^{-1})\sinh(\pi P' n b) \quad \text{if $m$ and $n$ are both even} \,,\\
        2 \sqrt{2} \cosh(\pi P' m b^{-1})\cosh(\pi P' n b) \quad \text{if $m$ and $n$ are both odd}\,.
        \label{psiroddodd}
    \end{cases}
\end{align}
Below, we will only need the case with $m=1$ and $n$ odd, which are the branes that will be relevant in the ungapped phase of the theory.

If $m+n$ is odd, the $(m,n)$ degenerate state is in the R sector, so we have to appropriately combine the states (\ref{openrp}) and (\ref{openrbarp}) as follows \cite{Fukuda:2002bv, Irie:2007mp}
\begin{align}
    \ket{(m,n)} = \begin{cases}
        \ket{\r, P_{m,n}} - \ket{\r, P_{m,-n}} \quad \text{if $m$ is odd and $n$ is even}\,, \\
        \ket{\r, P_{m,n}} - \ket{\overline{\r, P_{m,-n}}} \quad \text{if $m$ is even and $n$ is odd}\,.
    \end{cases}
    \label{fzztrcombinations}
\end{align}
This gives rise to the wavefunctions
\begin{align}
    \ket{(m,n)} &= \sqrt{2}\int_{0}^\infty \d P' \left( 
    \Psi^\ns_{m,n}(P') \,\vert \text{NS}, P', - \rrangle + 
    \Psi^\r_{m,n}(P') \,\vert \text{R}, P', - \rrangle \right)\, , \label{mnliouvneven}\\
    \ket{\overline{(m,n)}} &= \sqrt{2}\int_{0}^\infty \d P' \left( 
    \Psi^\ns_{m,n}(P') \,\vert \text{NS}, P', - \rrangle - 
    \Psi^\r_{m,n}(P') \,\vert \text{R}, P', - \rrangle \right)\, ,\\
    \Psi^\r_{m,n}(P') \, \Psi^\r_{1,1}(P')^* &= 
    \begin{cases}
    \i \, 2\sqrt{2} \cosh(\pi P' m b^{-1})\sinh(\pi P' n b)\quad \text{if $m$ is odd and $n$ is even}\,,\\
     \i \, 2\sqrt{2} \sinh(\pi P' m b^{-1})\cosh(\pi P' n b)\quad \text{if $m$ is even and $n$ is odd}\,,
    \end{cases}
    \label{mnliouvilleevenR}
\end{align}
and with $\Psi^\ns_{m,n}(P')$ still given by (\ref{psi-ns-mn}).
Below, we will only need the case with $m=1$ and $n$ even, which are the branes that will be relevant in the two-cut gapped phase.

\section{The input from string field theory}
\label{sftinput}

The worldsheet computation of annulus diagrams in the open string channel contains a sum over single-string states in Siegel gauge.
When we exponentiate the annulus to find the normalization constant that multiplies nonperturbative corrections, the exponentiated quantity is the Gaussian approximation to the path integral of the field theory on the D-brane worldvolume.

The annulus diagrams we will encounter contain two single-string states with $L_0=0$ that will lead to a divergence when we integrate over the modular parameter $t$ of the annulus.
When exponentiated, this divergence leads to the vanishing of the normalization constant.
The culprit states are the following two states in the NS sector \cite{Sen:2021tpp, Chakravarty:2022cgj}: 
\begin{align}
    \gamma_{-\frac{1}{2}} \, c_{1}\, \ket{-1} \,,\quad 
    \beta_{-\frac{1}{2}} \, c_{1}\, \ket{-1} \,.
    \label{problematicstates}
\end{align}
Here $\ket{-1} = e^{-\varphi}\ket{0}$ is the ``NS vacuum state'' in the $-1$ picture, with $\varphi$ being the scalar that appears in the bosonization of the $\beta \gamma$ system \cite{Polchinski:1998rr, deLacroix:2017lif}, and $\ket{0}$ being the OSp$(1\vert 2)$ invariant state.
We remind the reader that the $L_0$ eigenvalue of $e^{-\varphi}$ is $+\frac{1}{2}$.

For a D$p$-brane with $p \geq 0$, the states would also have a momentum label for directions parallel to the D-brane, and the two states in (\ref{problematicstates}) would represent gauge-fixing ghosts on the worldvolume gauge theory of the brane.\footnote{One simple check of this is that the ghost numbers of the two states in (\ref{problematicstates}) are $2$ and $0$, respectively. We remind the reader that $c$ and $\gamma$ have ghost number 1, while $b$ and $\beta$ have ghost number $-1$.}
In the string field theory language, the relevant gauge transformation parameter is the coefficient of the ghost number zero state $\beta_{-\frac{1}{2}} c_1 \ket{-1}$.
For a D-instanton, the worldvolume is just a point and this gauge symmetry is actually rigid.
This corresponds to the fact that the out-of-Siegel gauge, ghost number one state $\beta_{-\frac{1}{2}} c_0 c_1 \ket{-1}$ is in fact gauge invariant and cannot be set to zero by the gauge transformation just mentioned.

Thus, the fix is simply that we should ``un-Faddeev-Popov'' the worldvolume theory on the brane \cite{Sen:2021qdk}. 
That is, instead of integrating over the gauge-fixing ghosts, we should integrate over the gauge-invariant ghost-number one field (call it $\phi$) corresponding to the ghost number one single-string state $\beta_{-\frac{1}{2}} c_0 c_1 \ket{-1}$, and divide by the finite volume of the rigid ``gauge'' group.
This procedure has been carried out for the superstring in Refs.~\cite{Sen:2021tpp, Chakravarty:2022cgj} and the result is
\begin{align}
    \int \d p \, \d q \, e^{0 \cdot p q} \longrightarrow \frac{\int \d\phi \,  e^{-\frac{\phi^2}{4}}}{\int \d\theta}  = \frac{2\sqrt{\pi}}{4\pi/g_o} = \frac{1}{\sqrt{8\pi^3 T}}\, .
    \label{replacementrule}
\end{align}
Here $g_o$ is the coupling constant of the open string field theory on the D-brane (or, in other words, the D-brane worldvolume theory), and $T$ is the tension of the D-brane.
In the last step we used the fact that $T = \frac{1}{2\pi^2 g_o^2}$ \cite{Sen:1999xm, Sen:2021tpp}.

Apart from the modes with $L_0 =0 $ discussed above, typically one also finds modes with $L_0 < 0$ which are open-string tachyons.
The way to define the path integral over a tachyon field is to integrate over imaginary values of the field:
\begin{align}
    \int_{-\i\infty}^{\i \infty} \frac{\d \phi}{\sqrt{2\pi}} \,  e^{-\frac{1}{2} \, h  \phi^2 }
    = \frac{\i}{\sqrt{-h}} \,,\quad \text{ for } h < 0\, .
    \label{integratetachyon}
\end{align}
Here, we have chosen the steepest descent contour that runs from $-\i\infty$ to $\i \infty$ to match an analogous contour choice on the matrix integral side \cite{Eniceicu:2022nay, Eniceicu:2022dru}.

Depending on the exact problem being studied, we might need to choose a constant multiple of this steepest descent contour which lies along the imaginary axis. 
Some common examples of this constant multiple are $\pm 1$,  or $\pm \frac{1}{2}$, or even $0$ (in which case the saddle point under consideration does not contribute to the particular quantity being studied).
In other words, one needs to specify a defining contour and see what combination of steepest descent contours is homologous to the defining contour.

Fortunately, the duality between the matrix integral and string theory holds for genus-perturbation theory around each saddle point.
We need to specify a defining contour for the eigenvalues of the matrix integral that will pick out a particular linear combination of Lefshetz thimbles to sum over.
String theory also needs a corresponding defining contour in the complex plane of the open-string tachyon.
Thus,  there is a one to one correspondence between this set of choices on the string theory side and on the matrix integral side.

\section{The gapped phase of type 0B minimal superstring theory}
\label{sec:stringgapped}

The gapped phase of the matrix integral is dual to the type 0B minimal superstring with one sign of the Liouville bulk cosmological constant \cite{Klebanov:2003wg}.
Our convention is such that $\mu > 0$ corresponds to the gapped phase.
The spectral curve is the two-sheeted Riemann surface (\ref{spectralcurveweak}) and the eigenvalue instantons are located in the forbidden region $x\in (-1,1)$ \cite{Seiberg:2003nm}.

The first task is to identify the ZZ branes corresponding to these eigenvalue instantons.
For the type 0B $(2,4k)$ minimal superstring with $\mu > 0$, there are $2k-1$ ZZ branes, matching the number of one-eigenvalue instantons in the matrix integral, given in equation (\ref{xnweak}).
The $(m,n)$ labels of these ZZ branes are given by \cite{Seiberg:2003nm}
\begin{equation}
    m = 1 \, ,
    \quad
    n \in \{2,4,6,\ldots,4k-2\}\, .
\end{equation}
The ZZ brane with label $(1,n)$ corresponds on the matrix side to the eigenvalue instanton of the effective potential (\ref{veffweak}) located at $x_n = \cos \frac{\pi n}{4k}$.
In particular, $m+n$ is odd and thus the necessary boundary states are the ones appearing in (\ref{mnliouvneven}), with the wavefunctions given in (\ref{psi-ns-mn}) and the first line of (\ref{mnliouvilleevenR}).

As a basic check of the duality, let us compute the relative tensions of these ZZ branes.
The disk amplitude (with no vertex operator insertions) for the FZZT brane labeled by momentum $P$ is \cite{Seiberg:2003nm}
\begin{equation}
    D_P \,\, \propto \,\, 
    b^2 \sinh \pi b P \sinh \pi P b^{-1} - 
    \cosh \pi P b \cosh \pi P b^{-1}
    \, ,
\end{equation}
up to some $P$-independent constant.\footnote{The constant is somewhat convention dependent; it depends on $\mu, \mu_\text{B}$, and $b$.}
Thus, the ZZ disk amplitude (which determines the tension of the branes) is given by
\begin{equation}
    T_{(1,n)} =  D_{P_{1,n}} - D_{P_{1,-n}}
    \,\,
    \propto \,\,  (-1)^{k+n/2+1}  \sin \frac{\pi n}{4 k} \, .
\end{equation}
To get a number independent of the string coupling, we compute the ratio
\begin{align}
    \frac{T_{(1,n)}}{T_{(1,2k)}} = (-1)^\frac{n-2k}{2} \, \sin \frac{\pi n}{4k}\, .
\end{align}
This agrees with the matrix integral result in (\ref{vratioweak}).

\subsection{Annulus between two non-identical branes}

In this subsection, we compute the cylinder diagram between two different ZZ branes labeled by $(1,n)$ and $(1,n')$ where $n$ and $n'$ are both even.
We will first specify the contributions of the individual constituent CFTs and then integrate over the modular parameter of the cylinder.

\paragraph{Contribution from the NS-Ishibashi part of the state.}
The contribution of the Liouville NS-Ishibashi part of the boundary states is \cite{Ahn:2002ev}
\begin{align}
    Z^\ns_\text{Liouville}(q) = \frac{1}{2}
    \sum_{l = |n-n'|+1, 2}^{n+n'-1} \chi^\ns_{(1,l)}(q)
    =\frac{1}{2}
    \sqrt{\frac{\theta_3(q)}{\eta(q)^{3}}}
    \sum_{l = |n-n'|+1, 2}^{n+n'-1} \left( q^{- \frac{1}{8} (b^{-1} + l b)^2 } - q^{- \frac{1}{8} (b^{-1} - l b)^2 } \right) \, .
    \label{znsliouvweak}
\end{align}
Here, we have used the notation $\sum_{l = |n-n'|+1, 2}^{n+n'-1}$ to denote that the sum runs from $l = |n-n'|+1$ to $n+n'-1$ in steps of $2$. 
We have used (\ref{chinsmn}) in the second equality. 
This result can be derived using the explicit boundary state wavefunction $\Psi^\ns_{m,n}$ given in (\ref{psi-ns-mn}), (\ref{mnliouvneven}) and converting from the closed string channel to the open string channel via the modular S transformation.
Note that the index $l$ in the sum on the right side of (\ref{znsliouvweak}) is always an odd integer, which means that the label $(1,l)$ of the degenerate primaries appearing in the open string channel are in the NS sector.

The contribution from the NS part of the state of the $(2,4k)$ minimal SCFT is the NS identity character (\ref{chimatterns11}).
The contribution from the ghosts is $\frac{\eta(q)^{3}}{\theta_3(q)}$, which, as is usual in minimal string theories, serves to cancel the oscillator modes of the other sectors.

Multiplying these three contributions and integrating over the modular parameter $t$ with the correct measure, we get the string theory annulus\footnote{We use the standard convention that in the open string channel, the open string has length $\pi$ and the time circle has length $2\pi t$. In the closed string channel, the closed string has length $2\pi$ and the time interval has length $\pi/t = \pi s$. We also use the standard variables $q = e^{-2\pi t}$ and $\widetilde{q} = e^{-2\pi/t}$.}
\begin{multline}
    A^{\ns}_{(1,n),(1,n')} 
    = \frac{1}{2}\int_{0}^{\infty} \frac{\d t}{t} \sum_{j \in \mathbb{Z}} \sum_{l = |n-n'|+1, 2}^{n+n'-1} \Big( q^{\frac{(8k j + 4k + l - 1)(8k j - l - 1)}{16 k}} - q^{\frac{(8k j + 4k + l + 1)(8k j - l + 1)}{16 k}} \\
    - q^{\frac{(8k j + 4k - l - 1)(8k j + l - 1)}{16 k}} + q^{\frac{(8k j + 4k - l + 1)(8k j + l + 1)}{16 k}} \Big) \, .
    \label{znsweak}
\end{multline}
Note that we have written the measure factor as $\frac{\d t}{t}$ assuming that $n \neq n'$.
If $n = n'$, there is an additional symmetry factor of $\frac{1}{2}$.
When $n \neq n'$, we can straightforwardly evaluate (\ref{znsweak}) and get\footnote{One needs to use the following integral repeatedly:
\begin{align}
\label{eq:int_exp_over_t}
    \int_{0}^{\infty} \d t \, \frac{e^{- a t} - e^{-b t}}{t} 
    &= \log \frac{b}{a} \, ,
    \quad \quad  a,b > 0
    \, .
\end{align}
} 
\begin{align}
    A^{\ns}_{(1,n),(1,n')} &= \frac{1}{2}
    \sum_{l = |n-n'|+1,2}^{n+n'-1} 
    \log \left( \frac{ \cot^2 \frac{\pi (l+1)}{8k}}{\cot^2 \frac{\pi (l-1)}{8k}} \right) 
    = \frac{1}{2}\log \left( 
    \frac{ \cot^2 \frac{\pi (n+n')}{8k}}{\cot^2 \frac{\pi (n-n')}{8k}} \right) \, .
    \label{znsnnprime}
\end{align}

\paragraph{Contribution from the R-Ishibashi part of the state.}
The contribution of the Liouville R-Ishibashi part of the boundary state (\ref{mnliouvneven}) can be obtained using the explicit wavefunctions given in the first line of (\ref{mnliouvilleevenR}), 
\begin{equation}
\begin{split}
    Z^\r_\text{Liouville}(q) &= -
    \int_{0}^{\infty} \d P \, \frac{ \sqrt{2} \cosh( \pi P b^{-1} ) \sinh ( \pi P n b ) \sinh ( \pi P n' b ) }{\cosh (\pi P b)} \sqrt{\frac{\theta_2(\widetilde{q})}{2\eta(\widetilde{q})^{3}}} \,\widetilde{q}^{\, \frac{P^2}{2}}  \\
    &= -\frac{1}{2}\sqrt{\frac{\theta_4(q)}{\eta(q)^{3}}}
    \sum_{l = |n-n'|+1, 2}^{n+n'-1} (-1)^{(l - n - n' + 1)/2} \left( q^{- \frac{1}{8} (b^{-1} + l b)^2 } + q^{- \frac{1}{8} (b^{-1} - l b)^2 } \right) \, .
\end{split}
\label{zrliouville}
\end{equation}
The overall minus sign in the first line arises because we should really first compute a cylinder between two FZZT branes labeled by momenta $P_1$ and $P_2$:
The R-Ishibashi parts of these states, given in (\ref{openrp}) and (\ref{psirminus}), will give rise to a factor of $\sin (2\pi P P_1) \sin(2\pi P P_2)$.
We should then analytically continue $P_1$ and $P_2$ to the special purely imaginary values $\frac{\i}{2}(b^{-1} \pm n b)$ and $\frac{\i}{2}(b^{-1} \pm n' b)$, and combine the four terms so obtained according to (\ref{fzztrcombinations}).
The overall minus sign in the first line of (\ref{zrliouville}) then arises because of the identity $\sin(\i x)\sin(\i y) = - \sinh(x)\sinh(y)$.

Note also that in the second line of (\ref{zrliouville}), the summation index $l$ is always odd and the final expression is an alternating sum over $\nstilde$ characters in the open string channel (\ref{chinstildemn}).\footnote{The corresponding result in Ref.~\cite{Ahn:2002ev} is stated slightly incorrectly; it is missing the alternating signs between successive terms in the sum over $l$.}

The matter contribution in the open string channel is given by the $\nstilde$ character (\ref{chimatternstilde11}).
The ghost contribution is $-\frac{\eta(q)^{3}}{\theta_4(q)}$, which once again cancels all the oscillator modes of the matter and Liouville sectors.
Note the important minus sign in the ghost contribution \cite{Irie:2007mp, Polchinski:1998rr, Sen:2021tpp}.\footnote{
This minus sign arises because the operator $e^{-\varphi}$, that is part of the NS sector vertex operators in the $-1$ picture, has odd GSO parity.
This is the same minus sign that, in the torus and annulus amplitudes of type II superstring theory in ten dimensions, gives rise to the relative minus sign in the expression $\theta_3^4 - \theta_4^4$ that appears in annulus and torus amplitudes of type IIB string theory \cite{Polchinski:1998rr,Sen:2021tpp}.
\label{footnote:ghostminussign}
}

Multiplying the three contributions, we get the string theory RR annulus (for $n \neq n'$)
\begin{multline}
    A^{\r}_{(1,n),(1,n')} 
    = \frac{1}{2}\int_{0}^{\infty} \frac{\d t}{t} \sum_{j \in \mathbb{Z}}  \sum_{l = |n-n'|+1,2}^{n+n'-1} (-1)^{(l-n-n'+1)/2} \Big( q^{\frac{(8k j + 4k + l - 1)(8k j - l - 1)}{16 k}}  \\
   + q^{\frac{(8k j + 4k + l + 1)(8k j - l + 1)}{16 k}} + q^{\frac{(8k j + 4k - l - 1)(8k j + l - 1)}{16 k}} + q^{\frac{(8k j + 4k - l + 1)(8k j + l + 1)}{16 k}} \Big) \, .
   \label{zrweak}
\end{multline}
There are no subtleties for $n \neq n'$ and we can evaluate the above integrals to get
\begin{align}
    A^{\r}_{(1,n),(1,n')} = \frac{1}{2}\log \left( \frac{\sin^2 \frac{\pi (n-n')}{4k}}{\sin^2 \frac{\pi (n+n')}{4k}} \right)\, .
    \label{arnnprime}
\end{align}

\paragraph{Combining the two contributions.}
We now combine the contribution from the NS-Ishibashi part of the state (\ref{znsnnprime}) and the R-Ishibashi part of the state (\ref{arnnprime}) to get
\begin{equation}
\begin{split}
     A_{(1,n),(1,n')} 
    &=  
    A^{\ns}_{(1,n),(1,n')} + A^{\r}_{(1,n),(1,n')}
    = \log \left( \frac{\sin^2 \frac{\pi (n-n')} {8k}}{\sin^2 \frac{\pi (n+n')} {8k}} \right) \, .
    \label{annprimeweak}
\end{split}
\end{equation}
Exponentiating both sides, this result exactly matches with the matrix computation (\ref{crosscylinderweakmatrix}).

The cylinder between two different branes does not have divergences and we could also have obtained it using the FZZT annulus and the relationship between FZZT and ZZ boundary states \cite{Seiberg:2003nm, Irie:2007mp}, as shown in appendix \ref{app:FZZT_wc}.

\subsection{Normalization of the one-instanton contribution}

Consider a one-instanton contribution to the partition function, with the instanton labeled by $(1,n)$ (recall that $n$ is even in the gapped phase).
To evaluate the normalization constant, we need to compute the worldsheet cylinder with $\ket{(1,n)}$ boundary state on both ends.
Naively, the result (\ref{annprimeweak}) diverges when $n = n'$.
The reason for this divergence and how to cure it using string field theory was explained in section \ref{sftinput}, following Refs.~\cite{Sen:2021qdk, Sen:2021tpp, Chakravarty:2022cgj}.

Adding the contribution (\ref{znsweak}) from the NS-Ishibashi part and (\ref{zrweak}) for $n = n'$, and accounting for an additional symmetry factor of $\frac{1}{2}$ for two identical boundaries, we get
\begin{equation}
\begin{split}
    A_{(1,n)} = \frac{1}{2}\left( A^\ns + A^\r \right)
    &= \int_{0}^{\infty} \frac{\d t}{2 t} \sum_{j \in \mathbb{Z}}  \sum_{l = 3,4}^{2n-1} \Big( q^{\frac{(8k j + 4k + l - 1)(8k j - l - 1)}{16 k}} + q^{\frac{(8k j + 4k - l + 1)(8k j + l + 1)}{16 k}} \Big) \\
    & \hspace{0.3in} - \int_{0}^{\infty} \frac{\d t}{2 t} \sum_{j \in \mathbb{Z}}  \sum_{l = 1,4}^{2n-1} \Big( q^{\frac{(8k j + 4k + l + 1)(8k j - l + 1)}{16 k}} + q^{\frac{(8k j + 4k - l - 1)(8k j + l - 1)}{16 k}} \Big) \\
    &= \int_{0}^{\infty} \frac{\d t}{2 t} \sum_{j \in \mathbb{Z}}  \sum_{\ell = 1}^{n/2} \Big( q^{\frac{(4k j + 2k + 2\ell - 1)(2k j - \ell)}{2 k}} + q^{\frac{(4k j + 2k - 2\ell + 1)(2k j + \ell)}{2 k}}  \\
    & \hspace{1in} - q^{\frac{(4k j + 2k + 2\ell - 1)(2k j - \ell + 1)}{2 k}} - q^{\frac{(4k j + 2k - 2\ell + 1)(2k j + \ell - 1)}{2 k}} \Big) \, ,
\end{split}
\end{equation}
where we have reorganized the summation in going to the last line.

The $t$ integral has a divergence coming from the $t \to \infty$ end because of terms with $j=0$ in the integrand:
\begin{equation}
    q^{- \frac{(2k + 2\ell - 1) \ell}{2 k}} + q^{\frac{(2k - 2\ell + 1)\ell}{2 k}} - q^{- \frac{(2k + 2\ell - 1)(\ell - 1)}{2 k}} - q^{\frac{(2k - 2\ell + 1)(\ell - 1)}{2 k}} \, .
    \label{jequalszero}
\end{equation}
We can integrate the $j \neq 0$ terms using the integral (\ref{eq:int_exp_over_t}).
For $\ell > 1$ and $j=0$, we do not need any advanced methods from string field theory to deal with the divergences in (\ref{jequalszero}).
There are no zero modes and the tachyons are paired i.e., one bosonic and one fermionic, so we effectively just use \eqref{eq:int_exp_over_t} with $a,b < 0$.
So we isolate the only remaining terms with $\ell=1$ and $j=0$ to get
\begin{equation}
\begin{split}
    A_{(1,n)}
    &= \frac{1}{2} \log \left( \prod_{j \in \mathbb{Z}^*}  \prod_{\ell = 1}^{n/2} \frac{(2k j - \ell + 1)(2k j + \ell - 1)}{(2k j - \ell)(2k j + \ell)} \right) 
    \\ & \hspace{0.5in} 
    + \frac{1}{2} \log \left(  \prod_{\ell = 2}^{n/2} \frac{(\ell - 1)^2}{\ell^2} \right) + \int_{0}^{\infty} \frac{\d t}{2 t} \left( q^{- \frac{(2k+1)}{2 k}} - 2 + q^{\frac{(2k-1)}{2 k}} \right) \\
    &= \frac{1}{2} \log \left( \prod_{j \in \mathbb{Z}^*} \frac{1}{1 - \frac{(n/4k)^2}{j^2}} \right) + \log \left( \frac{2}{n} \right) + \int_{0}^{\infty} \frac{\d t}{2 t} \left( q^{- \frac{(2k+1)}{2 k}} - 2 + q^{\frac{(2k-1)}{2 k}} \right) \\
    &= \log \left( \frac{\frac{\pi} {2k}}{\sin 
    \frac{\pi n} {4k}} \right) + \int_{0}^{\infty} \frac{\d t}{2 t} \left( q^{- \frac{(2k+1)}{2 k}} - 2 + q^{\frac{(2k-1)}{2 k}} \right) \, .
\end{split}
\end{equation}

The remaining $t$-integral is ill-defined since it contains a tachyon and two zero modes.
The procedure to fix these issues using string field theory was explained in section \ref{sftinput}, following Refs.~\cite{Sen:2021tpp, Chakravarty:2022cgj}.
Carrying out the necessary steps, one finds
\begin{equation}
     \cN_{(1,n)}
    = \frac{\frac{\pi} {2k}}{\sin 
    \frac{\pi n} {4k}} \cdot \frac{2 \i k}{\sqrt{4k^2 - 1}} \cdot \frac{1}{(8 \pi^3 T_{(1,n)})^{1/2}}
    = \frac{\i}{\sqrt{2 \pi (16 k^2 - 4)}} \csc  \frac{\pi n} {4k}  T_{(1,n)}^{-1/2} \, .
    \label{n1nweak}
\end{equation}
The $\i$ appears because of (\ref{integratetachyon}) and the factor $\frac{1}{(8 \pi^3 T_{(1,n)})^{1/2}}$ arises because of (\ref{replacementrule}). 
This agrees precisely with the answer (\ref{matrixanswerweak}) computed using the matrix integral.

Since we have found an agreement between the matrix integral and string theory for the annulus diagram between two identical branes and also for the annulus diagram between two non-identical branes, the normalization of a general multi-instanton contribution is guaranteed to also agree.
See, for example, Ref.~\cite{Eniceicu:2022dru} for explicit expressions for a general multi-instanton configuration.

\section{The ungapped phase of type 0B minimal superstring theory}
\label{sec:stringungapped}

The ungapped, edgeless phase of the matrix integral is dual to type 0B minimal superstring theory with $\mu < 0$ in the conventions that we are using \cite{Klebanov:2003wg}.
The spectral curve is the \emph{reducible} algebraic curve given in (\ref{spectralcurvestrong}), the eigenvalue density is supported along the entire real axis, and the eigenvalue instantons are located on the imaginary axis.

Next, we identify the ZZ branes corresponding to these eigenvalue instantons.
For the type 0B $(2,4k)$ minimal superstring in this phase, there are $k$ pairs of ZZ branes with opposite R-charge \cite{Seiberg:2003nm}, matching the number of one-eigenvalue instantons in the matrix integral (\ref{xnstrong}).
The $(m,n)$ labels of these ZZ branes are given by 
\begin{equation}
    m = 1 \, ,
    \quad
    n \in \{1,3,5,\ldots,2k-1\}\, .
\end{equation} 
Since $m+n$ is even, the necessary boundary states $\ket{(m,n)}$ are the ones appearing in (\ref{cardy-mn}).
In addition to these $k$ branes, we also have the \emph{anti-branes} with the opposite R-charge, which are denoted by $\ket{\overline{(1,n)}}$ and are given in (\ref{cardy-mn-bar}).

The ZZ brane with label $\ket{(1,n)}$ corresponds, on the matrix side, to the eigenvalue instanton of the effective potential $V_{\eff}^+$ located at $x_n = \i\sin \frac{\pi n}{4k}$.
The anti-brane $\ket{\overline{(1,n)}}$ corresponds to the eigenvalue instanton of the effective potential $V_{\eff}^-$ located at $x_{\overline{n}} = -\i\sin \frac{\pi n}{4k}$.

In fact, the effective potential $V_\eff^+(x)$, which is initially defined in the upper half-plane, has $k$ additional extrema when analytically continued to the lower half-plane, which correspond to ``ghost'' instantons.
See Ref.~\cite{Eniceicu:2023cxn} for a detailed discussion of ghost instantons in the Gross-Witten-Wadia integral.
An identical comment applies for $V_\eff^-(x)$.
Therefore, there are $2k$ ghost instantons, $\ket{\overline{(1,n)}}^{\gh}$ and $\ket{(1,n)}^{\gh}$.
The boundary state of a ghost brane differs by an overall minus sign from the state of the corresponding real brane \cite{Douglas:2003up, Marino:2022rpz, Okuda:2006fb}.
See figure \ref{fig:strongcouplingspectralcurve} for a pictorial summary of the above facts.

As earlier, we can compute the relative tensions of these ZZ branes.
The disk amplitude for the FZZT brane labeled by momentum $P$ is \cite{Seiberg:2003nm}
\begin{equation}
    D_P \,\, \propto \,\, 
    b^2 \cosh \pi b P \cosh \pi P b^{-1} - 
    \sinh \pi P b \sinh \pi P b^{-1}
    \, ,
\end{equation}
up to some $P$-independent constant.
Note that this disk amplitude comes just from the NS sector, so both the ZZ brane and its anti-brane will have the same disk amplitude,
\begin{equation}
    T_{(1,n)} = 
    D_{P_{1,n}} - 
    D_{P_{1,-n}}
    \,\,
    \propto \,\,  (-1)^{k+(n-1)/2}  \cos \frac{\pi n}{4 k} \, .
\end{equation}
To get a number independent of the string coupling, we compute the ratio
\begin{align}
    \frac{T_{(1,n)}}{T_{(1,1)}} = (-1)^\frac{n-1}{2} \, \frac{\cos \frac{\pi n}{4k}}{\cos \frac{\pi}{4k}}\, .
\end{align}
This agrees with the matrix integral result in (\ref{eqn:ungapped_tension_ratios}).
Since $\ket{(1,n')}^\gh = - \ket{(1,n')}$, the tension of a ghost brane is $T_{(1,n)^\gh} = - T_{(1,n)}$.

\subsection{Normalization of an instanton/anti-instanton pair}

First, we consider a two-instanton contribution to the partition function with the two instantons being a charge conjugate pair.
More precisely, we consider a two-instanton contribution with one instanton described by the boundary state $\ket{(1,n)}$ and the other described by the boundary state $\ket{\overline{(1,n)}}$.
In this case, the RR terms from the four different pairs of cylinder boundary conditions cancel each other.
Other than the fact that $n$ and $n'$ are now both odd, the NS contribution to the cylinder diagram is identical to (\ref{znsweak}).
Taking into account combinatorial factors due to Chan-Paton factors and the GSO projection, we get
\begin{align}
    A_{(1,n),\overline{(1,n)}} &= 
    \int_{0}^{\infty} \frac{\d t}{t} \sum_{j \in \mathbb{Z}} \sum_{l = 1, 2}^{2n-1} \Big( 
    q^{\frac{(8k j + 4k + l - 1)(8k j - l - 1)}{16 k}} - q^{\frac{(8k j + 4k + l + 1)(8k j - l + 1)}{16 k}}  \nonumber\\
    &\hspace{1.5in} - q^{\frac{(8k j + 4k - l - 1)(8k j + l - 1)}{16 k}} + q^{\frac{(8k j + 4k - l + 1)(8k j + l + 1)}{16 k}} 
    \Big) \, .
\end{align}
As written, the integral is ill-defined since it has a divergence at $t \to \infty$ coming from the $j=0, l=1$ terms in the integrand $( q^{-1/2} - 2 + q^{1/2})$.
Following Refs.~\cite{Sen:2021qdk, Chakravarty:2022cgj}, we will regulate this term using string field theory; the procedure to do so was explained in section \ref{sftinput}.

Apart from these divergent terms, the integral can be computed using \eqref{eq:int_exp_over_t}.
The contribution from terms with $j \in \mathbb{Z}^*$ and $l=1$ is\footnote{Here we use the notation $\mathbb{Z}^* = \mathbb{Z} \backslash \{ 0 \}$.}
\begin{align}
    \log \left[ \prod_{j \in \mathbb{Z^*}} \frac{(8k j + 4k + 2)(8k j)(8k j + 4k - 2)(8k j)}{(8k j + 4k)(8k j - 2)(8k j + 4k)(8k j + 2)} \right] 
    = \log \left[ \prod_{j \in \mathbb{Z^*}} \frac{ (1 - \frac{(1/2k)^2}{(2j+1)^2}) }{(1 - \frac{(1/2k)^2}{(2j)^2})} \right] \nonumber \\
    & \hspace{-5in} = 2\log \left[ \frac{2k}{\sqrt{4k^2 - 1}} \prod_{j \in \mathbb{Z^+}} \frac{1 - \frac{(1/2k)^2}{(2j-1)^2} }{1 - \frac{(1/2k)^2}{(2j)^2} } \right] 
    = 2\log \left[ \frac{\pi}{ \sqrt{16k^2 - 4} }  \cot \frac{\pi}{4k} \right] \, .
\end{align}
Similarly, the contribution from terms in the integrand with $j \in \mathbb{Z}$ and $l > 1$ is
\begin{equation}
\begin{split}
    \log & \left[ \prod_{j \in \mathbb{Z}} \frac{(8k j + 4k + l + 1)(8k j - l + 1)(8k j + 4k - l - 1)(8k j + l - 1)}{(8k j + 4k + l - 1)(8k j - l - 1)(8k j + 4k - l + 1)(8k j + l + 1)} \right] = 2\log \left[ \frac{ \cot \frac{\pi (l+1)}{8k}}{\cot \frac{\pi (l-1)}{8k}} \right] \, .
\end{split}
\end{equation}
When we add these terms, the cotangents inside the logarithm telescope, and we get the result
\begin{align}
    A_{(1,n),\overline{(1,n)}} 
    &= 2\log \left[ \frac{\pi}{ \sqrt{16k^2 - 4} }  \cot \frac{\pi n}{4k} \right] + 2\int_{0}^{\infty} \frac{\d t}{2 t} \Big( q^{-1/2} - 2 + q^{1/2} \Big) \, .
    \label{logNwithdivnnbar}
\end{align}
Fixing the zero modes using (\ref{replacementrule}) and dealing with the tachyon using (\ref{integratetachyon}), we get the final result
\begin{equation}
    \cN_{(1,n),\overline{(1,n)}} = \frac{\pi^2}{ (16k^2 - 4) }  \cot^2 \frac{\pi n}{4k} \cdot (- 4) \cdot \frac{1}{8 \pi^3 T_n} = - \frac{\cot^2 \frac{\pi n}{4k}}{ 2 \pi (16k^2 - 4) }  \frac{1}{T_n} .
    \label{Nnnbar}
\end{equation}
This agrees precisely with the matrix computation in \eqref{eqn:ug_nm_rrs}.

\subsection{Normalization of other pairs of instantons}
Next, let us consider a more general two-instanton contribution. 
The two instantons need to have opposite R-charge, so we consider one instanton described by the boundary state $\ket{(1,n)}$ and the other by $\ket{\overline{(1,n')}}$.
We need to take into account the cylinder connecting a brane to itself, as well as the cylinder connecting the two branes.
Each of these has a piece coming from the NS-Ishibashi part of the state and a piece coming from the R-Ishibashi part of the state.
The parts of the amplitude coming from NS-Ishibashi states can be calculated using (\ref{logNwithdivnnbar}) and (\ref{znsnnprime})
\begin{equation}
\label{annnsstrong}
\begin{split}
    A^{\ns} 
    &= 
    \frac{1}{2} A^{\ns}_{(1,n),(1,n)} + 
    \frac{1}{2} A^{\ns}_{(1,n'),(1,n')} 
    + A^{\ns}_{(1,n),(1,n')} \\
    &= \frac{1}{2} \log \left[ \frac{\pi^2 \, \cot \frac{\pi n}{4 k} \, \cot \frac{\pi n'}{4 k} \, \cot^2 \frac{\pi (n+n')}{8k}}{\left(16 k^2 - 4\right) \, \cot^2 \frac{\pi |n-n'|}{8k} } \right] + \int_{0}^{\infty} \frac{\d t}{2 t} \Big( q^{-1/2} - 2 + q^{1/2} \Big) \, .
\end{split}
\end{equation}

The contribution from the Liouville R-Ishibashi states to the cylinder between the branes $\ket{(1,n)}$ and $\ket{(1,n')}$ with $n$ and $n'$ both odd is
\begin{equation}
\begin{split}
    \int_{0}^{\infty} \d P \, &\frac{\sqrt{2} \cosh( \pi P b^{-1} ) \cosh ( \pi P n b ) \cosh ( \pi P n' b ) }{\cosh (\pi P b)} \sqrt{\frac{\theta_2(\widetilde{q})}{2\eta(\widetilde{q})^{3}}} \widetilde{q}^{\, \frac{P^2}{2}}  \\
    &= \sqrt{\frac{\theta_4(q)}{\eta(q)^{3}}}
    \sum_{l = |n-n'|+1, 2}^{n+n'-1} (-1)^{(l - n - n' + 1)/2} \left( q^{- \frac{1}{8} (1/b + l b)^2 } + q^{- \frac{1}{8} (1/b - l b)^2 } \right) \, .
\end{split}
\end{equation}
Moreover, the matter contribution in the open string channel is given by (\ref{chimatternstilde11}), and the ghost contribution is $-\frac{\eta(q)^{3}}{\theta_4(q)}$, see footnote \ref{footnote:ghostminussign}.
Thus, the contribution to the annulus from the R-Ishibashi part of the state is
\begin{multline}
    A^{\r}_{(1,n),(1,n')} 
    = - \int_{0}^{\infty} \frac{\d t}{t} \sum_{j \in \mathbb{Z}}  \sum_{l = |n-n'|+1,2}^{n+n'-1} (-1)^{(l-n-n'+1)/2} \Big( q^{\frac{(8k j + 4k + l - 1)(8k j - l - 1)}{16 k}}  \\
   + q^{\frac{(8k j + 4k + l + 1)(8k j - l + 1)}{16 k}} + q^{\frac{(8k j + 4k - l - 1)(8k j + l - 1)}{16 k}} + q^{\frac{(8k j + 4k - l + 1)(8k j + l + 1)}{16 k}} \Big) \, .
\end{multline}
Note that this integrand has a divergence at $t = 0$, which will be removed once we add all the necessary contributions.
Taking into account the sign flip in the R component of the $\ket{\overline{(1,n')}}$ state, we find the following contribution to the normalization constant
\begin{equation}
\label{annrstrongdef}
    A^{\r} = 
    \frac{1}{2} A^{\r}_{(1,n),(1,n)} + 
    \frac{1}{2} A^{\r}_{(1,n'),(1,n')} 
    - A^{\r}_{(1,n),(1,n')} \, ,
\end{equation}
It is shown in appendix \ref{app:rstrong} that 
\begin{equation}
    A^{\r} 
    = \frac{1}{2} \log \left[ \frac{\pi^2 \, \sin \frac{\pi n}{2k} \, \sin \frac{\pi n'}{2k}}{ \left(4 k^2-1 \right) \sin^2 \frac{\pi(n+n')}{4k} \, \sin^2 \frac{\pi(n-n')}{4k}} \right] + \int_{0}^{\infty} \frac{\d t}{2 t} \Big( q^{-1/2} - 2 + q^{1/2} \Big) \, .
    \label{arstrongresult}
\end{equation}
We now combine the contributions coming from the NS-Ishibashi and the R-Ishibashi parts of the state,
\begin{align}
    A = A^\ns +  A^\r\, .
\end{align}
The divergent pieces on the right side are
\begin{equation}
    \log \mathcal{N}_{\text{div}} 
    = \int_{0}^{\infty} \frac{\d t}{2 t} \Big( q^{-1/2} - 2 + q^{1/2} \Big) + \int_{0}^{\infty} \frac{\d t}{2 t} \Big( q^{-1/2} - 2 + q^{1/2} \Big) \, ,
\end{equation}
where the first term comes from the NS-Ishibashi contribution and the second from the R-Ishibashi contribution.
We again use the string field theory technique explained in section \ref{sftinput} to replace \cite{Sen:2021tpp, Chakravarty:2022cgj}
\begin{equation}
     \mathcal{N}_\text{div} \to - 4 \, \frac{1}{8 \pi^3 T^{1/2}_{(1,n)} T^{1/2}_{(1,n')}} \, .
     \label{logNdiv}
\end{equation}
Note that half of the zero modes come from $Z_{(1,n),(1,n)}$ and the other half from $Z_{(1,n'),(1,n')}$, so there is no ambiguity about the powers of the tensions. 

Combining the various pieces in (\ref{annnsstrong}), (\ref{arstrongresult}), and (\ref{logNdiv}), we get the final answer
\begin{equation}
\begin{split}
    \cN_{(1,n), \overline{(1,n')}} 
    = - \frac{1}{2 \pi (16k^2-4) T^{1/2}_{(1,n)} T^{1/2}_{(1,n')}} \times \frac{ \cos \frac{\pi n}{4k} \, \cos \frac{\pi n'}{4k} }{\sin^2 \frac{\pi(n+n')}{8k} \cos^2 \frac{\pi (n-n')}{8k}}   \, .
\end{split}
\end{equation}
This matches the matrix integral result in \eqref{eqn:ug_nm_rrd}.

Next, consider the normalization for a two-instanton configuration described by $\ket{(1,n)}$ and $\ket{(1,n')}^{\gh}$.
Since $\ket{(1,n')}^\gh = - \ket{(1,n')}$, we can obtain this result from the one we just computed from 
\begin{equation}
    \log \mathcal{N}_{(1,n), (1,n')^{\gh}} 
    = \log \mathcal{N}_{(1,n), \overline{(1,n')}} - 2 A^{\ns}_{(1,n),(1,n')} 
    \, .
\end{equation}
The calculation yields
\begin{equation}
\label{normstrong_rg}
    \cN_{(1,n), (1,n')^{\gh}} 
    = - \frac{1}{2 \pi (16k^2-4) T^{1/2}_{(1,n)} T^{1/2}_{(1,n')^{\gh}}} \times \frac{ \cos \frac{\pi n}{4k} \, \cos \frac{\pi n'}{4k} }{\cos^2 \frac{\pi(n+n')}{8k} \sin^2 \frac{\pi (n-n')}{8k}} \, .
\end{equation}
This matches the matrix integral result in \eqref{eqn:ug_nm_rg}. 

Finally, note that if we change all labels to their ghost partners, the string theory answer for $\cN_{A,B} T_{A}^{1/2} T_{B}^{1/2}$ is unchanged. 
This completes the match by reproducing \eqref{eqn:ug_nm_gr} and \eqref{eqn:ug_nm_gg}.

It is also worth mentioning that the naive computation for the normalization of an instanton and its ghost partner yields $\cN_{(1,n),(1,n)^\gh} = 1$; see also Ref.~\cite{Okuda:2006fb}.
However, by choosing appropriate integration contours involving a specific principal value prescription, one obtains a non-trivial result \cite{Marino:2022rpz}.
We can pick analogous contours in the string field theory path integral to reproduce the result of Ref.~\cite{Marino:2022rpz}.

We have restricted to two-instanton configurations here. 
In appendix \ref{sec:appgeneralungapped}, we compute the normalization constant for the general multi-instanton configuration and match it to the matrix integral result.

\subsection{Vanishing of the one-instanton contribution}

In the ungapped phase, we expect that the one-instanton contribution to the partition function vanishes.
The leading nonperturbative effect is a two-instanton effect.
This is a well-known fact in the Gross-Witten-Wadia model \cite{Marino:2008ya}. 
From the matrix integral perspective, this is because the one-loop prefactor multiplying the one-instanton contribution vanishes \cite{Eniceicu:2023cxn}.
It is the purpose of this subsection to argue this independently on the string theory side.
Note that this argument can be easily generalized; if we have a differing number of positive- and negative-charged branes, the normalization vanishes.

In the string theory computation of the one-instanton effect due to the $(1,n)$ instanton, with $n$ odd, the cylinder diagram in the closed string channel contains the exchange of the $P=0$ Ramond sector ground state; see the second line of (\ref{psiroddodd}).
Explicitly, the holomorphic part of the vertex operator is $c_1 e^{-\frac{\varphi}{2}} \mathcal{O}_{1,2k} \, \sigma \, e^{\frac{Q}{2} \phi}$. 
Here, $\varphi$ is the boson that appears in the bosonization of the $\beta\gamma$ ghosts with the operator $e^{-\frac{\varphi}{2}}$ having scaling dimension $\frac{3}{8}$, 
the operator $\mathcal{O}_{1,2k}$ is the Ramond sector ground state operator from the minimal SCFT with Kac labels $(1,2k)$ that has scaling dimension $\frac{c_\text{Matter}}{24}$, 
and $\sigma \, e^{\frac{Q}{2} \phi}$ is the Ramond sector ground state operator from the Liouville sector with scaling dimension $\frac{1}{16} + \frac{Q^2}{8} = \frac{c_\text{Liouville}}{24}$.
Thus, the total scaling dimension of this operator vanishes.
The amplitude for the D-brane to emit this closed string state with $P=0$ is nonzero because the wavefunctions in the second line of (\ref{psiroddodd}) contain two factors of $\cosh$ and no $\sinh$.
The exchange of this state leads to a divergence in the moduli space integral from the $t = 0$ end, which is an IR divergence in the closed string channel.
The form of this divergence is $- \int^\infty \d s \int_0 \d P \, e^{- \pi s P^2} \sim - \int^\infty \frac{\d s}{\sqrt{s}}$, where $s = 1/t$ is the closed string time.
This divergence has an overall minus sign because of the minus sign discussed in footnote \ref{footnote:ghostminussign}.
Thus, when we exponentiate the cylinder diagram to compute the normalization constant, we get zero, as needed.

More physically, the $(1,n)$ instanton with $n$ odd carries RR charge, and we should have a neutral brane-antibrane combination to get a nonzero answer.
This is the same underlying physics as the IR divergences discussed in the $\widehat{c}=1$ case in Refs.~\cite{DeWolfe:2003qf, Balthazar:2022apu, Sen:2022clw}.
This argument generalizes to show that if we have a differing number of positive- and negative-charged branes, the normalization vanishes.
In S-matrix computations, one needs to consider semi-inclusive cross-sections to get a non-zero result. 
The physical intuition is that summing over an infinite number of closed string states can capture the change in the RR flux.
The analog of this in the matrix integral is that if we compute an observable like $\det U$ in the ungapped phase, it does have a one-instanton correction \cite{Ahmed:2017lhl, Eniceicu:2023cxn}.

\section{Discussion and future directions}
\label{sec:discussion}

In this work, we have calculated, to one-loop order, the non-perturbative corrections to the partition function in both phases of the $(2,4k)$ type 0B minimal superstring theory.
The potential divergences in the annulus diagrams between various super-ZZ branes were cured using insights from string field theory, leading to a perfect agreement with results from the dual matrix integral, which we also computed explicitly for all $k$ and in both phases.

One immediate avenue for future work is to perform the same analysis in the type 0A minimal superstring theory, which is dual to a matrix integral over complex matrices \cite{Klebanov:2003wg, Morris:1990cq, Dalley:1991qg, Dalley:1991vr, Dalley:1991yi, Dalley:1992br}.

The $k \to \infty$ limit of the $(2,4k)$ series of type 0B minimal superstring theory is expected to be the same as the type 0B $\mathcal{N}=1$ super-JT gravity theory \cite{Stanford:2019vob}.
In the super-JT variables, only the ungapped phase is well understood, with the density of eigenvalues $\rho(x) \, \propto\, \cosh(x)$.
Starting from the ungapped phase, Ref.~\cite{Rosso:2021orf} presented a deformation of super-JT gravity that gives rise to the phase transition, provided the deformation parameter is large enough.
However, the gapped phase of the type 0B $\mathcal{N}=1$ super-JT theory is not well-understood, and there is no simple bulk theory that is known to capture the main physics.
One hint might be that, in the associated supersymmetric SYK model \cite{Fu:2016vas}, supersymmetry would have to be spontaneously broken at order one, to reproduce the gap in the density of eigenvalues.
It would be interesting if this possibility could be explicitly realized in super-SYK; we are not aware of any work along these lines.

From the point of view of the gapped phase of the $(2,4k)$ minimal superstring theory, the difficulty lies in the fact that the one-eigenvalue effective potential has $2k-1$ extrema in the interval $[-1,1]$, and so, in the $k \to \infty$ limit, the $k$-dependent rescaling that is needed sends one or both cut end-points to infinity.
If we take the limit while zooming in near the end-point $x = 1$, we believe that the theory will just become \emph{bosonic} JT gravity.
In this regard, consider the spectral density in \eqref{dosweak},
\begin{equation}
    \rho(x) \, \propto \, \sinh \left( 2k \acosh |x| \right) \Theta (|x|-1) \, .
\end{equation}
If we send $k \to \infty$ and zoom in around $x = 1$ as $x = 1 + \varepsilon E$, while holding $k^2 \varepsilon = \pi^2/2$ fixed, we obtain the JT spectral density \cite{Stanford:2017thb, Saad:2019lba}
\begin{equation}
    \rho(E) = \sinh \left( 2 \pi \sqrt{E} \right) \Theta (E) \, .
\end{equation}

It should also be noted that, in our calculations, the R-Ishibashi part of the boundary state gives nonzero contributions.
In the path integral language, this contribution arises from the spin-structure that is periodic around the $S^1$ direction of the cylinder.
From the string theory point of view, this is not surprising since D-branes carry Ramond-Ramond charge and therefore a closed string with periodic boundary conditions for fermions can be emitted and absorbed by a D-brane.
However, for readers familiar with the super-JT gravity literature, this non-vanishing might sound a bit surprising, as it was argued in Ref.~\cite{Stanford:2019vob} that the contribution from this spin-structure to the path integral of $\mathcal{N}=1$ JT gravity vanishes.\footnote{The precise statement of Ref.~\cite{Stanford:2019vob} is that the path integral of $\mathcal{N}=1$ super-JT gravity vanishes on a general two-manifold with (possibly disconnected) boundary, even if there is a single boundary circle with Ramond spin structure. See also Ref.~\cite{Iliesiu:2021are}.}
These two facts are, in fact, consistent, since the boundary Hamiltonian in type 0B super-JT gravity is $H = Q^2$. 
In the language of the type 0B minimal superstring, one is computing correlators of $\log(x-Q)$ and so we have to take appropriate linear combinations using the identity $\log (x^2 - Q^2) =  \log(x-Q) +  \log (x+Q)$ to get the correlators that are relevant for $H$.
It can be easily shown using the FZZT amplitudes computed in Refs.~\cite{Okuyama:2005rn, Irie:2007mp} for the ungapped, edgeless phase that the R-Ishibashi contributions vanish when we take the linear combinations.

\paragraph{Acknowledgments.}
We would like to thank Ashoke Sen, Shu-Heng Shao, Douglas Stanford and Gustavo J. Turiaci for discussions, and Shreya Vardhan for comments on the draft.
D.S.E.~is supported by the Shoucheng Zhang Graduate Fellowship.
C.M.~is supported by the DOE through DESC0013528 and the QuantISED grant DE-SC0020360.
C.M.~performed this work in part at Kavli Institute for Theoretical Physics (KITP), which is supported by NSF grant PHY-2309135 and also at Aspen Center for Physics, which is supported by NSF grant PHY-2210452.

\appendix
\addtocontents{toc}{\protect\setcounter{tocdepth}{1}}

\section{The gapped phase of unitary matrix integrals}
\label{app:weak}
We will focus on unitary matrix integrals of the form
\begin{equation}
    Z\left(N,g,t_l^\pm\right)=\int\frac{\d U}{\text{vol } U(N)}\exp\left(\frac{N}{g}\Tr(W(U))\right)\, , \label{eqn:definition_general_unitary}
\end{equation}
with
\begin{equation}
    W(z)=\sum_{l=1}^\infty\left(\frac{t_l^+}{l}z^l+\frac{t_l^-}{l}z^{-l}\right)\,.
\end{equation}
Here, the integral is performed over $N\times N$ unitary matrices $U$ using the standard Haar measure on $U(N)$, and the parameters $t_l^\pm$ are complex conjugates of one another, $t_l^+=\left(t_l^-\right)^\dagger$ in order to guarantee the reality of the potential $W(z)$. For the models of interest, namely the $(2,4k)$ minimal superstring theories, we will restrict the summation to the first $k$ terms, i.e. $t_l^\pm$ will be set to 0 for $l>k$. The 't Hooft coupling parameter $g$, which we take to be real and positive can be absorbed into the definitions of the coupling constants $t_l^\pm$, but we choose to keep it explicit in order to study the behavior of the unitary matrix integral under the simultaneous scaling of all of the couplings.
\subsection{Fundamental properties of the gapped phase}
The gapped (also known as the one-cut, or weak-coupling) phase of such matrix integrals was studied extensively in previous work \cite{Gross:1980he, Marino:2008ya, Periwal:1990gf, Oota:2021qky, Jurkiewicz:1982iz, Migdal:1983qrz, Mandal:1989ry}, and corresponds to the case where the couplings $t_l^\pm$ are chosen such that the leading saddle in the large-$N$ limit of the matrix integral (\ref{eqn:definition_general_unitary}) has an associated density of eigenvalues $\rho(\theta)$ which is supported on a single arc of the unit circle, but does not cover the circle entirely. For convenience, in this subsection, we reproduce the expressions found in the aforementioned references which are relevant to our calculation.

The integral (\ref{eqn:definition_general_unitary}) can be recast as an integral over the eigenvalues $e^{i\theta_i}$ of a unitary matrix after gauge-fixing to the diagonal basis,
\begin{equation}
    Z(N,g,t_l^\pm)=\frac{1}{N!}\int\prod_{i=1}^N\frac{\d\theta_i}{2\pi}\,\left[\prod_{1\leq i<j\leq N}4\sin^2\left(\frac{\theta_i-\theta_j}{2}\right)\right]\,\exp\left(\frac{N}{g}\sum_{i=1}^NW(e^{i\theta_i})\right)\,.\label{eqn:definition_general_unitary_eigenvalues}
\end{equation}
Here, the integrals range from $\theta_i=0$ to $\theta_i=2\pi$ for each index $i$. In the large-$N$ limit, the leading saddle of the matrix integral (\ref{eqn:definition_general_unitary}) is determined in terms of a continuous density of eigenvalues $\rho(\theta)$ which we choose to normalize by requiring
\begin{equation}
    \int_{\text{cut}}\d\theta\,\rho(\theta)=1\,.\label{eqn:rho_constraint}
\end{equation}
Here, the integral is taken over the entire cut, i.e.~over the entire support of $\rho(\theta)$.

\subsubsection{The planar free energy and the large-\texorpdfstring{$N$}{N} eigenvalue density}
Before restricting to the one-cut situation, let us review the expressions for the planar free energy associated to the matrix integral (\ref{eqn:definition_general_unitary}) and the saddle-point equation which determines the eigenvalue density in the large-$N$ limit. We shall take the definition of the planar free energy in the large-$N$ limit to be given by
\begin{equation}
    F_{0}(g,t_l^\pm)=\lim_{N\rightarrow\infty}\frac{g^2}{N^2}\log Z(N,g,t_l^\pm)\,.\label{eqn:planar_free_energy}
\end{equation}
The large-$N$ saddle-point limit associated to an eigenvalue density $\rho(\theta)$ allows us to rewrite sums of functions of eigenvalues as integrals,
\begin{align}
    \sum_{i=1}^Nf(\theta_i)&\rightarrow N\int_{\text{cut}}\d\theta\,\rho(\theta)f(\theta)\,,\label{eqn:saddle_sum_integral}\\
    \sum_{1\leq i<j\leq N}f(\theta_i,\theta_j)&\rightarrow\frac{N^2}{2}\mathcal{P}\int_{\text{cut}}\d\theta\,\rho(\theta)\int_{\text{cut}}\d\theta'\,\rho(\theta')\,
    f(\theta,\theta')\,,
\end{align}
where the notation `$\mathcal{P}$' represents the Cauchy principal value prescription, which takes into account the fact that terms with $i =j$ are not present on the left hand side. 
Using this rewriting, the planar free energy takes the form
\begin{equation}
    F_0(g,t_l^\pm)=g\int_{\text{cut}}\d\theta\,\rho(\theta)W(e^{i\theta})+\frac{g^2}{2}\mathcal{P}\int_{\text{cut}}\d\theta\,\rho(\theta)\int_{\text{cut}}\d\theta'\,\rho(\theta')\,\log\left(4\sin^2\left(\frac{\theta-\theta'}{2}\right)\right)\,.\label{eqn:planar_free_energy_formula}
\end{equation}
The eigenvalue density $\rho(\theta)$ is determined by extremizing the free energy with respect to $\rho(\theta)$, subject to the constraint (\ref{eqn:rho_constraint}). This constraint can be implemented by introducing a Lagrange multiplier $\Lambda$, and extremizing the combination
\begin{equation}
    \mathcal{F}(\{\rho(\theta)\},\Lambda)=F_0(g,t_l^\pm)-g\Lambda\left(\int_{\text{cut}}\d\theta\,\rho(\theta)-1\right)\,.
\end{equation}
Since
\begin{equation}
    \delta\mathcal{F}(\{\rho(\theta)\},\Lambda)=g\int_{\text{cut}}\d\theta\,\delta\rho(\theta)\left(W(e^{i\theta})+g\,\mathcal{P}\int_{\text{cut}}\d\theta'\,\rho(\theta')\,\log\left(4\sin^2\left(\frac{\theta-\theta'}{2}\right)\right)-\Lambda\right)\,,
\end{equation}
the eigenvalue density must satisfy
\begin{equation}
    \Lambda=W(e^{i\theta})+g\,\mathcal{P}\int_{\text{cut}}\d\theta'\,\rho(\theta')\,\log\left(4\sin^2\left(\frac{\theta-\theta'}{2}\right)\right)\,,\,\forall\theta\in\text{supp}(\rho)\,.\label{eqn:rho_eom_with_Lambda}
\end{equation}
Taking a derivative of the previous equation with respect to $\theta$ leads to the saddle-point equation which determines $\rho(\theta)$,
\begin{equation}
    0=ie^{i\theta}W'(e^{i\theta})+g\,\mathcal{P}\int_{\text{cut}}\d\theta'\,\rho(\theta')\,\cot\left(\frac{\theta-\theta'}{2}\right)\,,\,\forall\theta\in\text{supp}(\rho)\,.\label{eqn:Wprime}
\end{equation}
One can also remove the Lagrange multiplier from (\ref{eqn:rho_eom_with_Lambda}) by taking the difference between it and a copy of itself at another angle $\theta_0\in\text{supp}(\rho)$, to obtain
\begin{equation}
    W(e^{i\theta})=W(e^{i\theta_0})+g\,\mathcal{P}\int_{\text{cut}}\d\theta'\,\rho(\theta')\,\left[\log\left(\sin^2\left(\frac{\theta_0-\theta'}{2}\right)\right)-\log\left(\sin^2\left(\frac{\theta-\theta'}{2}\right)\right)\right]\,.
\end{equation}
This relation can then be used to remove the double integral in the expression (\ref{eqn:planar_free_energy_formula}) for the free energy,
\begin{equation}
    F_0(g,t_l^\pm)=\frac{g}{2}W(e^{i\theta_0})+\frac{g}{2}\int_{\text{cut}}\d\theta\,\rho(\theta)\,W(e^{i\theta})+\frac{g^2}{2}\mathcal{P}\int_{\text{cut}}\d\theta'\,\rho(\theta')\,\log\left(4\sin^2\left(\frac{\theta_0-\theta'}{2}\right)\right)\,,\label{eqn:F0_final_expression_B14}
\end{equation}
where $\theta_0$ denotes the angle corresponding to an arbitrary point in the support of the eigenvalue density.
\subsubsection{The planar resolvent and the one-cut saddle}
An important quantity in the study of large-$N$ asymptotics of matrix integrals is the planar resolvent,
\begin{equation}
\label{eqn:planar_resolvent}
    \omega_0(z) := \lim_{N\rightarrow\infty}\frac{1}{N}\ev{\Tr\left(\frac{1}{z-U}\right)} = \int_{\text{cut}}\d\theta\,\frac{\rho(\theta)}{z-e^{i\theta}}\,.
\end{equation}
Defining
\begin{equation}
    f(z) :=\lim_{N\rightarrow\infty}\frac{1}{N}\ev{\Tr\left(\frac{W'(z)-W'(U)-g/z+gU^{-1}}{z-U}\right)}\,,
\end{equation}
it follows that
\begin{equation}
    g\omega_0(z)^2+(W'(z)-g/z)\omega_0(z)=f(z)\,.\label{eqn:omega0f}
\end{equation}
To shows this result, first note that
\begin{equation}
    f(z)=\int_{\text{cut}}\d\theta\frac{\rho(\theta)}{z-e^{i\theta}}\left(W'(z)-W'(e^{i\theta})-\frac{g}{z}+\frac{g}{e^{i\theta}}\right)\,.
\end{equation}
Additionally,
\begin{align}
    \omega_0(z)^2&=\int_{\text{cut}}\d\theta\int_{\text{cut}}\d\theta'\frac{\rho(\theta)}{z-e^{i\theta}}\frac{\rho(\theta')}{z-e^{i\theta'}}\,,\\
    (W'(z)-g/z)\omega_0(z)&=\int_{\text{cut}}\d\theta'\frac{\rho(\theta')}{z-e^{i\theta'}}\left(W'(z)-\frac{g}{z}\right)\,,
\end{align}
and we can replace the derivative of $W(e^{i\theta})$ using (\ref{eqn:Wprime}):
\begin{equation}
    W'(e^{i\theta})=\frac{g}{e^{i\theta}}-2g\,\mathcal{P}\int_{\text{cut}}\d\theta'\frac{\rho(\theta')}{e^{i\theta}-e^{i\theta'}}\,.
\end{equation}
We deduce that
\begin{align}
    \int_{\text{cut}}\d\theta\frac{\rho(\theta)}{z-e^{i\theta}}\left(W'(e^{i\theta})-\frac{g}{e^{i\theta}}\right)&=-2g\,\mathcal{P}\int_{\text{cut}}\d\theta\int_{\text{cut}}\d\theta'\frac{\rho(\theta)\rho(\theta')}{(z-e^{i\theta})(e^{i\theta}-e^{i\theta'})}\nonumber\\
    &=-g\,\mathcal{P}\int_{\text{cut}}\d\theta\int_{\text{cut}}\d\theta'\frac{\rho(\theta)\rho(\theta')}{e^{i\theta}-e^{i\theta'}}\left(\frac{1}{z-e^{i\theta}}-\frac{1}{z-e^{i\theta'}}\right)\nonumber\\
    &=-g\int_{\text{cut}}\d\theta\int_{\text{cut}}\d\theta'\frac{\rho(\theta)\rho(\theta')}{(z-e^{i\theta})(z-e^{i\theta'})}\nonumber\\
    &=-g\omega_0(z)^2\,,
\end{align}
from which we conclude that (\ref{eqn:omega0f}) holds. Finally, from this we can deduce that the planar resolvent can be expressed as
\begin{equation}
    \omega_0(z)=\frac{1}{2g}\left(\frac{g}{z}-W'(z)+\sqrt{\left(W'(z)-\frac{g}{z}\right)^2+4gf(z)}\right)\,.
\end{equation}
The one-cut saddle of the unitary matrix integral (\ref{eqn:definition_general_unitary}) is obtained by requiring
\begin{equation}
    \sqrt{\left(W'(z)-\frac{g}{z}\right)^2+4gf(z)}=\frac{M(z)}{z}\sqrt{\sigma(z)}\,,\qquad\sigma(z)=(z-\alpha)(z-\beta)\,,
\end{equation}
where $\alpha$ and $\beta$ denote the endpoints of the cut, and $M(z)$ is a Laurent polynomial in $z$ \cite{Oota:2021qky}.

\subsubsection{The Schwinger-Dyson equations for the planar resolvent}
We now present the derivation of the Schwinger-Dyson equations satisfied by the planar resolvent which will be useful in our calculation. The starting point is the expression for the planar resolvent presented above:
\begin{equation}
    \frac{2g\zeta\omega_0(\zeta)}{\sqrt{\sigma(\zeta)}}=\frac{g}{\sqrt{\sigma(\zeta)}}-\frac{\zeta W'(\zeta)}{\sqrt{\sigma(\zeta)}}+M(\zeta)\,.\label{eqn:starting_SD}
\end{equation}
Let $z\in\mathbb{C}^*\,\backslash\,\text{supp}(\rho)$, and let $C_{\text{cut}}$ be a contour surrounding the cut, but excluding the origin and the point $z$. Dividing the previous equation by $\zeta-z$ and taking a contour integral on both sides, we have
\begin{equation}
    \oint_{C_{\text{cut}}}\frac{\d \zeta}{2\pi i}\frac{2g\zeta\omega_0(\zeta)}{(\zeta-z)\sqrt{\sigma(\zeta)}}=\oint_{C_{\text{cut}}}\frac{\d\zeta}{2\pi i}\frac{g}{(\zeta-z)\sqrt{\sigma(\zeta)}}-\oint_{C_{\text{cut}}}\frac{\d\zeta}{2\pi i}\frac{\zeta W'(\zeta)}{(\zeta-z)\sqrt{\sigma(\zeta)}}\,,\label{eqn:first_oint}
\end{equation}
where the term involving $M(\zeta)$ vanished since $M(\zeta)$ is a Laurent polynomial and is therefore holomorphic away from the origin. The contour $C_{\text{cut}}$ can be expressed as a combination of contours
\begin{equation}
    C_{\text{cut}}=C_{\infty}-C_z-C_0\,,
\end{equation}
where $C_{\infty}$ is a contour enclosing the cut, the origin, and the point $z$, $C_z$ is a contour enclosing $z$, that is away from the cut and excludes the origin, and $C_0$ is a contour enclosing the origin, that is away from the cut and excludes the point $z$. 

Since $\omega_0(\zeta)\sim1/\zeta$ as $\zeta\rightarrow\infty$, the integral of the left side of (\ref{eqn:first_oint}) along the contour $C_{\infty}$ vanishes. 
Additionally, $\omega_0(\zeta)$ does not have a singularity at the origin.
Therefore, we have
\begin{equation}
    \oint_{C_{\text{cut}}}\frac{\d\zeta}{2\pi i}\frac{2g\zeta\omega_0(\zeta)}{(\zeta-z)\sqrt{\sigma(\zeta)}}=-\oint_{C_{z}}\frac{\d\zeta}{2\pi i}\frac{2g\zeta\omega_0(\zeta)}{(\zeta-z)\sqrt{\sigma(\zeta)}}=-\frac{2gz\omega_0(z)}{\sqrt{\sigma(z)}}\,.
\end{equation}
Similarly, we have
\begin{equation}
    \oint_{C_{\text{cut}}}\frac{\d\zeta}{2\pi i}\frac{g}{(\zeta-z)\sqrt{\sigma(\zeta)}}=-\oint_{C_z}\frac{\d\zeta}{2\pi i}\frac{g}{(\zeta-z)\sqrt{\sigma(\zeta)}}=-\frac{g}{\sqrt{\sigma(z)}}\,.
\end{equation}
Thus, we have
\begin{equation}
    \omega_0(z)=\frac{1}{2z}+\frac{1}{2gz}\oint_{C_{\text{cut}}}\frac{\d\zeta}{2\pi i}\frac{\zeta W'(\zeta)}{(\zeta-z)}\frac{\sqrt{\sigma(z)}}{\sqrt{\sigma(\zeta)}}\,,\qquad\forall z\in\mathbb{C}^*\,\backslash\,\text{supp}(\rho)\,.\label{eqn:firstSD}
\end{equation}
We could have also started with a version of (\ref{eqn:starting_SD}) where we first divided both sides by $\zeta$ before performing the contour integrals,
\begin{equation}
    \oint_{C_{\text{cut}}}\frac{\d\zeta}{2\pi i}\frac{2g\omega_0(\zeta)}{(\zeta-z)\sqrt{\sigma(\zeta)}}=\oint_{C_{\text{cut}}}\frac{\d\zeta}{2\pi i}\frac{g}{\zeta(\zeta-z)\sqrt{\sigma(\zeta)}}-\oint_{C_{\text{cut}}}\frac{\d\zeta}{2\pi i}\frac{W'(\zeta)}{(\zeta-z)\sqrt{\sigma(\zeta)}}\,.\label{eqn:second_oint}
\end{equation}
The previous argument can be repeated, with the only relevant modification being that we need to take into account the extra pole at $\zeta=0$ in the first term on the right side of the identity. It follows that
\begin{equation}
    \frac{\omega_0(z)}{\sqrt{\sigma(z)}}=\frac{1}{2z\sqrt{\sigma(z)}}-\frac{1}{2z\sqrt{\sigma(0)}}+\frac{1}{2g}\oint_{C_{\text{cut}}}\frac{\d\zeta}{2\pi i}\frac{W'(\zeta)}{(\zeta-z)}\frac{1}{\sqrt{\sigma(\zeta)}}\,,\qquad\forall z\in\mathbb{C}^*\,\backslash\,\text{supp}(\rho)\,.\label{eqn:secondSD}
\end{equation}
This latter identity is useful since it can be expanded in powers of $1/z$ for $|z|>1$ to derive two equations which we will need in our calculation. At first order in $z^{-1}$, we have
\begin{equation}
    \oint_{C_{\text{cut}}}\frac{\d\zeta}{2\pi i}W'(\zeta)\frac{\sqrt{\sigma(0)}}{\sqrt{\sigma(\zeta)}}=-g\,, \label{eqn:SD_firstorder}
\end{equation}
and at second order in $z^{-1}$, we have
\begin{equation}
    \oint_{C_{\text{cut}}}\frac{\d\zeta}{2\pi i}\frac{\zeta W'(\zeta)}{\sqrt{\sigma(\zeta)}}=-g\,.\label{eqn:SD_secondorder}
\end{equation}

\subsection{The derivatives of the free energy and of the effective potential}

The goal of this subsection is to evaluate the first two derivatives of the planar free energy $F_0(g,t_l^\pm)$, as well as the first derivative of the effective potential $V_{\text{eff}}(z;g)$ with respect to the coupling $g$. Our starting point consists of the definitions of the unitary matrix integral (\ref{eqn:definition_general_unitary}) and of the planar free energy (\ref{eqn:planar_free_energy}). Taking a derivative of (\ref{eqn:planar_free_energy}) with respect to $g$, we have,
\begin{align}
    \partial_gF_0(g,t_l^\pm)&=\frac{2}{g}F_0(g,t_l^\pm)+\lim_{N\rightarrow\infty}\frac{g^2}{N^2}\partial_g\log Z(N,g,t_l^\pm)\nonumber\\
    &=\frac{2}{g}F_0(g,t_l^\pm)-\lim_{N\rightarrow\infty}\frac{1}{N}\ev{\Tr(W(U))}\nonumber\\
    &=\frac{2}{g}F_0(g,t_l^\pm)-\int_{\text{cut}}\d\theta\,\rho(\theta)W(e^{i\theta})\,,
\end{align}
where the second line follows from taking the derivative of the definition (\ref{eqn:definition_general_unitary}), and the third follows from (\ref{eqn:saddle_sum_integral}). Recall the expression (\ref{eqn:F0_final_expression_B14}) for the planar free energy,
\begin{equation}
    F_0(g,t_l^\pm)=\frac{g}{2}W(e^{i\theta_0})+\frac{g}{2}\int_{\text{cut}}\d\theta\,\rho(\theta)\,W(e^{i\theta})+\frac{g^2}{2}\mathcal{P}\int_{\text{cut}}\d\theta'\,\rho(\theta')\,\log\left(4\sin^2\left(\frac{\theta_0-\theta'}{2}\right)\right)\,,
\end{equation}
where $\theta_0$ is the angle corresponding to an arbitrary point on the cut. Together, the previous two relations allow us to write,
\begin{equation}
    \partial_gF_0(g,t_l^\pm)=W(e^{i\theta_0})+g\,\mathcal{P}\int_{\text{cut}}\d\theta'\,\rho(\theta')\,\log\left(4\sin^2\left(\frac{\theta_0-\theta'}{2}\right)\right)\,.
\end{equation}
The effective potential an eigenvalue feels as a result of both the external potential $W(z)$ and the eigenvalue repulsion contribution, both of which are analytically continued away from the unit circle, takes the form,
\begin{align}
    V_{\text{eff}}(z;g)&=-W(z)-\lim_{N\rightarrow\infty}\frac{g}{N}\ev{\Tr\log\left(-\frac{(z-U)^2}{z}U^{-1}\right)}\nonumber\\
    &=-W(z)-g\int_{\text{cut}}\d\theta\;\rho(\theta)\log\left(-\frac{(z-e^{i\theta})^2}{ze^{i\theta}}\right)\,.\label{eqn:Veff}
\end{align}
On the cut, the effective potential is real and is given by the following formula involving the principal value
\begin{equation}
    V_{\text{eff}}(z;g)=-W(z)-g\,\mathcal{P}\int_{\text{cut}}\d\theta\;\rho(\theta)\log\left(-\frac{(z-e^{i\theta})^2}{ze^{i\theta}}\right)\,.\label{eqn:Veff_principal}
\end{equation}
We note that the first derivative of the planar free energy can therefore be expressed in the form,
\begin{equation}
    \partial_gF_0(g,t_l^\pm)=-V_{\text{eff}}(e^{i\theta_0};g)\,,\label{eqn:dgF0}
\end{equation}
where $\theta_0$ denotes an arbitrary angle on the cut. 
Since for $z$ away from the cut, from (\ref{eqn:planar_resolvent}), we get
\begin{equation}
    2\omega_0(z)-\frac{1}{z}=\partial_z\int_{\text{cut}}\d\theta'\,\rho(\theta')\log\left(-\frac{(z-e^{i\theta'})^2}{ze^{i\theta'}}\right)\,,
\end{equation}
it follows from (\ref{eqn:Veff}) that
\begin{equation}
    \partial_gV_{\text{eff}}'(z;g)=-\partial_g\left(2g\omega_0(z)-\frac{g}{z}\right)\,.\label{eqn:dgVeff}
\end{equation}
To evaluate the right side of this equality, we can use the Schwinger-Dyson equations (\ref{eqn:firstSD}), (\ref{eqn:SD_firstorder}), and (\ref{eqn:SD_secondorder}), along with
\begin{equation}
    \frac{\zeta}{z(\zeta-z)}\partial_g\left(\frac{\sqrt{\sigma(z)}}{\sqrt{\sigma(\zeta)}}\right)=\frac{1}{z}\frac{\sqrt{\sigma(0)}}{\sqrt{\sigma(z)}}\partial_g\left(\frac{\sqrt{\sigma(0)}}{\sqrt{\sigma(\zeta)}}\right)-\frac{\zeta}{\sqrt{\sigma(z)}}\partial_g\left(\frac{1}{\sqrt{\sigma(\zeta)}}\right)\,,
\end{equation}
to obtain
\begin{align}
    \partial_g\left(2g\omega_0(z)-\frac{g}{z}\right)&=\frac{1}{z}\frac{\sqrt{\sigma(0)}}{\sqrt{\sigma(z)}}\,\partial_g\oint_{C_{\text{cut}}}\frac{\d\zeta}{2\pi i}W'(\zeta)\frac{\sqrt{\sigma(0)}}{\sqrt{\sigma(\zeta)}}-\frac{1}{\sqrt{\sigma(z)}}\,\partial_g\oint_{C_{\text{cut}}}\frac{\d\zeta}{2\pi i}\frac{\zeta W'(\zeta)}{\sqrt{\sigma(\zeta)}}\nonumber\\
    &=\frac{1}{\sqrt{\sigma(z)}}-\frac{1}{z}\frac{\sqrt{\sigma(0)}}{\sqrt{\sigma(z)}}\,.
\end{align}
To obtain $\partial_gV_{\text{eff}}(z;g)$, we need to integrate (\ref{eqn:dgVeff}) with respect to $z$ and impose the appropriate boundary condition. The boundary condition can be determined by analyzing the behavior of $V_{\text{eff}}(z;g)+V_{\text{eff}}(1/z;g)$ at large $|z|$. Since for $|z|>1$,
\begin{equation}
    V_{\text{eff}}(z;g)+V_{\text{eff}}(1/z;g)=-W(z)-W(1/z)-g\,\int_{\text{cut}}\d\theta'\,\rho(\theta')\log\left(z^2(1-e^{i\theta'}/z)^2(1-e^{-i\theta'}/z)^2\right)\,,
\end{equation}
for large $|\Lambda|$, we have
\begin{equation}
    V_{\text{eff}}(\Lambda;g)+V_{\text{eff}}(1/\Lambda;g)=-W(\Lambda)-W(1/\Lambda)-g\log(\Lambda^2)+O(1/\Lambda)\,.
\end{equation}
Therefore,
\begin{equation}
    \partial_gV_{\text{eff}}(\Lambda;g)+\partial_gV_{\text{eff}}(1/\Lambda;g)=-\log(\Lambda^2)+O(1/\Lambda)\,,
\end{equation}
which allows us to conclude that
\begin{equation}
    \partial_g V_{\text{eff}}(z;g)=\log\left(\frac{4z\sqrt{\sigma(0)}}{\left(z+\sqrt{\sigma(0)}+\sqrt{\sigma(z)}\right)^2}\right) \,.
\label{eqn:dgVeff_final}
\end{equation}
We shall take $\theta_0$ to denote one of the ends of the cut in order to simplify the expressions. For instance, this allows us to remove the principal value prescription from the relevant integrals. However, the more important reason for doing so is to leverage the fact that
\begin{equation}
    V_{\text{eff}}'(e^{i\theta_0};g)=-W'(e^{i\theta_0})-2g\omega_0(e^{i\theta_0})+\frac{g}{e^{i\theta_0}}=0\,,
\end{equation}
which follows since
\begin{equation}
    2g\omega_0(z)+W'(z)-\frac{g}{z}=\frac{M(z)}{z}\sqrt{\sigma(z)}\,,
\end{equation}
and $\sigma(z)=0$ for a cut endpoint, $z=e^{i\theta_0}$. Thus,
\begin{equation}
    \partial_g^2F_0(g,t_l^\pm)=-\partial_gV_{\text{eff}}(e^{i\theta_0};g)\,,
\end{equation}
where we used the previous observation to conclude that the right side will not contain a term coming from the derivative with respect to $g$ acting on the location of the cut endpoint. Thus,
\begin{equation}
    \partial_g^2 F_0(g,t_l^\pm) = \log\left(\frac{ \alpha + \beta + 2\sqrt{\sigma(0)}}{4 \sqrt{\sigma(0)}} \right)
    \,.\label{eqn:dg2F}
\end{equation}

As a consistency check, let us consider two particular limits of this formula.
First, consider a small eigenvalue cut that goes from $\alpha = e^{-\i \delta}$ to $\beta =e^{\i \delta}$. 
In this case, $\sqrt{\sigma(0)} = - 1$, which follows from $\sqrt{\sigma(z)} \approx z$ for $|z| \gg 1$ and $\sqrt{\sigma(z)}$ having a branch cut on the eigenvalue cut. 
Thus,
\begin{equation}
    \partial_g^2 F_0(g,t_l^\pm) \approx 2 \log \frac{\delta}{2} \, ,
\end{equation}
which matches the analogous result for a one-cut solution of the Hermitian matrix integral \cite{Marino:2007te}.
Next, consider an eigenvalue cut covering almost the entire unit circle that goes from $\alpha = e^{\i (\pi - \delta)}$ to $\beta = e^{\i (\pi + \delta)}$.
In this case, we also have $\sqrt{\sigma(0)} = -1$. 
It follows that
\begin{equation}
    \partial_g^2 F_0(g,t_l^\pm) = O(\delta) \, .
\end{equation}
This says that the transition between the gapped and ungapped phases is at least third order \cite{Oota:2021qky}, as $\partial_g^2 F_0(g,t_l^\pm) = 0$ for the ungapped phase.

\subsection{The connected double-trace correlator}

The last important piece of the calculation of the instanton contributions is the connected double-trace correlator,
\begin{equation}
    A_{0,2}(z_1,z_2;g)=\frac{1}{4}\lim_{N\rightarrow\infty}\ev{\Tr\log\left(-\frac{(z_1-U)^2}{z_1}U^{-1}\right)\Tr\log\left(-\frac{(z_2-U)^2}{z_2}U^{-1}\right)}_c\,.\label{eqn:A02expression}
\end{equation}
The goal of this subsection is to derive a simple expression for this quantity in the one-cut phase of the unitary matrix integral, for generic values of the coupling constants $t_l^\pm$. As we shall see, the final answer depends only on $z_1$, $z_2$, and the locations of the endpoints of the cut. Our starting point is the evaluation of the connected double-resolvent correlator,
\begin{equation}
    R_{0,2}(z_1,z_2)=\lim_{N\rightarrow\infty}\ev{\Tr\left(\frac{1}{z_1-U}\right)\Tr\left(\frac{1}{z_2-U}\right)}_c\, .
\end{equation}
For $|z_2|>1$, we can write
\begin{equation}
    R_{0,2}(z_1,z_2)=\sum_{m=1}^\infty z_2^{-m-1}\lim_{N\rightarrow\infty}\ev{\Tr\left(\frac{1}{z_1-U}\right)\Tr\left(U^m\right)}_c\,.
\end{equation}
Each of the connected correlators in the sum above can be obtained by taking derivatives of $\omega_0(z_1)$ with respect to the couplings $t_l^\pm$ since
\begin{align}
    \partial_{t_m^\pm}\omega_0(z_1)&=\lim_{N\rightarrow\infty}\frac{1}{N}\partial_{t_m^\pm}\left(\frac{1}{Z(N,t_l^\pm)}\int\frac{\d U}{\text{vol } U(N)}\exp\left(\frac{N}{g}\Tr(W(U))\right)\Tr\left(\frac{1}{z_1-U}\right)\right)\nonumber\\
    &=\frac{1}{gm}\lim_{N\rightarrow\infty}\left(\ev{\Tr(U^{\pm m})\Tr\left(\frac{1}{z_1-U}\right)}-\ev{\Tr(U^{\pm m})}\ev{\Tr\left(\frac{1}{z_1-U}\right)}\right)\nonumber\\
    &=\frac{1}{gm}\lim_{N\rightarrow\infty}\ev{\Tr(U^{\pm m})\Tr\left(\frac{1}{z_1-U}\right)}_c\,.
\end{align}
Thus,
\begin{equation}
    R_{0,2}(z_1,z_2)=g\sum_{m=1}^\infty m\,z_2^{-m-1}\partial_{t_m^+}\omega_0(z_1)\,.
\end{equation}
We can determine $\partial_{t_m^+}\omega_0(z_1)$ using the Schwinger-Dyson equation (\ref{eqn:firstSD}),
\begin{equation}
    \omega_0(z)=\frac{1}{2z}+\frac{1}{2gz}\oint_{C_{\text{cut}}}\frac{\d\zeta}{2\pi i}\frac{\zeta W'(\zeta)}{\zeta-z}\frac{\sqrt{\sigma(z)}}{\sqrt{\sigma(\zeta)}}\,.
\end{equation}
We have
\begin{equation}
    \partial_{t_m^+}W'(\zeta)=\zeta^{m-1}\Rightarrow\sum_{m=1}^\infty mz_2^{-m-1}\partial_{t_m^+}W'(\zeta)=\frac{1}{(\zeta-z_2)^2}\,,
\end{equation}
where the last equation follows since we can take $\zeta$ arbitrarily close to the cut, i.e. arbitrarily close to having absolute value $|\zeta|=1$, and we are restricting to $|z_2|>1$. Thus,
\begin{equation}
    R_{0,2}(z_1,z_2)=\frac{1}{2z_1}\left(\oint_{C_{\text{cut}}}\frac{\d\zeta}{2\pi i}\frac{\zeta}{(\zeta-z_1)(\zeta-z_2)^2}\frac{\sqrt{\sigma(z_1)}}{\sqrt{\sigma(\zeta)}}+\sum_{m=1}^\infty m\,z_2^{-m-1}\oint_{C_{\text{cut}}}\frac{\d\zeta}{2\pi i}\frac{\zeta W'(\zeta)}{\zeta-z_1}\partial_{t_m^+}\left(\frac{\sqrt{\sigma(z_1)}}{\sqrt{\sigma(\zeta)}}\right)\right)
\end{equation}
Since
\begin{equation}
    \partial_{t_m^+}\sqrt{\sigma(z)}=\frac{1}{2\sqrt{\sigma(z)}}\partial_{t_m^+}\left(\alpha\beta-\left(\alpha+\beta\right)z\right)\,,
\end{equation}
we have
\begin{equation}
    \partial_{t_m^+}\left(\frac{\sqrt{\sigma(z_1)}}{\sqrt{\sigma(\zeta)}}\right)=\frac{\zeta-z_1}{2\sigma(\zeta)^{3/2}\sqrt{\sigma(z_1)}}\left((\alpha\beta-\zeta z_1)\partial_{t_m^+}(\alpha+\beta)+(\zeta+z_1-(\alpha+\beta))\partial_{t_m^+}(\alpha\beta)\right)\,.
\end{equation}
Taking a derivative with respect to $t_m^+$ of the Schwinger-Dyson equations (\ref{eqn:SD_firstorder}) and (\ref{eqn:SD_secondorder}),
\begin{align}
    \oint_{C_{\text{cut}}}\frac{\d\zeta}{2\pi i}W'(\zeta)\frac{\sqrt{\sigma(0)}}{\sqrt{\sigma(\zeta)}}&=-g\,,\\
    \oint_{C_{\text{cut}}}\frac{\d\zeta}{2\pi i}\frac{W'(\zeta)\zeta}{\sqrt{\sigma(\zeta)}}&=-g\,,
\end{align}
leads to
\begin{align}
    &\oint_{C_{\text{cut}}}\frac{\d\zeta}{2\pi i}\frac{\zeta W'(\zeta)}{2\sigma(\zeta)^{3/2}}\left(\alpha\beta\partial_{t_m^+}(\alpha+\beta)+\left(\zeta-(\alpha+\beta)\right)\partial_{t_m^+}(\alpha\beta)\right)=-\sigma(0)\oint_{C_{\text{cut}}}\frac{\d\zeta}{2\pi i}\frac{\partial_{t_m^+}W'(\zeta)}{\sqrt{\sigma(\zeta)}}\,,\\
    &\oint_{C_{\text{cut}}}\frac{\d\zeta}{2\pi i}\frac{\zeta W'(\zeta)}{2\sigma(\zeta)^{3/2}}\partial_{t_m^+}\left(-(\alpha+\beta)\zeta+\alpha\beta\right)=\oint_{C_{\text{cut}}}\frac{\d\zeta}{2\pi i}\frac{\zeta\partial_{t_m^+}W'(\zeta)}{\sqrt{\sigma(\zeta)}}\,.
\end{align}
Therefore,
\begin{equation}
    \oint_{C_{\text{cut}}}\frac{\d\zeta}{2\pi i}\frac{\zeta W'(\zeta)}{\zeta-z_1}\partial_{t_m^+}\left(\frac{\sqrt{\sigma(z_1)}}{\sqrt{\sigma(\zeta)}}\right)=\frac{1}{\sqrt{\sigma(z_1)}}\oint_{C_{\text{cut}}}\frac{\d\zeta}{2\pi i}\frac{(\zeta z_1-\sigma(0))}{\sqrt{\sigma(\zeta)}}\partial_{t_m^+}W'(\zeta)\,.
\end{equation}
We can use this relation to rewrite the double-resolvent correlator as
\begin{equation}
    R_{0,2}(z_1,z_2)=\frac{1}{2z_1}\left(\oint_{C_{\text{cut}}}\frac{\d\zeta}{2\pi i}\frac{\zeta}{(\zeta-z_1)(\zeta-z_2)^2}\frac{\sqrt{\sigma(z_1)}}{\sqrt{\sigma(\zeta)}}+\frac{1}{\sqrt{\sigma(z_1)}}\oint_{C_{\text{cut}}}\frac{\d\zeta}{2\pi i}\frac{(\zeta z_1-\sigma(0))}{\sqrt{\sigma(\zeta)}}\frac{1}{(\zeta-z_2)^2}\right)\,,
\end{equation}
which further simplifies to
\begin{equation}
    R_{0,2}(z_1,z_2)=\frac{1}{2\sqrt{\sigma(z_1)}}\oint_{C_{\text{cut}}}\frac{\d\zeta}{2\pi i}\frac{\sqrt{\sigma(\zeta)}}{(\zeta-z_1)(\zeta-z_2)^2}\,.
\end{equation}
The integrand is holomorphic away from the cut and the points $z_1$ and $z_2$. Thus, the contour integral can be evaluated as a difference between the corresponding integral along a simple contour $C_\infty$ enclosing the cut and both $z_1$ and $z_2$, and the two contours $C_{z_1}$ and $C_{z_2}$ enclosing $z_1$ and $z_2$, respectively,
\begin{equation}
    C_{\text{cut}}=C_{\infty}-C_{z_1}-C_{z_2}\,.
\end{equation}
The contour $C_\infty$ can be deformed on the Riemann sphere under the map $\zeta\rightarrow1/\zeta$ to evaluate to the corresponding residue at infinity. Since the integrand can be holomorphically extended to infinity, the integral along this contour vanishes. Alternatively, one can note that the absolute value of the integrand decays as $|\zeta|^{-2}$, and therefore the integral vanishes as the contour is pushed to infinity. The integrals along the contours enclosing the points $z_1$ and $z_2$ evaluate to the following expressions:
\begin{align}
    \oint_{C_{z_1}}\frac{\d\zeta}{2\pi i}\frac{\sqrt{\sigma(\zeta)}}{(\zeta-z_1)(\zeta-z_2)^2}&=\frac{\sqrt{\sigma(z_1)}}{(z_2-z_1)^2}\,,\\
    \oint_{C_{z_2}}\frac{\d\zeta}{2\pi i}\frac{\sqrt{\sigma(\zeta)}}{(\zeta-z_1)(\zeta-z_2)^2}&=\frac{(\alpha+\beta)(z_1+z_2)/2-z_1z_2-\alpha\beta}{(z_2-z_1)^2\sqrt{\sigma(z_2)}}\,.
\end{align}
Putting these together, we find
\begin{equation}
\begin{split}
    R_{0,2}(z_1,z_2)
    &= \frac{1}{4\sqrt{\sigma(z_1)}\sqrt{\sigma(z_2)}}\left(\frac{\left(\sqrt{\sigma(z_1)}-\sqrt{\sigma(z_2)}\right)^2}{(z_2-z_1)^2}-1\right) \\
    &= \frac{1}{2 (z_2-z_1)^2} \left(\frac{2 \alpha \beta + 2 z_1 z_2 - (\alpha + \beta)(z_1 + z_2)}{2 \sqrt{\sigma(z_1)} \sqrt{\sigma(z_2)}}-1\right) 
    \,.
    \label{eqn:R02}
\end{split}
\end{equation}
We see that the double-resolvent correlator depends only on the locations $z_1$, $z_2$, and the locations of the cut endpoints, which themselves are determined by the choice of couplings $t_l^\pm$. Although the derivation we presented here assumed $|z_2|>1$, the case $|z_2|<1$ can be carried out analogously to obtain the same expression.
The same result can also be obtained by converting the unitary matrix model to a Hermitian matrix model following Ref.~\cite{Mizoguchi:2004ne} and using the Hermitian result for $R_{0,2}$.

We now focus on determining the double-trace correlator $A_{0,2}(z_1,z_2;g)$. Note that
\begin{equation}
    \partial_{z_2}A_{0,2}(z_1,z_2;g)=\frac{1}{2}\lim_{N\rightarrow\infty}\ev{\Tr\log\left(-\frac{(z_1-U)^2}{z_1}U^{-1}\right)\Tr\left(\frac{1}{z_2-U}\right)}_c\,,\label{eqn:partial_A02}
\end{equation}
and subsequently that
\begin{equation}
    \partial_{z_1}\partial_{z_2}A_{0,2}(z_1,z_2;g)=\lim_{N\rightarrow\infty}\ev{\Tr\left(\frac{1}{z_1-U}\right)\Tr\left(\frac{1}{z_2-U}\right)}_c=R_{0,2}(z_1,z_2)\,.
\end{equation}
Thus, the difficulty in determining $A_{0,2}(z_1,z_2;g)$ comes solely from determining the relevant integration constants. Integrating equation (\ref{eqn:R02}) with respect to $z_1$, we find
\begin{equation}
    \partial_{z_2}A_{0,2}(z_1,z_2;g)=F(z_2)+\frac{1}{2(z_1-z_2)}\left(1-\frac{\sqrt{\sigma(z_1)}}{\sqrt{\sigma(z_2)}}\right)\,,\label{eqn:dzA02}
\end{equation}
where $F(z_2)$ is a function of $z_2$ which should be determined by boundary conditions. We can find this function by analyzing the behavior of the expression (\ref{eqn:partial_A02}) for $\partial_{z_2}A_{0,2}(z_1,z_2;g)$ at $z_1\rightarrow0$ and at $z_1\rightarrow\infty$. Concretely, consider the combination
\begin{multline}
    \partial_{z_2}A_{0,2}(z_1,z_2;g)+\partial_{z_2}A_{0,2}(z_1^{-1},z_2;g)\\
    =\frac{1}{2}\lim_{N\rightarrow\infty}\ev{\Tr\log\left(z_1^2(\mathbb{1}-z_1^{-1}U)^2(\mathbb{1}-z_1^{-1}U^{-1})^2\right)\Tr\left(\frac{1}{z_2-U}\right)}_c \, .
\end{multline}
Since this expression involves a connected correlator, the factor of $|z_1|^2$ inside the logarithm can be dropped without affecting the result. In the $z_1\rightarrow\infty$ limit, the argument of the logarithm will then become independent of $U$. More precisely, it will be proportional to the identity up to a phase that is independent of $U$, so the connected correlator will vanish. Thus,
\begin{equation}
    \lim_{z_1\rightarrow\infty}\partial_{z_2}\left(A_{0,2}(z_1,z_2;g)+A_{0,2}(z_1^{-1},z_2;g)\right)=0\,.
\end{equation}
We can use this to determine the function $F(z_2)$:
\begin{equation}
    F(z_2)=\frac{1}{4\sqrt{\sigma(z_2)}}+\frac{1}{4z_2}\left(1-\frac{\sqrt{\sigma(0)}}{\sqrt{\sigma(z_2)}}\right)\,.
\end{equation}
Integrating (\ref{eqn:dzA02}) with respect to $z_2$, we find
\begin{align}
    A_{0,2}(z_1,z_2;g)&=-\frac{1}{4}\log\left(\frac{(\alpha-\beta)^4}{\left((\alpha+\beta)-2\sqrt{\sigma(0)}\right)\left(z_2+\sqrt{\sigma(z_2)}+\sqrt{\sigma(0)}\right)^2}\right)\nonumber\\
    &-\frac{1}{2}\log\left(\frac{(z_2-z_1)^2}{\left(\sqrt{\sigma(z_2)}-\sqrt{\sigma(z_1)}\right)^2-(z_2-z_1)^2}\right)+G(z_1)\,.
\end{align}
By a similar argument as above, we require that
\begin{equation}
    \lim_{z_2\rightarrow\infty}\left(A_{0,2}(z_1,z_2;g)+A_{0,2}(z_1,z_2^{-1};g)\right)=0\,,
\end{equation}
from which we can deduce $G(z_1)$:
\begin{equation}
    G(z_1)=\frac{1}{4}\log\left(-\frac{\left(z_1+\sqrt{\sigma(z_1)}+\sqrt{\sigma(0)}\right)^2}{4\sqrt{\sigma(0)}}\right)\,.
\end{equation}
Finally, we have
\begin{equation}
    A_{0,2}(z_1,z_2;g)=\frac{1}{4}\log\left[\frac{2\sqrt{\sigma(0)}-(\alpha+\beta)}{4\sqrt{\sigma(0)}}\frac{\left(z_1+\sqrt{\sigma(z_1)}+\sqrt{\sigma(0)}\right)^2\left(z_2+\sqrt{\sigma(z_2)}+\sqrt{\sigma(0)}\right)^2}{\left(\left(\sqrt{\sigma(z_1)}+\sqrt{\sigma(z_2)}\right)^2-(z_2-z_1)^2\right)^2}\right]\,.\label{eqn:A02_final}
\end{equation}

\subsection{Leading multi-instanton contributions in the gapped phase}

The goal of this subsection is to study the leading-order eigenvalue instanton contributions to the unitary matrix integral (\ref{eqn:definition_general_unitary}) in the gapped phase. These correspond to saddles which involve removing eigenvalues from the cut and moving them to extrema of the effective potential, where they are integrated along their steepest descent contours.
We first write the matrix integral (\ref{eqn:definition_general_unitary}) as an integral over its eigenvalues instead of the corresponding angular variables,
\begin{align}
    Z(N,g,t_l^\pm)&=\frac{1}{N!}\prod_{i=1}^N\oint\frac{\d z_i}{2\pi iz_i}\exp\left(\frac{N}{g}\sum_{i=1}^NW(z_i)+\sum_{1\leq i<j\leq N}\log\left(-\frac{(z_i-z_j)^2}{z_iz_j}\right)\right)\,.
\end{align}
The different extrema of the effective potential correspond to different types of instantons, which we will index by $n$. We will analyze the general multi-instanton case, where $s_n$ eigenvalues are removed from the cut and placed at instanton location $z_n^\star$. We denote by $S$ the total number of eigenvalues removed from the cut, i.e.~$S=\sum_ns_n$. The contribution associated to this case involves integrating the displaced eigenvalues along their steepest descent contours while evaluating the integrals over the remaining eigenvalues in the large-$N$ limit,
\begin{align}
    Z^{(\{s_n\})}&(N,g,t_l^\pm)=\prod_\alpha\frac{1}{s_n!}\prod_{i=1}^S\int_{C_{z_i}}\frac{\d z_i}{2\pi iz_i}\exp\left(\frac{N}{g}\sum_{i=1}^S W(z_i)+\sum_{1\leq i<j\leq S}\log\left(-\frac{(z_i-z_j)^2}{z_iz_j}\right)\right)\cdot\nonumber\\
    &\cdot\frac{1}{(N-S)!}\prod_{i=S+1}^N\oint\frac{\d z_i}{2\pi iz_i}\exp\left(\frac{N-S}{g(1-S/N)}\sum_{i=S+1}^NW(z_i)+\sum_{S+1\leq i<j\leq N}\log\left(-\frac{(z_i-z_j)^2}{z_iz_j}\right)\right)\cdot\nonumber\\
    &\cdot\exp\left(\sum_{i=1}^S\sum_{j=S+1}^N\log\left(-\frac{(z_i-z_j)^2}{z_iz_j}\right)\right)\,.
\end{align}
Here, $C_{z_i}$ denotes the steepest descent contour for the $i$th eigenvalue, and depends only on the index $n$ associated to it, i.e.~$C_{z_i}$ will be the steepest descent contour passing through $z_n^\star$. We have also included the combinatorial factor $\binom{N}{S}$ to account for all the different choices of subsets of eigenvalues we choose to displace. We recognize the second and third line of the previous equation as an expectation value in the unitary matrix integral (\ref{eqn:definition_general_unitary}) with $N$ replaced by $N-S$ and $g$ replaced by $g(1-S/N)$:
\begin{align}
    Z^{(\{s_n\})}&(N,g,t_l^\pm)=Z^{(0)}(N-S,\,g(1-S/N),t_l^\pm)\prod_n\frac{1}{s_n!}\prod_{i=1}^S\int_{C_{z_i}}\frac{\d z_i}{2\pi iz_i}\prod_{1\leq i<j\leq S}\left(-\frac{(z_i-z_j)^2}{z_iz_j}\right)\cdot\nonumber\\
    &\cdot \ev{\exp\left(\frac{N}{g}\sum_{i=1}^S W(z_i)+\sum_{i=1}^S\Tr\log\left(-\frac{(z_i-U)^2}{z_i}U^{-1}\right)\right)}_{(N-S,\,g(1-S/N))}\,.\label{eqn:instanton_correction_ev}
\end{align}
In the large-$N$ limit, we have
\begin{align}
    \log\left(\frac{Z^{(0)}(N-S,\,g(1-S/N),t_l^\pm)}{Z^{(0)}(N,g,t_l^\pm)}\right)&\approx\frac{N^2}{g^2}\left(F_0(g(1-S/N),t_l^\pm)-F_0(g,t_l^\pm)\right)\\
    &=-\frac{SN}{g}\partial_gF_0(g,t_l^\pm)+\frac{S^2}{2}\partial_g^2F_0(g,t_l^\pm)+O(1/N)\,.
\end{align}
An expectation value of the exponential of a quantity such as the one in equation (\ref{eqn:instanton_correction_ev}),
\begin{equation}
    \mathcal{O}=\sum_{i=1}^S W(z_i)+\frac{g}{N}\sum_{i=1}^S\Tr\log\left(-\frac{(z_i-U)^2}{z_i}U^{-1}\right)\,,
\end{equation}
in the ensemble of size $N$, with coupling $g$, can be written in the large-$N$ limit as
\begin{equation}
    \ev{\exp\left(\frac{N}{g}\mathcal{O}\right)}_{(N,g)}=\exp\left[\ev{\frac{N}{g}\mathcal{O}}_{(N,g)}+\frac{1}{2}\ev{\frac{N^2}{g^2}\mathcal{O}^2}_{(N,g),\,c}+\dots\right]\,.
\end{equation}
We express the first term using the definition (\ref{eqn:Veff}) for the effective potential,
\begin{equation}
    \ev{\frac{N}{g}\mathcal{O}}_{(N,g)}=-\frac{N}{g}\sum_{i=1}^SV_{\text{eff}}(z_i;g)\,,
\end{equation}
and the second term as the sum of connected double-trace correlators,
\begin{equation}
    \ev{\frac{N^2}{g^2}\mathcal{O}^2}_{(N,g),\,c}=4\sum_{i=1}^S\sum_{j=1}^S A_{0,2}(z_i,z_j;g)\,,\label{eqn:second_term_A02}
\end{equation}
where we recall the expression (\ref{eqn:A02expression}) for the connected double-trace correlator. Since we are only interested in the multi-instanton contributions to one-loop order (i.e.~neglecting terms that are suppressed in powers of $1/N$), we can approximate the first term in the ensemble of size $N-S$, with coupling $g(1-S/N)$, as
\begin{equation}
    \ev{\frac{N}{g}\mathcal{O}}_{(N-S,\,g(1-S/N))}\approx-\frac{N}{g}\sum_{i=1}^S\left(V_{\text{eff}}(z_{n_i}^\star;g)+\frac{1}{2}V_{\text{eff}}''(z_{n_i}^\star;g)(z_i-z_{n_i}^\star)^2\right)+S\sum_{i=1}^S\partial_gV_{\text{eff}}(z_{n_i}^\star;g)\,.\\
\end{equation}
Here, we are denoting by $z_{n_i}^\star$ the corresponding saddle location of the $i$th eigenvalue.

Additionally, all the factors of $1/z_i$ appearing outside the expectation value in (\ref{eqn:instanton_correction_ev}), as well as the connected double-trace correlators appearing in (\ref{eqn:second_term_A02}) can be approximated by their corresponding values at coupling $g$, at the saddle locations,
\begin{equation}
    1/z_i\approx 1/z_{n_i}^\star\,,\qquad  A_{0,2}(z_i,z_j;g(1-S/N))\approx A_{0,2}(z_{n_i}^\star,z_{n_j}^\star;g)\,.
\end{equation}
Finally, the Vandermonde factors $(z_i-z_j)^2$ can be approximated by their saddle values when $z_i$ and $z_j$ correspond to different instanton types, but should not be approximated when $z_i$ and $z_j$ correspond to the same saddle location $z_n^\star$. The integrals can then be organized as independent $s_n\times s_n$ Gaussian matrix integrals associated to the saddles $z_n^\star$, where $n$ ranges over all the different instanton types. Each Gaussian matrix integral evaluates to
\begin{equation}
\label{gaussianintegral_n}
    \frac{1}{s_n!}\prod_{i\in\Omega_n}\int_{C_n}\frac{\d z_i}{2\pi}\,\prod_{\substack{i,j\in\Omega_n\\i<j}}(z_i-z_j)^2\prod_{i\in\Omega_n}\exp\left[-\frac{N}{2g}V_{\text{eff}}''(z_n^\star;g)(z_i-z_n^\star)^2\right]=\frac{G_2(s_n+1)}{(2\pi)^{s_n(s_n+1)/2}}\left(\sqrt{\frac{2\pi g}{NV_{\text{eff}}''(z_n^\star)}}\right)^{s_n^2}\,,
\end{equation}
where $\Omega_n$ denotes the subset of eigenvalue indices associated to the saddle $z_n^*$, and $G_2$ denotes the Barnes $G$-function. The one-loop multi-instanton contributions to the matrix integral (\ref{eqn:definition_general_unitary}) therefore take the simple form,
\begin{align}
    \frac{Z^{(\{s_n\})}(N,g,t_l^\pm)}{Z^{(0)}(N,g,t_l^\pm)}&=\prod_n\left[\frac{G_2(s_n+1)}{(2\pi)^{s_n(s_n+1)/2}}\mathcal{B}_n^{s_n^2}\exp\left(-\frac{N}{g}s_n\mathcal{A}_n\right)\right]\prod_{n<n'}\mathcal{C}_{n,n'}^{s_ns_n'}\,,
\end{align}
where we have defined
\begin{align}
    \mathcal{A}_n&=V_{\text{eff}}(z_n^\star;g)+\partial_gF_0(g,t_l^\pm)\,,\\
    \mathcal{B}_n&=-\frac{i}{z_n^\star}\sqrt{\frac{2\pi g}{NV_{\text{eff}}''(z_n^\star)}}\exp\left[\frac{1}{2}\partial_g^2F_0(g,t_l^\pm)+\partial_gV_{\text{eff}}(z_n^\star;g)+2A_{0,2}(z_n^\star,z_n^\star;g)\right]\,,\\
    \mathcal{C}_{n,n'}&=-\frac{(z_n^\star-z_{n'}^\star)^2}{z_n^\star z_{n'}^\star}\exp\left[\partial_g^2F_0(g,t_l^\pm)+\partial_gV_{\text{eff}}(z_n^\star;g)+\partial_gV_{\text{eff}}(z_{n'}^\star;g)+4A_{0,2}(z_n^\star,z_{n'}^\star;g)\right]\,.
\end{align}
We can use the results derived in the previous subsections to present simple expressions for these quantities. Using (\ref{eqn:dgF0}), we have
\begin{equation}
    \mathcal{A}_n=\int_{\alpha}^{z_n^\star}\d z\,V_{\text{eff}}'(z;g)\,.\label{eqn:mathcal_A}
\end{equation}
Combining (\ref{eqn:dgVeff_final}), (\ref{eqn:dg2F}), and (\ref{eqn:A02_final}), we find
\begin{align}
    \mathcal{B}_n&=-\frac{i}{4\sigma(z_n^\star)}\sqrt{-(\alpha-\beta)^2}\sqrt{\frac{2\pi g}{NV_{\text{eff}}''(z_n^\star)}}\,,\label{eqn:mathcal_B}\\
    \mathcal{C}_{n,n'}&=\frac{(\alpha-\beta)^2(z_n^\star-z_{n'}^\star)^2}{\left(\left(\sqrt{\sigma(z_n^\star)}+\sqrt{\sigma(z_{n'}^\star)}\right)^2-(z_n^\star-z_{n'}^\star)^2\right)^2}\,.\label{eqn:mathcal_C}
\end{align}

\subsection{The double-scaling limit}
The spectral curve of the unitary matrix integral (\ref{eqn:definition_general_unitary}) is related to the derivative of the effective potential (\ref{eqn:Veff}) and takes the form,
\begin{equation}
    y(z)=-V_{\text{eff}}'(z;g)=\frac{M(z)}{z}\sqrt{\sigma(z)}\,.
\end{equation}
In this subsection, we will restrict to the case of positive real couplings $t_l^+=t_l^-$. 
In the gapped phase, the large-$N$ eigenvalue density will cover a portion of the unit circle around $\theta=0$, and will be symmetric across the real axis. The endpoints of the cut will be given by
\begin{equation}
    \alpha=e^{\i\theta_0},\qquad\beta=e^{-\i\theta_0}\,,
\end{equation}
where $0<\theta_0<\pi$. As one tunes the couplings to transition between the gapped phase and the ungapped phase, one will see the endpoints of the cut coalesce at $\theta=\pi$, i.e.~$z=-1$. 
In order to reproduce the spectral curves associated to the conformal background of the $(2,4k)$ minimal superstring theories \cite{Seiberg:2003nm},\footnote{Note that, in order to agree with our analysis of the non-double scaled case, the square root is defined such that $\sqrt{1-x^2}=-|1-x^2|^{1/2}$ for $x\in[-1,1]$.}
\begin{equation}
    \Tilde{y}(x)=U_{2k-1}(x)\sqrt{1-x^2}\,,
\end{equation}
we will need to zoom in on the region near the point $z=-1$,
\begin{equation}
    z=-1+2\i\varepsilon x\,,
\end{equation}
and tune the couplings $t_l^\pm$ accordingly around their values at the $k$th multi-critical point \cite{Oota:2021qky},
\begin{equation}
    t_l^\pm = \frac{g(k!)^2}{(k+l)!(k-l)!}\left(1+\sum_{m=1}^ka_{l,m}\varepsilon^{2m}\right)\, . \label{eqn:multicrit}
\end{equation}
Here, $\varepsilon$ is the expansion parameter associated to double-scaling, and $a_{l,m}$ are finite constants which can be fixed by requiring that
\begin{equation}
    y(z)=2\i g{\binom{2k}{k}}^{-1}\varepsilon^{2k}\Tilde{y}(x)+O(\varepsilon^{2k+1})\, .\label{eqn:spectral_id}
\end{equation}
The double-scaling limit involves taking $N\rightarrow\infty$ and $\varepsilon\rightarrow0$, while holding fixed the quantity
\begin{equation}
    \dsfixedquantity={\binom{2k}{k}}^{-1}N\varepsilon^{2k+1}\,.
\end{equation}
The edges of the cut will then correspond to the angle
\begin{equation}
    \theta_0=\pi-2\varepsilon\,.
\end{equation}
The goal of this subsection is to obtain expressions for the quantities (\ref{eqn:mathcal_A}), (\ref{eqn:mathcal_B}), and (\ref{eqn:mathcal_C}) associated with the multi-instanton contributions found in the previous subsection in the double-scaling limit. In the dual $(2,4k)$ minimal superstring theory, these quantities are related to the tension of the ZZ brane with label $n$,
\begin{equation}
    \mathcal{T}_n=\frac{N}{g}\mathcal{A}_n\,,
\end{equation}
the exponential of the cylinder diagram connecting the ZZ brane with label $n$ to itself,
\begin{equation}
    \mathcal{N}_n=\frac{\mathcal{B}_n}{2\pi}\,,
    \label{nbrelation}
\end{equation}
and the exponential of the cylinder diagram connecting the ZZ brane with label $n$ to the one with label $n'$. The tension can be computed as follows:
\begin{align}
    \mathcal{T}_n&=-\frac{N}{g}\int_{e^{i\theta_0}}^{z_n^\star}\d z\,y(z)=4\dsfixedquantity\int_{\pm1}^{x_n^\star}\d x\,U_{2k-1}(x)\sqrt{1-x^2}\nonumber\\
    &=-\frac{4\dsfixedquantity}{(4k^2-1)}\left(-2kT_{2k}(x_n^\star)+x_n^\star U_{2k-1}(x_n^\star)\right)\sqrt{1-x_n^{\star2}}\,.
\end{align}
Since
\begin{equation}
    V_{\text{eff}}''(z;g)=2gk{\binom{2k}{k}}^{-1}\varepsilon^{2k-1}\frac{T_{2k}(x)}{\sqrt{1-x^2}}\,,
\end{equation}
the exponential of the cylinder diagram connecting identical boundary states takes the form (see (\ref{eqn:mathcal_B}) and (\ref{nbrelation}))
\begin{equation}
    \mathcal{N}_n=-\frac{\i}{8}\frac{\dsfixedquantity^{-1/2}}{\sqrt{\pi k\,T_{2k}(x_n^\star)}}\frac{(1-x_n^{\star2})^{1/4}}{1-x_n^{\star2}}\,,
\end{equation}
while the cross-cylinder takes the form (see (\ref{eqn:mathcal_C})),
\begin{equation}
    \mathcal{C}_{n,n'}=\frac{(x_n^\star-x_{n'}^\star)^2}{\left(1-x_n^\star x_{n'}^\star+\sqrt{1-x_n^{\star2}}\sqrt{1-x_{n'}^{\star2}}\right)^2}\,.
\end{equation}
The extrema of the effective potential correspond to the zeros of the Chebyshev polynomial $U_{2k-1}(x)$,
\begin{equation}
    x_n^\star=\cos\frac{\pi n}{4k},\qquad n\in\{2,4,\dots,2k-2\}\,.
\end{equation}
For these values, the quantities associated with the multi-instanton contributions take the forms:
\begin{align}
    \mathcal{T}_n&= \kappa\,(-1)^{\frac{n}{2}+1}\,\frac{8k}{4k^2-1}\sin\frac{\pi n}{4k}\,,\\
    \mathcal{N}_n&=-\frac{\i}{8}\left(\pi k(-1)^{n/2}\dsfixedquantity\right)^{-1/2}\cdot\frac{\sqrt{-\sin\frac{\pi n}{4k}}}{\sin^2\frac{\pi n}{4k}}\,,\\
    \mathcal{C}_{n,n'} &=\frac{\sin^2\frac{\pi(n-n')}{8k}}{\sin^2\frac{\pi(n+n')}{8k}}\,.
\end{align}
We find that the corresponding tensions are therefore related by
\begin{equation}
    \frac{\mathcal{T}_n}{\mathcal{T}_{2k}}=(-1)^{\frac{n-2k}{2}}\sin\frac{\pi n}{4k}\,,
\end{equation}
while the tensions and the exponentials of the cylinder diagrams with identical boundaries are related by
\begin{equation}
    \mathcal{N}_n\mathcal{T}_n^{1/2}=\frac{\i}{\sqrt{2\pi(16k^2-4)}}\frac{1}{\sin\frac{\pi n}{4k}}\,.
\end{equation}

\section{The ungapped phase of unitary matrix integrals}
\label{app:strong}
The ungapped or strong-coupling phase of the unitary matrix integrals,
\begin{equation}
    Z\left(N,g,t_l^\pm\right)=\int\frac{\d U}{\text{vol } U(N)}\exp\left(\frac{N}{g}\Tr(W(U))\right)\, ,
\end{equation}
with
\begin{equation}
    W(z)=\sum_{l=1}^k\left(\frac{t_l^+}{l}z^l+\frac{t_l^-}{l}z^{-l}\right)\,,
\end{equation}
corresponds to the case where the couplings $t_l^\pm$ are chosen such that the leading saddle in the large-$N$ limit has an associated density of eigenvalues $\rho(\theta)$ which is supported on the entire unit circle,
\begin{equation}
    \rho(\theta) = \frac{1}{2\pi} \left( 1 + \frac{1}{g} \sum_{l=1}^{k} \left( t^{+}_{l} e^{\i l\theta} + t^{-}_{l} e^{-\i l\theta} \right) \right) \, . 
\end{equation}
See \cite{Eniceicu:2023cxn, Oota:2021qky} for some recent expositions. The leading instanton contributions in this phase were studied in Ref.~\cite{Eniceicu:2023cxn}, and we shall refer the reader interested in a detailed derivation to appendix C of that reference. The goal of the current appendix is to list the main results found therein in the conventions of the present paper along with their counterparts in the double-scaling limit.

In contrast to the situation in the gapped phase, in the ungapped phase, there are two distinct notions of effective potential, $V_{\text{eff}}^\pm$ which are naturally defined starting from outside ($+$) and inside ($-$) the unit circle, respectively. 
These effective potentials are not analytic continuations of each other, but define two separate analytic functions in the full complex plane.
Additionally, as noted in Ref.~\cite{Eniceicu:2023cxn}, following \cite{Marino:2008ya, Ahmed:2017lhl} for the $k=1$ case, the leading nonperturbative contribution to the partition function is a two-instanton effect, where one instanton is associated with $V_{\text{eff}}^+$\,, and the other is associated with $V_{\text{eff}}^-$\,. 
Due to these features of the ungapped phase, it is important to distinguish between different types of instanton contributions, namely real/ghost and positive/negative-charge instantons, which we define in the following subsection. This is shown pictorially in figure \ref{fig:strongcouplingspectralcurve}.

\subsection{Naming conventions}
We will refer to branes as real/ghost and positive/negative-charge depending on the analogous properties of the matrix instantons which correspond to them. Concretely, we refer to a brane as having positive charge if the matrix instanton associated with it comes from $V_{\eff}^+$. 
Similarly, a brane will have negative charge if the matrix instanton associated with it comes from $V_{\eff}^-$.

Additionally, we will call a brane real if its corresponding matrix instanton location is on the same side of the unit circle as indicated by the sign of $V_{\eff}^\pm$ associated with it. 
For example, if the corresponding matrix instanton depends on $V_{\eff}^+(z)$ and $|z|>1$, then we shall call the associated brane a real brane. 
Instead, we will refer to it as a ghost brane if the corresponding matrix instanton location is on the other side of the unit circle compared to the one indicated by the sign of $V_{\eff}^\pm$.
For example, a matrix instanton depending on $V_{\eff}^-(z)$ for $|z|>1$ will be associated with a ghost brane. 
Note that this definition has the effect of the real branes having positive and negative tension in an alternating pattern. The same will be true for the ghost branes.

We will use the term ``charge-conjugate partner'' of a brane $A$ to denote the brane $\overline{A}$ associated with the matrix instanton obtained by flipping the indicated sign of $V_{\eff}^\pm$ and taking $z \rightarrow 1/z^\dagger$.\footnote{Therefore, all one-loop instanton contributions calculated in Ref.~\cite{Eniceicu:2023cxn} correspond to cylinders connecting a positive-charge real brane to a negative-charge real brane, or connecting a positive-charge ghost brane to a negative-charge ghost brane.}

We will use the term ``ghost partner'' of a brane $A$ to denote the brane $A^{\gh}$ corresponding to the instanton in the matrix integral associated with the same location $z$ as the instanton which $A$ corresponds to, but with opposite charge. 
For instance, if brane $A$ corresponds to the instanton associated with $V_{\eff}^+(z)$, with $|z|>1$, then $A$ is a positive-charge real brane, whose ghost partner will be the ghost instanton associated with $V_{\eff}^-(z)$. 
In this case, the ghost partner of $A$ is a negative-charge ghost brane.

\subsection{Leading instanton contributions in the ungapped phase}
\subsubsection{The effective potential and the instanton locations}
The effective potential an eigenvalue feels, i.e.~the combination of the external potential and the eigenvalue repulsion coming from the Vandermonde determinant has distinct expressions depending on whether it is defined outside ($+$) or inside ($-$) the unit circle \cite{Eniceicu:2023cxn}. 
This is because the expression for the resolvent,
\begin{equation}
    \omega_0(z)=\int_0^{2\pi}\d\theta\frac{\rho(\theta)}{z-e^{i\theta}}\,,
\end{equation}
takes different forms in the two regions,
\begin{equation}
    \omega_0^+(z)=\frac{1}{z}+\frac{1}{g}\sum_{l=1}^kt_l^-z^{-l-1}\,,\qquad\omega_0^-(z)=-\frac{1}{g}\sum_{l=1}^kt_l^+z^{l-1}\,.
\end{equation}
The derivative of the effective potential defined in the previous appendix takes the form,
\begin{equation}
    V_{\text{eff}}^\pm{'}(z;g)=-W'(z)-2g\omega_0^\pm(z)+\frac{g}{z}\,,
\end{equation}
and therefore evaluates to
\begin{align}
    V_{\text{eff}}^+{'}(z;g)&=-\frac{g}{z}-\sum_{l=1}^k\left(t_l^+z^{l-1}+t_l^-z^{-l-1}\right)\,,\label{eqn:Veffprime_strong_+}\\
    V_{\text{eff}}^-{'}(z;g)&=\frac{g}{z}+\sum_{l=1}^k\left(t_l^+z^{l-1}+t_l^-z^{-l-1}\right)\,.\label{eqn:Veffprime_strong_-}
\end{align}
Integrating with respect to $z$, we find
\begin{align}
    V_{\text{eff}}^+(z;g)&=-g\log z-\sum_{l=1}^k\left(\frac{t_l^+}{l}z^l-\frac{t_l^-}{l}z^{-l}\right)\,,\\
    V_{\text{eff}}^-(z;g)&=g\log z+\sum_{l=1}^k\left(\frac{t_l^+}{l}z^l-\frac{t_l^-}{l}z^{-l}\right)\,,
\end{align}
and we note that these expressions satisfy the identities,
\begin{equation}
    V_{\text{eff}}^-(z;g)+V_{\text{eff}}^+(z;g)=0,\qquad V_{\text{eff}}^-\left(1/z^\dagger;g\right)=\left(V_{\text{eff}}^+\left(z;g\right)\right)^\dagger\,,
\end{equation}
where the dagger symbol denotes complex conjugation. We also note that the large-$N$ limit of the unitary matrix integral takes a  simple form in the ungapped phase and receives no additional perturbative corrections,
\begin{equation}
\label{Z0_strong}
    Z^{(0)}(N,g,t_l^\pm)=\exp\left(\frac{N^2}{g^2}\sum_{l=1}^k\frac{t_l^+t_l^-}{l}\right)\,,
\end{equation}
and the planar free energy is therefore independent of $g$,
\begin{equation}
    F_0(g,t_l^\pm)=\sum_{l=1}^k\frac{t_l^+t_l^-}{l}\,.
\end{equation}
As shown in Ref.~\cite{Eniceicu:2023cxn}, in the large-$N$ limit, for $|z_1|>1$, $|z_2|<1$,
\begin{equation}
    \ev{e^{\Tr\log\left[(1-z_1^{-1}U)^2(1-z_2U^{-1})^2\right]}}\approx\frac{1}{(1-z_2/z_1)^4}\exp\left[-2 \frac{N}{g}\sum_{l=1}^k\left(\frac{t_l^-}{l}z_1^{-l}+\frac{t_l^+}{l}z_2^l\right)\right]\,,\label{eqn:ungapped_correlator_id}
\end{equation}
up to one-loop.
The eigenvalue instanton locations are given by the common zeros of (\ref{eqn:Veffprime_strong_+}) and (\ref{eqn:Veffprime_strong_-}), i.e.~the locations $z^\star$ satisfying the relation
\begin{equation}
    g+\sum_{l=1}^k\left(t_l^+(z^{\star})^l+t_l^-(z^\star)^{-l}\right)=0\,.\label{eqn:ungapped_locations}
\end{equation}
There are $2k$ solutions to this equation and they come in pairs, with one location inside, and one outside the unit circle --- if $z^\star$ is a solution, so is $1/(z^\star)^\dagger$.
\subsubsection{The instanton pair contributions}
The leading nonperturbative contributions to the partition function correspond to removing two eigenvalues, $z_1$ and $z_2$, from the cut and placing them at locations $\zeta^\star_1$ and $\zeta^\star_2$ satisfying (\ref{eqn:ungapped_locations}), where they are integrated along their steepest descent contours. As explained in Ref.~\cite{Eniceicu:2023cxn}, the calculation should be done assuming one of the locations is inside the unit circle, while the other is outside. Without loss of generality, we will pick $|\zeta_1^\star|>1$ and $|\zeta_2^\star|<1$. Doing so will produce the contribution associated to the pair of a real positive-charge brane and a real negative-charge brane. However, once obtained, the answer can then be analytically extended beyond the original domain of definition of $\zeta_1^\star$ and $\zeta_2^\star$ to obtain ghost instanton contributions.

The leading-order pair contribution takes the form
\begin{align}
\label{Z11_strong}
    Z^{(1,1)}(N,g,t_l^\pm)
    &=\int_{C_1}\frac{\d z_1}{2\pi}\int_{C_2}\frac{\d z_2}{2\pi}\exp\left[\frac{N}{g}\left(W(z_1)+W(z_2)\right)\right]\times\frac{z_1^{N-4}}{z_2^N}(z_1-z_2)^2\nonumber\\
    &\qquad\qquad\times Z^{(0)}(N-2,g(1-2/N),t_l^\pm)\ev{e^{\Tr\log\left[(1-z_1^{-1}U)^2(1-z_2U^{-1})^2\right]}}_{(N-2,\,g(1-2/N))}\nonumber\\
    &\approx Z^{(0)}(N,g,t_l^\pm)\int_{C_1}\frac{\d z_1}{2\pi}\int_{C_2}\frac{\d z_2}{2\pi}\exp\left[-\frac{N}{g}\left(V_{\text{eff}}^+(z_1;g)+V_{\text{eff}}^-(z_2;g)\right)\right]\frac{1}{(z_1-z_2)^2} \,,
\end{align}
where, to obtain the second equality, we used (\ref{eqn:ungapped_correlator_id}) as well as the fact that $Z^{(0)}(N-2,g(1-2/N),t_l^\pm)=Z^{(0)}(N,g,t_l^\pm)$, which is specific to the ungapped case.

The one-loop contribution therefore takes the form
\begin{equation}
    \frac{Z^{(1,1)}(N,g,t_l^\pm)}{Z^{(0)}(N,g,t_l^\pm)}\approx-\frac{g}{2\pi N}\frac{1}{\sqrt{-V_{\text{eff}}^+{''}(\zeta_1^\star;g)}\sqrt{-V_{\text{eff}}^-{''}(\zeta_2^\star;g)}}\frac{1}{(\zeta_1^\star-\zeta_2^\star)^2}e^{-\frac{N}{g}\left(V_{\text{eff}}^+(\zeta_1^\star;g)+V_{\text{eff}}^-(\zeta_2^\star;g)\right)}\,.
\end{equation}
There are four cases which can all be obtained immediately from the previous expression by plugging in the appropriate locations, $\zeta_1^\star$ and $\zeta_2^\star$:

\begin{itemize}
    \item  The contribution associated to the pair comprised of a real positive-charge brane and a real negative-charge brane --- obtained for $|\zeta_1^\star|>1$ and $|\zeta_2^\star|<1$,

\item The contribution associated to the pair comprised of a real positive-charge brane and a ghost negative-charge brane --- obtained for $|\zeta_1^\star|>1$ and $|\zeta_2^\star|>1$, as long as $\zeta_2^\star\neq\zeta_1^\star$,

\item The contribution associated to the pair comprised of a ghost positive-charge brane and a real negative-charge brane --- obtained for $|\zeta_1^\star|<1$ and $|\zeta_2^\star|<1$, as long as $\zeta_2^\star\neq\zeta_1^\star$,

\item The contribution associated to the pair comprised of a ghost positive-charge brane and a ghost negative-charge brane --- obtained for $|\zeta_1^\star|<1$ and $|\zeta_2^\star|>1$.
\end{itemize}

The one-loop determinant for $\zeta_1^\star=1/(\zeta_2^\star)^\dagger$ is associated with the exponential of a cylinder diagram with identical boundary conditions (either real-real or ghost-ghost), while the one-loop determinant for a case with both $\zeta_1^\star\neq\zeta_2^\star$ and $\zeta_1^\star\neq1/(\zeta_2^\star)^\dagger$ is associated with the exponential of a cross-cylinder. We will not comment on the more subtle case when $\zeta_1^\star=\zeta_2^\star$ which would correspond to a cylinder between a real brane and its ghost partner, but will instead refer the reader to \cite{Marino:2022rpz} and \cite{Schiappa:2023ned} where such cases were considered in the context of hermitian matrix integrals.
\subsection{The double-scaling limit}
The spectral curve of the unitary matrix integral in the ungapped phase is related to the derivative of the effective potential (\ref{eqn:Veffprime_strong_+}), (\ref{eqn:Veffprime_strong_-}),
\begin{equation}
    y^\pm(z)=-V_{\text{eff}}^\pm{'}(z;g)=\pm\left[\frac{g}{z}+\sum_{l=1}^k\left(t_l^+z^{l-1}+t_l^-z^{-l-1}\right)\right]\,.
\end{equation}
We will restrict to positive real coupling constants $t_l^\pm$. In order to reproduce the components of the spectral curves associated to the conformal background of the $(2,4k)$ minimal superstring theories \cite{Seiberg:2003nm}
\begin{equation}
    \Tilde{y}^\pm = \pm \i(-1)^k\, T_{2k}(\i x)\,,
\end{equation}
we will need to zoom in on the region near the point $z=-1$ where the gap closes,
\begin{equation}
    z=-1+2\i\varepsilon x\,,
\end{equation}
and tune the couplings $t_l^\pm$ accordingly around their values at the $k$th multi-critical point \cite{Oota:2021qky},
\begin{equation}
    t_l^\pm = \frac{g(k!)^2}{(k+l)!(k-l)!}\left(1+\sum_{m=1}^ka_{l,m}\varepsilon^{2m}\right)\, .
\end{equation}
Here, $\varepsilon$ is the expansion parameter associated to double-scaling, and $a_{l,m}$ are finite constants which can be fixed by requiring that
\begin{equation}
    y^\pm(z)=2\i g{\binom{2k}{k}}^{-1}\varepsilon^{2k}\Tilde{y}^\pm(x)+O(\varepsilon^{2k+1})\, .\label{eqn:spectral_id_ungapped}
\end{equation}
The double-scaling limit involves taking $N\rightarrow\infty$ and $\varepsilon\rightarrow0$, while holding fixed the quantity
\begin{equation}
    \dsfixedquantity={\binom{2k}{k}}^{-1}N\varepsilon^{2k+1}\,.
\end{equation}
In the dual $(2,4k)$ minimal superstring theory, the tensions of the ZZ branes comprising the pair under consideration are given by
\begin{equation}
    \mathcal{T}^\pm_{\zeta_i^\star}=\frac{N}{g}V_{\text{eff}}^\pm(\zeta_i^\star;g)\,,\qquad i\in\{1,2\}\,,
\end{equation}
while the exponential of the sum of cylinder diagrams connecting the branes is given by
\begin{equation}
    \mathcal{N}_{\zeta_1^\star,\zeta_2^\star}=-\frac{g}{2\pi N}\frac{1}{\sqrt{-V_{\text{eff}}^+{''}(\zeta_1^\star;g)}\sqrt{-V_{\text{eff}}^-{''}(\zeta_2^\star;g)}}\frac{1}{(\zeta_1^\star-\zeta_2^\star)^2}\,.
\end{equation}
The tensions can be obtained as follows,
\begin{align}
    \mathcal{T}^\pm_{\zeta_i^\star}&=\frac{N}{g}\int_{-1}^{\zeta_i^\star}\d z\,V_{\text{eff}}^\pm{'}(z;g)=\pm 4\i\dsfixedquantity(-1)^k\int_{0}^{x_i^\star}\d x\,T_{2k}(\i x)\nonumber\\
    &=\pm 4\dsfixedquantity(-1)^k\left(\frac{2k}{4k^2-1}T_{2k+1}(\i x_i^\star)-\frac{\i x_i^\star}{2k-1}T_{2k}(\i x_i^\star)\right)\,.
\end{align}
The extrema of the effective potential correspond to the zeros of the Chebyshev polynomial $T_{2k}(x)$,
\begin{equation}
    x_n^{\star\pm}=\pm\i\sin\frac{\pi n}{4k}\,,\qquad n\in\{1,3,\dots,2k-1\}\,,
\end{equation}
and we note that in the double-scaling limit, the symmetry, $z\rightarrow1/(z^\dagger)$ becomes $x\rightarrow x^\dagger$. The four types of branes therefore correspond to the following locations and have the following tensions:

\begin{itemize}
    \item 
The positive-charge real branes:
\begin{equation}
    x_n=+\i\sin\frac{\pi n}{4k}\,,\qquad\mathcal{T}_n=\frac{8k}{4k^2-1}(-1)^{(n+1)/2}\cos \frac{\pi n}{4k} \dsfixedquantity\, .
\end{equation}

\item The positive-charge ghost branes:
\begin{equation}
    x_{\overline{n}^\gh}=-\i\sin\frac{\pi n}{4k}\,,\qquad\mathcal{T}_{\overline{n}^\gh}=-\frac{8k}{4k^2-1}(-1)^{(n+1)/2}\cos \frac{\pi n}{4k}\dsfixedquantity\,.
\end{equation}

\item The negative-charge ghost branes:
\begin{equation}
    x_{n^\gh}=+\i\sin\frac{\pi n}{4k}\,,\qquad\mathcal{T}_{n^\gh}=-\frac{8k}{4k^2-1}(-1)^{(n+1)/2}\cos \frac{\pi n}{4k}\dsfixedquantity\,.
\end{equation}

\item The negative-charge real branes:
\begin{equation}
    x_{\overline{n}}=-\i\sin\frac{\pi n}{4k}\,,\qquad\mathcal{T}_{\overline{n}}=\frac{8k}{4k^2-1}(-1)^{(n+1)/2}\cos \frac{\pi n}{4k}\dsfixedquantity\,. 
\end{equation}
\end{itemize}

Thus, we see that the positive-charge real brane and negative-charge real brane associated with an index $n$ have the same tension, while the positive-charge ghost brane and negative-charge ghost brane associated with the same index $n$ have opposite tension compared to their real counterparts. The ratios of tensions of the real branes take the form
\begin{equation}
    \frac{\mathcal{T}_n}{\mathcal{T}_{1}}=(-1)^{(n-1)/2}\times\frac{\cos \frac{\pi n}{4k}}{\cos \frac{\pi}{4k}}\,.
\end{equation}
Since the second derivative of the effective potential evaluated at the instanton locations satisfies
\begin{align}
    -V_{\text{eff}}^\pm{''}(z_n^{\star+};g)&=\frac{g}{4\dsfixedquantity}{\binom{2k}{k}}^{-1}\varepsilon^{2k-1}\frac{4k^2-1}{\cos^2 \frac{\pi n}{4k}}\,\mathcal{T}_{x_n^{\star+}}^\pm\,,\\
    -V_{\text{eff}}^\pm{''}(z_n^{\star-};g)&=\frac{g}{4\dsfixedquantity}{\binom{2k}{k}}^{-1}\varepsilon^{2k-1}\frac{4k^2-1}{\cos^2 \frac{\pi n}{4k}}\,\mathcal{T}_{x_n^{\star-}}^\pm\,,
\end{align}
the normalization factors will be related to the corresponding tensions as follows,
\begin{equation}
    \mathcal{N}_{x_n^{\star\pm},x_{n'}^{\star\pm}}=\frac{1}{2\pi(4k^2-1)}\times\frac{\cos \frac{\pi n}{4k}\cos \frac{\pi n'}{4k}}{\left(x_n^{\star\pm}-x_{n'}^{\star\pm}\right)^2}\times\frac{1}{\sqrt{\mathcal{T}^+_{x_n^{\star\pm}}}\sqrt{\mathcal{T}^-_{x_{n'}^{\star\pm}}}}\,,
\end{equation}
and will take the following forms in the four cases of interest:

\begin{itemize}
    \item 

Real positive-charge brane $n$ and real negative-charge brane $\overline{n}'$:
\begin{equation}
    \mathcal{N}_{n,\overline{n}'}\,\mathcal{T}_n^{1/2}\,\mathcal{T}_{\overline{n}'}^{1/2}=-\frac{1}{2\pi(16k^2-4)}\times\frac{\cos \frac{\pi n}{4k}\cos \frac{\pi n'}{4k}}{\sin^2 \frac{\pi (n+n')}{8k}\cos^2 \frac{\pi(n-n')}{8k}}\,.
\end{equation}

For $n=n'$, this simplifies to
\begin{equation}
    \mathcal{N}_{n,\overline{n}}\,\mathcal{T}_n=-\frac{\cot^2\frac{\pi n}{4k}}{2\pi(16k^2-4)}\,.
\end{equation}

\item Real positive-charge brane $n$ and ghost negative-charge brane $n'{}^\gh$:
\begin{equation}
    \mathcal{N}_{n,n'{}^{\text{gh}}}\,\mathcal{T}_n^{1/2}\,\mathcal{T}_{n'{}^{\text{gh}}}^{1/2}=-\frac{1}{2\pi(16k^2-4)}\times\frac{\cos \frac{\pi n}{4k}\cos \frac{\pi n'}{4k}}{\cos^2 \frac{\pi (n+n')}{8k}\sin^2 \frac{\pi (n-n')}{4k}}\,.
\end{equation}

\item  Ghost positive-charge brane $\overline{n}^\gh$ and real negative-charge brane $\overline{n}'$:
\begin{equation}
    \mathcal{N}_{\overline{n}^{\text{gh}},\overline{n}'}\,\mathcal{T}_{\overline{n}^{\text{gh}}}^{1/2}\,\mathcal{T}_{\overline{n}'}^{1/2}=-\frac{1}{2\pi(16k^2-4)}\times\frac{\cos \frac{\pi n}{4k}\cos \frac{\pi n'}{4k}}{\cos^2 \frac{\pi (n+n')}{8k}\sin^2 \frac{\pi (n-n')}{4k}}\,.
\end{equation}

\item Ghost positive-charge brane $\overline{n}^\gh$ and ghost negative-charge brane $n'{}^\gh$:
\begin{equation}
    \mathcal{N}_{\overline{n}^{\text{gh}},n'{}^{\text{gh}}}\,\mathcal{T}_{\overline{n}^{\text{gh}}}^{1/2}\,\mathcal{T}_{n'{}^{\text{gh}}}^{1/2}=-\frac{1}{2\pi(16k^2-4)}\times\frac{\cos \frac{\pi n}{4k} \cos \frac{\pi n'}{4k}}{\sin^2 \frac{\pi (n+n')}{8k} \cos^2 \frac{\pi(n-n')}{8k} }\,.
\end{equation}

For $n=n'$, this simplifies to
\begin{equation}
    \mathcal{N}_{\overline{n}^{\text{gh}},n{}^{\text{gh}}}\,\mathcal{T}_{\overline{n}^{\text{gh}}}=-\frac{\cot^2\frac{\pi n}{4k}}{2\pi(16k^2-4)}\,.
\end{equation}
\end{itemize}

\section{Deriving Eq. (\ref{annprimeweak}) using the FZZT annulus}
\label{app:FZZT_wc}

The annulus between two FZZT branes labeled by momenta $P$ and $P'$ is \cite{Okuyama:2005rn, Irie:2007mp}
\footnote{The result in Ref.~\cite{Okuyama:2005rn} can be generalized to the $(2,4k)$ case by modifying their equation (2.9) to $\theta = \pi b P = \frac{\pi P}{2\sqrt{2k}}$.}
\begin{equation}
    A_{P, P'} 
    = - \log \left(\cosh \frac{\pi (P + P')}{2\sqrt{2k}}  \right) \, ,
\end{equation}
up to an additive constant which is ambiguous and convention-dependent in the double-scaling limit.
These factors and also the extra term in the boundary state argued for in Ref.~\cite{Seiberg:2004at} will cancel when we take combinations of FZZT branes to get the results for the ZZ brane.
This implies that
\begin{equation}
    A_{P_{1,n}, P_{1,n'} }  
    =  - \log \left(\sin  \frac{\pi (n+n')}{8k}  \right)  \, ,
\end{equation}
where the special momenta $P_{m,n}$ were defined in (\ref{defpmn}) and we recall that $b = \frac{1}{\sqrt{2k}}$.
We can take appropriate differences to get the ZZ annulus amplitude
\begin{equation}
\begin{split}
    A_{(1,n),(1,n')} 
    &= A_{P_{1,n} P_{1,n'}} - A_{P_{1,-n} P_{1,n'}} - A_{P_{1,n} P_{1,-n'}}
    + A_{P_{1,-n} P_{1,-n'}}
    = \log \left( \frac{\sin^2  \frac{\pi (n-n')}{8k}}{\sin^2  \frac{\pi (n+n')}{8k}} \right) \, .
\end{split}
\end{equation}
This matches the result in \eqref{annprimeweak}.

\section{R-Ishibashi contribution in the ungapped phase}
\label{app:rstrong}

In this appendix, we derive Eq.(\ref{arstrongresult}) for the R-Ishibashi contribution $A^\r$ in the ungapped phase.
Let us define
\begin{align}
    I_l := 
    \int_{0}^{\infty} \frac{\d t}{4 t} \sum_{j \in \mathbb{Z}} \Big( & q^{\frac{(8k j + 4k + l - 1)(8k j - l - 1)}{16 k}} + q^{\frac{(8k j + 4k + l + 1)(8k j - l + 1)}{16 k}} \nonumber \\
    &\qquad + q^{\frac{(8k j + 4k - l - 1)(8k j + l - 1)}{16 k}} + q^{\frac{(8k j + 4k - l + 1)(8k j + l + 1)}{16 k}} \Big) \, .
\end{align}
Then we can write $A^\r$ from (\ref{annrstrongdef}) as
\begin{equation}
\label{eq:ar_sum_il}
\begin{split}
    A^{\r} 
    &= 2 \sum_{\ell = |n-n'|+1, 2}^{n+n'-1} (-1)^{(l-n-n'+1)/2} I_{\ell} - \sum_{\ell=1, 2}^{2n-1} (-1)^{(l-1)/2} I_{\ell} - \sum_{\ell=1, 2}^{2n'-1} (-1)^{(l-1)/2} I_{\ell} \\
    &= 2 \sum_{\ell = n-n'+3, 2}^{n+n'-1} (-1)^{(l-n-n'+1)/2} I_{\ell} + \sum_{\ell=5, 2}^{2n-3} (-1)^{(l+1)/2} I_{\ell} + \sum_{l=5, 2}^{2n'-3} (-1)^{(l+1)/2} I_{\ell} \\
    & \qquad\qquad + 2(I_3 - I_1) + 2 I_{|n-n'|+1} - I_{2n-1} - I_{2n'-1} \, .
\end{split}
\end{equation}
Using the integral in \eqref{eq:int_exp_over_t}, we have for $l > 1$,
\begin{equation}
\begin{split}
    I_{l+2} - I_{l} 
    &= \frac{1}{4} \log \left[ \prod_{j \in \mathbb{Z}} \frac{(8k j + 4k + l - 1)(8k j - l + 1)(8k j + l - 1)(8k j + 4k - l + 1)}{(8k j - l - 3)(8k j + 4k + l + 3)(8k j + 4k - l - 3)(8k j + l + 3)} \right] \\
    &= \frac{1}{2} \log \left[ \frac{l-1}{l+3} \prod_{j \in \mathbb{Z}^+} \frac{(4k j)^2 - (l-1)^2}{(4k j)^2 - (l+3)^2} \right]
    = \frac{1}{2} \log \left[ \frac{\sin \frac{\pi(l-1)}{4k}}{\sin \frac{\pi(l+3)}{4k}} \right] \, .
\end{split}
\end{equation}
Moreover, it follows that for $l > 1$,
\begin{equation}
\begin{split}
    I_{l + 2p} - I_l
    &= \sum_{\ell = l, 2}^{l + 2p -2} \left( I_{\ell+2} - I_{\ell} \right) = \frac{1}{2} \log \left[ \frac{\sin \frac{\pi(l-1)}{4k} \sin \frac{\pi(l+1)}{4k}}{\sin \frac{\pi(l+2p-1)}{4k} \sin \frac{\pi(l+2p+1)}{4k}} \right] \, .
\end{split}
\end{equation}
We need to handle $(I_3 - I_1)$ separately because the $j=0$ terms in $I_1$ are problematic.
We can sum up the remaining terms
\begin{equation}
\begin{split}
    (I_{3} - I_{1})_{j \neq 0} 
    &= \frac{1}{4} \log \left[ \prod_{j \in \mathbb{Z}^*} \frac{(8k j + 4k)(8k j)(8k j)(8k j + 4k)}{(8k j - 4)(8k j + 4k + 4)(8k j + 4k - 4)(8k j + 4)} \right] \\
    &= \frac{1}{4} \log \left[ \prod_{j \in \mathbb{Z}^*} \frac{1}{(1 - \frac{(1/k)^2}{(2j)^2})(1 - \frac{(1/k)^2}{(2j+1)^2})} \right] 
    = \frac{1}{4} \log \left[ \frac{ \pi^2 \left(k^2 - 1 \right)}{k^4 \sin^2 \frac{\pi}{k}} \right] \, .
\end{split}
\end{equation}
We reorganize the $j=0$ terms in $(I_3 - I_1)$ as follows
\begin{equation}
\begin{split}
    \int_{0}^{\infty} \frac{\d t}{4 t} \Big( & q^{\frac{2k-1}{2k}}+q^{-\frac{2k+1}{2k}}+q^{\frac{k-1}{2 k}}+q^{-\frac{k+1}{2k}} \Big) - \int_{0}^{\infty} \frac{\d t}{4 t} \Big( q^{-1/2} + 2 + q^{1/2} \Big) \\
    &= \int_{0}^{\infty} \frac{\d t}{4 t} \Big( q^{-1/2} - 2 + q^{1/2} \Big) + \int_{0}^{\infty} \frac{\d t}{4 t} \Big( q^{\frac{2k-1}{2k}}+q^{-\frac{2k+1}{2k}}+q^{\frac{k-1}{2 k}}+q^{-\frac{k+1}{2k}} - 2q^{-1/2} - 2 q^{1/2} \Big) \\
    &= \int_{0}^{\infty} \frac{\d t}{4 t} \Big( q^{-1/2} - 2 + q^{1/2} \Big) + \frac{1}{4} \log \left[ \frac{k^4}{(4k^2-1)(k^2-1)} \right] \, .
\end{split}
\end{equation}
Thus,
\begin{equation}
    I_3 - I_1 = \frac{1}{4} \log \left[ \frac{ \pi^2 }{\left(4 k^2 - 1 \right) \sin^2 \frac{\pi}{k}} \right] + \int_{0}^{\infty} \frac{\d t}{4 t} \Big( q^{-1/2} - 2 + q^{1/2} \Big) \, .
\end{equation}

Using these results, we find that the three sums in \eqref{eq:ar_sum_il} telescope, so
\begin{equation}
\begin{split}
    A^{\r} 
    &= \log \left[ \frac{\sin \frac{\pi(|n-n'|+2)}{4k}}{ \sin \frac{\pi(n+n')}{4k}} \right] + \frac{1}{2} \log \left[ \frac{\sin \frac{\pi}{k}}{ \sin \frac{\pi (n-1)}{2k}} \right] + \frac{1}{2} \log \left[ \frac{\sin \frac{\pi}{k}}{ \sin \frac{\pi (n'-1)}{2k}} \right] \\
    & \hspace{0.5in} + \frac{1}{2} \log \left[ \frac{ \pi^2 }{\left(4 k^2 - 1 \right) \sin^2 \frac{\pi}{k}} \right] + \int_{0}^{\infty} \frac{\d t}{2 t} \Big( q^{-1/2} - 2 + q^{1/2} \Big) \\
    & \hspace{1in} + \frac{1}{2} \log \left[ \frac{\sin \frac{\pi (n-1)}{2k} \sin \frac{\pi n}{2k}}{\sin \frac{\pi|n-n'|}{4k} \sin \frac{\pi(|n-n'|+2)}{4k}} \right] + \frac{1}{2} \log \left[ \frac{\sin \frac{\pi (n'-1)}{2k} \sin \frac{\pi n'}{2k}}{\sin \frac{\pi|n-n'|}{4k} \sin \frac{\pi(|n-n'|+2)}{4k}} \right] \\
    &= \frac{1}{2} \log \left( \frac{\pi^2 \, \sin \frac{\pi n}{2k} \, \sin \frac{\pi n'}{2k}}{\left( 4 k^2-1 \right) \sin^2 \frac{\pi(n+n')}{4k} \, \sin^2 \frac{\pi(n-n')}{4k}} \right) + \int_{0}^{\infty} \frac{\d t}{2 t} \Big( q^{-1/2} - 2 + q^{1/2} \Big) \, .
\end{split}
\end{equation}
This is the claimed result in (\ref{arstrongresult}).

\section{General multi-instanton contribution in the ungapped phase}
\label{sec:appgeneralungapped}

In this appendix, we compute the normalization for a general multi-instanton configuration in the ungapped phase.
We will be rather brief, since the computation technique is, by now, familiar.
Consider the configuration consisting of $\ell_{\alpha}$ positively charged ZZ branes of type $(1,n_{\alpha})$, and $\ell_{\beta}$ negatively  charged ZZ branes of type $\overline{(1,n_{\beta})}$, where $\alpha$ and $\beta$ vary over the possible brane types. 
Charge conservation requires that
\begin{equation}
    \sum_{\alpha} \ell_{\alpha} = \sum_{\beta} \ell_{\beta} =: L \, .
\end{equation}

First, we perform the computation on the matrix integral side.
The generalization of (\ref{Z11_strong}) is
\begin{multline}
\label{ZLL_strong}
    Z^{(L,L)}(N,g,t_l^\pm)
    = \int_{C_i} \frac{\d z_i}{2\pi} \int_{C'_i} \frac{\d w_i}{2\pi} \, \exp\left[\frac{N}{g} \left( \sum_{i=1}^{L} W(z_i) + W(w_i) \right) \right] \times \frac{\prod_{i=1}^L z_i^{N-4 L}}{ \prod_{i=1}^L w_i^N} \times \prod_{i,j=1}^{L} (z_i-w_j)^2 \\
    \times \prod_{1 \leq i<j\leq L}(z_i-z_j)^2 (w_i-w_j)^2 \times Z^{(0)}(N-2L,g(1-2L/N),t_l^\pm) \\
    \times \ev{ e^{ \Tr \log \left[ \prod_{i=1}^{N} (1-z_i^{-1}U)^2 (1-w_i U^{-1})^2 \right]} }_{(N-2L, \, g(1-2L/N))} \, ,
\end{multline}
where we have taken $|z_i| > 1$ and $|w_i| < 1$.
Using \eqref{Z0_strong}, it follows that $Z^{(0)}(N-2L,g(1-2L/N),t_l^\pm) = Z^{(0)}(N,g,t_l^\pm)$.
Also, a calculation similar to that in Ref.~\cite{Eniceicu:2023cxn} gives us that, in the large-$N$ limit,
\begin{equation}
    \ev{ e^{ \Tr \log \left[ \prod_{i=1}^{N} (1 - z_i^{-1} U)^2 (1 - w_i U^{-1})^2 \right]} } 
    \approx \frac{1}{\prod_{i,j=1}^{L} (1-w_j / z_i)^4}  \exp \left[-2 \frac{N}{g} \sum_{l=1}^{k} \sum_{i=1}^{L} \left( \frac{t_l^-}{l} z_i^{-l} + \frac{t_l^+}{l} w_i^l \right) \right] \,,
\end{equation}
up to one-loop order.
Substituting these into (\ref{ZLL_strong}), 
\begin{align}
    \frac{Z^{(L,L)}(N,g,t_l^\pm)}{Z^{(0)}(N,g,t_l^\pm)}
    &\approx \int_{C_i} \frac{\d z_i}{2\pi} \int_{C'_i} \frac{\d w_i}{2\pi} \, \exp \left[-  \frac{N}{g}  \sum_{i=1}^{L} \left( V_{\eff}^{+}(z_i; g) + V_{\eff}^{-}(w_i; g) \right) \right] \nonumber \\
    &\hspace{1.2in} 
    \times \frac{\prod_{1 \leq i<j\leq L}(z_i-z_j)^2 (w_i-w_j)^2 }{ \prod_{i,j=1}^{L} (z_i-w_j)^2 } \, .
\end{align}
This result assumes that all the $z_i \, ,w_i$ are integrated along distinct contours, but if that is not the case, we need to include the appropriate symmetry factors.
We are interested in the configuration wherein after double-scaling, $\ell_{\alpha}$ of the $z_i$'s are integrated along the steepest descent contour passing through the saddle point 
\begin{equation}
\label{appe_xna}
    x_{n_{\alpha}} = + \i \sin \frac{\pi n_{\alpha}}{4k} \, ,
\end{equation}
and $\ell_{\beta}$ of the $w_i$'s are integrated along the steepest descent contour passing through the saddle point
\begin{equation}
\label{appe_xnb}
    x_{\overline{n_{\beta}}} = - \i \sin \frac{\pi n_{\beta}}{4k} \, .
\end{equation}
Recall that
\begin{align}
\label{appe_veff}
    V_{\eff}^+(x_n) = V_{\eff}^-(x_{\overline{n}}) 
    &= (-1)^{(n+1)/2} \frac{2k \kappa}{4k^2 - 1} \cos \frac{\pi n}{4k}\,,\\
\label{appe_veffpp}
    V_{\eff}^+{''}(x_n) = V_{\eff}^-{''}(x_{\overline{n}}) 
    &= (-1)^{(n+1)/2} 2k \kappa \sec \frac{\pi n}{4k} \, .
\end{align}
Including the symmetry factors and performing the Gaussian integrals around the saddle point using (\ref{gaussianintegral_n}), we get the final result
\begin{multline}
    \frac{Z^{(L,L)}(N,g,t_l^\pm)}{Z^{(0)}(N,g,t_l^\pm)} = \prod_{\alpha} \frac{G_2(\ell_{\alpha}+1)}{(2\pi)^{\ell_{\alpha}/2}} \frac{\exp \left[ - \ell_{\alpha} V^{+}_{\eff}(x_{n_{\alpha}}) \right]}{\left( {V^{+}_{\eff}{''} (x_{n_{\alpha}}) } \right)^{\ell_{\alpha}^2/2}} 
    \prod_{\beta} \frac{G_2(\ell_{\beta}+1)}{(2\pi)^{\ell_{\beta}/2}} \frac{\exp \left[- \ell_{\beta} V^{-}_{\eff}(x_{\overline{n_{\beta}}}) \right]}{\left( {V^{-}_{\eff}{''}(x_{\overline{n_{\beta}}})} \right)^{\ell_{\beta}^2/2}} \\
    \times \frac{ \prod_{\alpha \neq \alpha'} \left( x_{n_{\alpha}} - x_{n_{\alpha'}} \right)^{2 \ell_{\alpha} \ell_{\alpha'}} \prod_{\beta \neq \beta'} \left( x_{\overline{n_{\beta}}} -  x_{\overline{n_{\beta'}}} \right)^{2 \ell_{\beta} \ell_{\beta'}} }{\prod_{\alpha,\beta} \left( x_{n_{\alpha}} - x_{\overline{n_{\beta}}} \right)^{2 \ell_{\alpha} \ell_{\beta}} } \, ,  
\end{multline}
where $G_2(\ell+1) = \prod_{i=0}^{\ell-1} i!$ is the Barnes G-function.

The normalization $\cN$ is defined via the relationship
\begin{equation}
    \frac{Z^{(L,L)}(N,g,t_l^\pm)}{Z^{(0)}(N,g,t_l^\pm)} = \cN \exp \left[ - \sum_{\alpha} \ell_{\alpha} V^{+}_{\eff}(x_{n_{\alpha}}) - \sum_{\beta} \ell_{\beta} V^{-}_{\eff}(x_{\overline{n_{\beta}}}) \right] \, .
\end{equation}
Using (\ref{appe_xna})$-$(\ref{appe_veffpp}), the normalization is
\begin{align}
    \cN &= \prod_{\alpha} \frac{G_2(\ell_{\alpha}+1)}{(2 \pi)^{\ell_{\alpha}/2}} \left( \frac{1}{\sqrt{4 k^2 - 1}} \frac{\cos \frac{\pi n_{\alpha}}{4 k}}{\sqrt{V_{\eff}^+(x_{n_{\alpha}})}} \right)^{ \ell_{\alpha}^2}
    \times \prod_{\beta} \frac{G_2(\ell_{\beta}+1)}{(2 \pi)^{ \ell_{\beta}/2} }  \left( \frac{1}{\sqrt{4 k^2 - 1}} \frac{\cos \frac{\pi n_{\beta}}{4 k}}{\sqrt{V_{\eff}^-(x_{\overline{n_{\beta}}})}} \right)^{ \ell_{\beta}^2} \nonumber\\
    &\quad \times \frac{ \prod_{\alpha \neq \alpha'} \left( 2 \i \cos \frac{\pi(n_{\alpha} + n_{\alpha'})}{4k} \sin \frac{\pi(n_{\alpha} - n_{\alpha'})}{4k} \right)^{2 \ell_{\alpha} \ell_{\alpha'}} \prod_{\beta \neq \beta'} \left(2 \i  \cos \frac{\pi(n_{\beta} + n_{\beta'})}{4k} \sin \frac{\pi(n_{\beta} - n_{\beta'})}{4k} \right)^{2 \ell_{\beta} \ell_{\beta'}} }{\prod_{\alpha,\beta} \left(2 \i \sin \frac{\pi(n_{\alpha} + n_{\beta})}{4k} \cos \frac{\pi(n_{\alpha} - n_{\beta})}{4k} \right)^{2 \ell_{\alpha} \ell_{\beta}} } \nonumber \\   
    &= \prod_{\alpha} \frac{G_2(\ell_{\alpha}+1)}{(2 \pi)^{\ell_{\alpha}/2}} \left( \frac{\i}{\sqrt{16 k^2 - 4}} \frac{\cos \frac{\pi n_{\alpha}}{4 k}}{\sqrt{V_{\eff}^+(x_{n_{\alpha}})}} \right)^{ \ell_{\alpha}^2}
    \times \prod_{\beta} \frac{G_2(\ell_{\beta}+1)}{(2 \pi)^{ \ell_{\beta}/2} }  \left( \frac{\i}{\sqrt{16 k^2 - 4}} \frac{\cos \frac{\pi n_{\beta}}{4 k}}{\sqrt{V_{\eff}^-(x_{\overline{n_{\beta}}})}} \right)^{ \ell_{\beta}^2} \nonumber \\
    &\quad \times \frac{ \prod_{\alpha \neq \alpha'} \left( \cos^2 \frac{\pi(n_{\alpha} + n_{\alpha'})}{4k} \sin^2 \frac{\pi(n_{\alpha} - n_{\alpha'})}{4k} \right)^{\ell_{\alpha} \ell_{\alpha'}} \prod_{\beta \neq \beta'} \left( \cos^2 \frac{\pi(n_{\beta} + n_{\beta'})}{4k} \sin^2 \frac{\pi(n_{\beta} - n_{\beta'})}{4k} \right)^{\ell_{\beta} \ell_{\beta'}} }{\prod_{\alpha,\beta} \left( \sin^2 \frac{\pi(n_{\alpha} + n_{\beta})}{4k} \cos^2 \frac{\pi(n_{\alpha} - n_{\beta})}{4k} \right)^{\ell_{\alpha} \ell_{\beta}} } \, . 
    \label{appe_norm_mm}
\end{align}

Now we proceed to the string theory computation.
The contribution from the NS-Ishibashi states is
\begin{multline}
    A^{\ns} = \frac{1}{2} \sum_{\alpha} \ell_{\alpha}^2 A^{\ns}_{(1,n_{\alpha}),(1,n_{\alpha})} + \frac{1}{2} \sum_{\beta} \ell_{\beta}^2 A^{\ns}_{(1,n_{\beta}),(1,n_{\beta})} \\
    + \sum_{\alpha \neq \alpha'} \ell_{\alpha} \ell_{\alpha'} A^{\ns}_{(1,n_{\alpha}),(1,n_{\alpha'})} + \sum_{\beta \neq \beta'} \ell_{\beta} \ell_{\beta'} A^{\ns}_{(1,n_{\alpha}),(1,n_{\beta})} + \sum_{\alpha, \beta} \ell_{\alpha} \ell_{\beta} A^{\ns}_{(1,n_{\alpha}),(1,n_{\beta})} \, .  
\end{multline}
Using (\ref{logNwithdivnnbar}) and (\ref{znsnnprime}), we get
\begin{multline}
    A^{\ns} = \frac{1}{2} \log \left[ \prod_{\alpha} \left( \frac{\pi}{\sqrt{16 k^2 - 4}} \cot \frac{\pi n_{\alpha}}{4 k} \right)^{\ell_{\alpha}^2}\prod_{\beta} \left( \frac{\pi}{\sqrt{16 k^2 - 4}} \cot \frac{\pi n_{\beta}}{4 k} \right)^{\ell_{\beta}^2} \right] \\
    + \frac{1}{2} \log \left[ \prod_{\alpha \neq \alpha'} \left( \frac{\cot^2 \frac{\pi(n_{\alpha} + n_{\alpha'})}{4k}}{\cot^2 \frac{\pi(n_{\alpha} - n_{\alpha'})}{4k}} \right)^{\ell_{\alpha} \ell_{\alpha'}} \prod_{\beta \neq \beta'} \left( \frac{\cot^2 \frac{\pi(n_{\beta} + n_{\beta'})}{4k}}{\cot^2 \frac{\pi(n_{\beta} - n_{\beta'})}{4k}} \right)^{\ell_{\beta} \ell_{\beta'}} \prod_{\alpha,\beta} \left( \frac{\cot^2 \frac{\pi(n_{\alpha} + n_{\beta})}{4k}}{\cot^2 \frac{\pi(n_{\alpha} - n_{\beta})}{4k}} \right)^{\ell_{\alpha} \ell_{\beta}} \right] \\
    + \sum_{\alpha} \frac{\ell_{\alpha}^2}{2} \int_{0}^{\infty} \frac{\d t}{2 t} \left( q^{-1/2} - 2 + q^{1/2} \right) + \sum_{\beta} \frac{\ell_{\beta}^2}{2} \int_{0}^{\infty} \frac{\d t}{2 t} \left( q^{-1/2} - 2 + q^{1/2} \right) \, .  
\end{multline}
The contribution from the R-Ishibashi state is
\begin{multline}
    A^{\r} = \frac{1}{2} \sum_{\alpha} \ell_{\alpha}^2 A^{\r}_{(1,n_{\alpha}),(1,n_{\alpha})} + \frac{1}{2} \sum_{\beta} \ell_{\beta}^2 A^{\r}_{(1,n_{\beta}),(1,n_{\beta})} \\
    + \sum_{\alpha \neq \alpha'} \ell_{\alpha} \ell_{\alpha'} A^{\r}_{(1,n_{\alpha}),(1,n_{\alpha'})} + \sum_{\beta \neq \beta'} \ell_{\beta} \ell_{\beta'} A^{\r}_{(1,n_{\alpha}),(1,n_{\beta})} -\sum_{\alpha, \beta} \ell_{\alpha} \ell_{\beta} A^{\r}_{(1,n_{\alpha}),(1,n_{\beta})} \, .  
\end{multline}
A computation similar to that in appendix \ref{app:rstrong} gives us
\begin{multline}
    A^{\r} = \frac{1}{2} \log \left[ \prod_{\alpha} \left( \frac{2 \pi}{\sqrt{16 k^2 - 4}} \sin \frac{\pi n_{\alpha}}{2 k} \right)^{\ell_{\alpha}^2}\prod_{\beta} \left( \frac{2 \pi}{\sqrt{16 k^2 - 4}} \sin \frac{\pi n_{\beta}}{2 k} \right)^{\ell_{\beta}^2} \right] \\
    + \frac{1}{2} \log \left[ \frac{ \prod_{\alpha \neq \alpha'} \left( \sin^2 \frac{\pi(n_{\alpha} + n_{\alpha'})}{2k} \sin^2 \frac{\pi(n_{\alpha} - n_{\alpha'})}{2k} \right)^{\ell_{\alpha} \ell_{\alpha'}} \prod_{\beta \neq \beta'} \left( \sin^2 \frac{\pi(n_{\beta} + n_{\beta'})}{2k} \sin^2 \frac{\pi(n_{\beta} - n_{\beta'})}{2k} \right)^{\ell_{\beta} \ell_{\beta'}} }{\prod_{\alpha,\beta} \left( \sin^2 \frac{\pi(n_{\alpha} + n_{\beta})}{2k} \sin^2 \frac{\pi(n_{\alpha} - n_{\beta})}{2k} \right)^{\ell_{\alpha} \ell_{\beta}} } \right] \\
    + \sum_{\alpha} \frac{\ell_{\alpha}^2}{2} \int_{0}^{\infty} \frac{\d t}{2 t} \left( q^{-1/2} - 2 + q^{1/2} \right) + \sum_{\beta} \frac{\ell_{\beta}^2}{2} \int_{0}^{\infty} \frac{\d t}{2 t} \left( q^{-1/2} - 2 + q^{1/2} \right) \, .  
\end{multline}
Combining the NS- and R-sector contributions, we get
\begin{multline}
    A = \log \left[ \prod_{\alpha} \left( \frac{\pi}{\sqrt{16 k^2 - 4}} \cos \frac{\pi n_{\alpha}}{4 k} \right)^{\ell_{\alpha}^2}\prod_{\beta} \left( \frac{\pi}{\sqrt{16 k^2 - 4}} \cos \frac{\pi n_{\beta}}{4 k} \right)^{\ell_{\beta}^2} \right] \\
    + \log \left[ \frac{ \prod_{\alpha \neq \alpha'} \left( \cos^2 \frac{\pi(n_{\alpha} + n_{\alpha'})}{4k} \sin^2 \frac{\pi(n_{\alpha} - n_{\alpha'})}{4k} \right)^{\ell_{\alpha} \ell_{\alpha'}} \prod_{\beta \neq \beta'} \left( \cos^2 \frac{\pi(n_{\beta} + n_{\beta'})}{4k} \sin^2 \frac{\pi(n_{\beta} - n_{\beta'})}{4k} \right)^{\ell_{\beta} \ell_{\beta'}} }{\prod_{\alpha,\beta} \left( \sin^2 \frac{\pi(n_{\alpha} + n_{\beta})}{4k} \cos^2 \frac{\pi(n_{\alpha} - n_{\beta})}{4k} \right)^{\ell_{\alpha} \ell_{\beta}} } \right] \\
    + \sum_{\alpha} \ell_{\alpha}^2 \int_{0}^{\infty} \frac{\d t}{2 t} \left( q^{-1/2} - 2 + q^{1/2} \right) + \sum_{\beta} \ell_{\beta}^2 \int_{0}^{\infty} \frac{\d t}{2 t} \left( q^{-1/2} - 2 + q^{1/2} \right) \, .  
\end{multline}
In this case, the string field theory replacement rule for the fermionic zero modes gets modified to account for the Chan-Paton factors and the volume of the rigid $U(\ell)$ gauge group \cite{Eniceicu:2022dru, Sen:2021jbr}. 
The result is
\begin{align}
    \exp \left[ \ell^2 \int_{0}^{\infty} \frac{\d t}{2 t} \left(  q^{-1/2} - 2 + q^{1/2} \right) \right] 
    &\to (2\i)^{\ell^2} \cdot (2\sqrt{\pi})^{\ell^2} \cdot \frac{G_2(\ell+1)}{(2\pi)^{(\ell^2 + \ell)/2} 2^{\ell^2}} \cdot (2 \pi^2 T)^{-\frac{\ell^2}{2}} \\ 
    &= \left(\frac{\i}{\pi \, T^\frac{1}{2}} \right)^{\ell^2} 
    \cdot \frac{G_2(\ell+1)}{(2 \pi)^{\ell/2}} \, .
\end{align}
Therefore, the normalization is
\begin{multline}
    \cN = \prod_{\alpha} \frac{G_2(\ell_{\alpha}+1)}{(2 \pi)^{\ell_{\alpha}/2}} \left( \frac{\i}{\sqrt{16 k^2 - 4}} \frac{\cos \frac{\pi n_{\alpha}}{4 k}}{T_{ (1,n_{\alpha}) }^{1/2}} \right)^{ \ell_{\alpha}^2} 
    \times \prod_{\beta} \frac{G_2(\ell_{\beta}+1)}{(2 \pi)^{ \ell_{\beta}/2} }  \left( \frac{\i}{\sqrt{16 k^2 - 4}} \frac{\cos \frac{\pi n_{\beta}}{4 k}}{T_{ (1,n_{\beta}) }^{1/2}} \right)^{ \ell_{\beta}^2}  \\
    \times \frac{ \prod_{\alpha \neq \alpha'} \left( \cos^2 \frac{\pi(n_{\alpha} + n_{\alpha'})}{4k} \sin^2 \frac{\pi(n_{\alpha} - n_{\alpha'})}{4k} \right)^{\ell_{\alpha} \ell_{\alpha'}} \prod_{\beta \neq \beta'} \left( \cos^2 \frac{\pi(n_{\beta} + n_{\beta'})}{4k} \sin^2 \frac{\pi(n_{\beta} - n_{\beta'})}{4k} \right)^{\ell_{\beta} \ell_{\beta'}} }{\prod_{\alpha,\beta} \left( \sin^2 \frac{\pi(n_{\alpha} + n_{\beta})}{4k} \cos^2 \frac{\pi(n_{\alpha} - n_{\beta})}{4k} \right)^{\ell_{\alpha} \ell_{\beta}} } \, .  
\end{multline}
This matches the matrix integral result in (\ref{appe_norm_mm}).
The result can be generalized to the case where we also have ghost instantons; in practice, all one does is replace $n_{\alpha} \to -n_{\alpha}$ whenever there is a ghost.

\small
\bibliographystyle{apsrev4-1long}
\bibliography{main}

\end{document}